\RequirePackage{etex}
\documentclass{aa}  
\usepackage[version=4]{mhchem}
\usepackage{amsmath}
\usepackage[labelfont=bf]{caption}
\usepackage{subcaption}        
\usepackage{amssymb}
\usepackage{graphicx}
\usepackage{txfonts}
\usepackage{lipsum}
\usepackage{makecell}
\usepackage{tabularx}
\usepackage{array}
\usepackage{booktabs}
\usepackage{afterpage}
\usepackage{stfloats}
\usepackage{threeparttable}
\usepackage{soul}
\usepackage{xcolor}
\sethlcolor{yellow}
\setcellgapes{5pt}
\usepackage{tikz}
\usepackage{subcaption}         
\usepackage{lscape}             
\usepackage{placeins}          
\usepackage{natbib}

\begin{document}

   \title{Sulfur chemistry in temperate N$_2$-dominated exoplanetary atmospheres investigated through experimental simulations: \\ Volatile characterization}

   \author{L. Maratrat\inst{1}\corrauth{louis.maratrat@latmos.ipsl.fr}        
        \and A.Y. Jaziri \inst{1}\email{yassin.jaziri@latmos.ipsl.fr}
        \and L. Vettier\inst{1}\email{ludovic.vettier@latmos.ipsl.fr}
        \and A. Chatain\inst{1}\email{audrey.chatain@latmos.ipsl.fr}
        \and M. Millan\inst{1}\email{maeva.millan@latmos.ipsl.fr}
        \and N. Carrasco\inst{1,2}\email{nathalie.carrasco@latmos.ipsl.fr}}

   \institute{LATMOS/IPSL, UVSQ Université Paris-Saclay, Sorbonne Université, CNRS, Guyancourt France
   \and ENS Paris-Saclay, 91190, Gif-sur-Yvette, France}

   \date{Received June 2026}

  \abstract 
   {}
   {This study addresses through an experimental approach an understudied aspect of sulfur atmospheric chemistry which is the coupled and complex reactivity existing between sulfur, carbon, and nitrogen elements. Experimental simulations have been used to better characterize the volatile products resulting from the coupling between these three elements and discuss the potential observability of such species. Sulfur chemistry appears as a source of diversified signatures which need to be characterized for the current and the future exploration of exoplanetary atmospheres.} 
   {SO$_2$ and H$_2$S reactivity was explored experimentally in plasma discharges using mixtures of either pure N$_2$ or composed of N$_2$ and CH$_4$. The volatile species produced in such conditions were analyzed and quantified by \textit{in situ} mass spectrometry.}
   {SO$_2$ reactivity in the presence of N$_2$ alone or combined with methane led to the detection of six sulfur species: H$_2$S, NHSO, S$_2$O, CS$_2$, OCS, and CH$_3$SH. Under the same conditions, the reactivity of H$_2$S resulted in the observation of CS$_2$, CS, and CH$_3$SH. The detection of these species highlights the need to establish a consistent photochemical network describing this sulfur chemistry. Furthermore, the study addresses the detectability of these sulfur compounds via transit spectroscopy. Among them, OCS and CS$_2$, exhibit several detectable bands in the mid-infrared, and constitute promising targets to probe sulfur active chemistry with space-telescopes. In contrast, the detection of the other sulfur volatiles observed appears more challenging at low spectral resolution. }
   {}

\authorrunning{L. Maratrat et al.}
   \titlerunning{Sulfur chemistry in N$_2$-dominated exoplanet atmospheres}
   \maketitle
\nolinenumbers

\section{Introduction}
Sulfur volatiles are widespread in the atmospheres of planetary bodies. In the solar system, significant concentrations of SO$_2$ up to ppm (part per millions) are found in the atmospheres of Venus \citep{marcq_composition_2018} and the Earth \citep{oppenheimer_sulfur_2011}. In extra-solar systems, signatures attributed to sulfur gases have also been observed with the James Webb Space Telescope (JWST) in a wide diversity of exoplanets. SO$_2$ bands have been detected in several bodies such as the Hot-Jupiters WASP-39 b \citep{rustamkulov_early_2023,powell_sulphur_2024} and WASP-15 b \citep{kirk_bowie-align_2025}, the Warm Neptunes WASP-107 b \citep{dyrek_so2_2024} and GJ-3470 b \citep{beatty_sulfur_2024}, the Warm Super-Neptune HAT-P-26 b \citep{gressier_jwst-tst_2025}, and the Super-Earth L98-59 d \citep{banerjee_atmospheric_2024}. Moreover, a band attributed to CS$_2$ has been identified in the transit spectrum of the  temperate Sub-Neptune TOI-270 d \citep{felix_competing_2025, holmberg_possible_2024}.
\par So far, several of these detections are considered to result from an active sulfur chemistry in these exoplanetary environments.  CS$_2$ observed in TOI-270 d could originate from H$_2$S chemistry in the deeper and unobservable part of the atmosphere \citep{felix_competing_2025}, whereas the abundances retrieved for SO$_2$ in the atmospheres of WASP-107 b  and WASP-39 b are too high to result solely from a thermochemical equilibrium \citep{dyrek_so2_2024,tsai_photochemically_2023}. These observations highlight the important role played by sulfur volatiles and sulfur chemistry in hot and temperate H$_2$ worlds probed by JWST.
\par To date, only H$_2$-rich atmospheres have been successfully characterized through transit spectroscopy \citep{ducrot_combined_2025,zieba_no_2023,maurel_constraints_2025}. However, these limited types of exoplanets analyzed is expected to quickly expand and diversify with the ongoing observations of the JWST, and the upcoming launches of space telescopes such as Plato \citep{rauer_plato_2025}, Ariel \citep{tinetti_chemical_2018}, or Nancy Grace Roman \citep{bartusek_nancy_2022}. Moreover, mission projects dedicated to exoplanets, such as Habitable World Observatory (HWO) \citep{feinberg_habitable_2024} or Large Interferometer For Exoplanets (LIFE) \citep{glauser_large_2024}, are emerging to prepare the future of exoplanetary space observation by the 2040s. Within the framework of these missions, secondary temperate atmospheres having a high-metallicity will be privileged targets in the search for habitable worlds and, among them, N$_2$-dominated atmospheres. Nevertheless, as suggested by the first observations of the JWST, sulfur chemistry can lead to numerous observable signatures in the infrared domain probed by transit spectroscopy. The consideration of sulfur chemistry in these N$_2$-dominated temperate atmospheres therefore appears relevant in the preparation of future observations.
\par In this article we adopt a forward exploration approach to investigate understudied aspects of sulfur chemistry in the context of N$_2$-dominated atmospheres. In such exoplanetary environments, the coupling between carbon, nitrogen, and sulfur chemistry may be particularly efficient and be an interesting source of diversity within the volatiles and the hazes composing the atmosphere. Our goal is therefore to better characterize the products resulting from this coupling and to consider the potential observable signatures in infrared spectroscopy resulting from the reactivity between these three elements.
\par In the literature previous experiments  investigated H$_2$S chemistry in very different kind of environments such as warm CO$_2$-dominated Super-Earth (800K) \citep{he_sulfur-driven_2020} or under conditions analogous to Archean Early-Earth \citep{reed_impact_2020,reed_trace_2022}. Former experiments also considered SO$_2$ reactivity in the Archean atmosphere \citep{dewitt_formation_2010}. All of these studies highlighted an efficient gas phase reactivity \citep{he_sulfur-driven_2020,reed_abiotic_2024}, and the production of complex organo-sulfur aerosols \citep{reed_impact_2020,reed_trace_2022,dewitt_formation_2010,vuitton_h_2021}. The analysis of both gas and solid phases revealed the presence of organic sulfur species, which attests to the existence of a coupling still poorly explored between the chemistry of sulfur, carbon, and nitrogen.
\par In this work we considered experimental conditions to study the coupling between C, N, and S elements. The reactivity between these three elements was investigated in  mixtures comprising three main reactants: N$_2$, the main component and nitrogen source; CH$_4$, a source of reduced carbon, and a sulfur input introduced as SO$_2$ or H$_2$S. The objective was not to experimentally simulate the global atmospheric sulfur chemistry existing in these N$_2$ secondary atmospheres, but rather to highlight new processes or pathways that could be active in such environments. The selected mixtures should therefore be considered as model conditions, used to further investigate the coupled reactivity of C, N, and S, which differs from thermodynamical equilibrium predictions for temperate secondary N$_2$-dominated atmospheres. In particular, mixtures comprising CH$_4$ and SO$_2$ are not realistic from a thermodynamic point of view, due to the high contrast between the redox states of these two species, which are unlikely to coexist alone in a given N$_2$-dominated atmosphere. However, they constitute an interesting analytical case to characterize an extended diversity of sulfur volatiles products and reaction pathways starting from a simple initial mixture.
The reactions were performed within plasma discharges, which initiate complex chemistry leading to the efficient formation of volatile compounds as well as aerosols containing heavier species. This article focuses specifically on the characterization of sulfur volatiles generated in these discharges. The detectability of these sulfur compounds in transit spectroscopy is also addressed by considering their main infrared absorption bands and comparing with the signatures of other common sulfur and non-sulfur-bearing volatiles.

\section{Methods}
\subsection{Experimental set-up and mixture compositions}
\subsubsection{VULCAIN reactor}
\begin{figure}[ht!]  
\centering
\includegraphics[width=\textwidth/2]{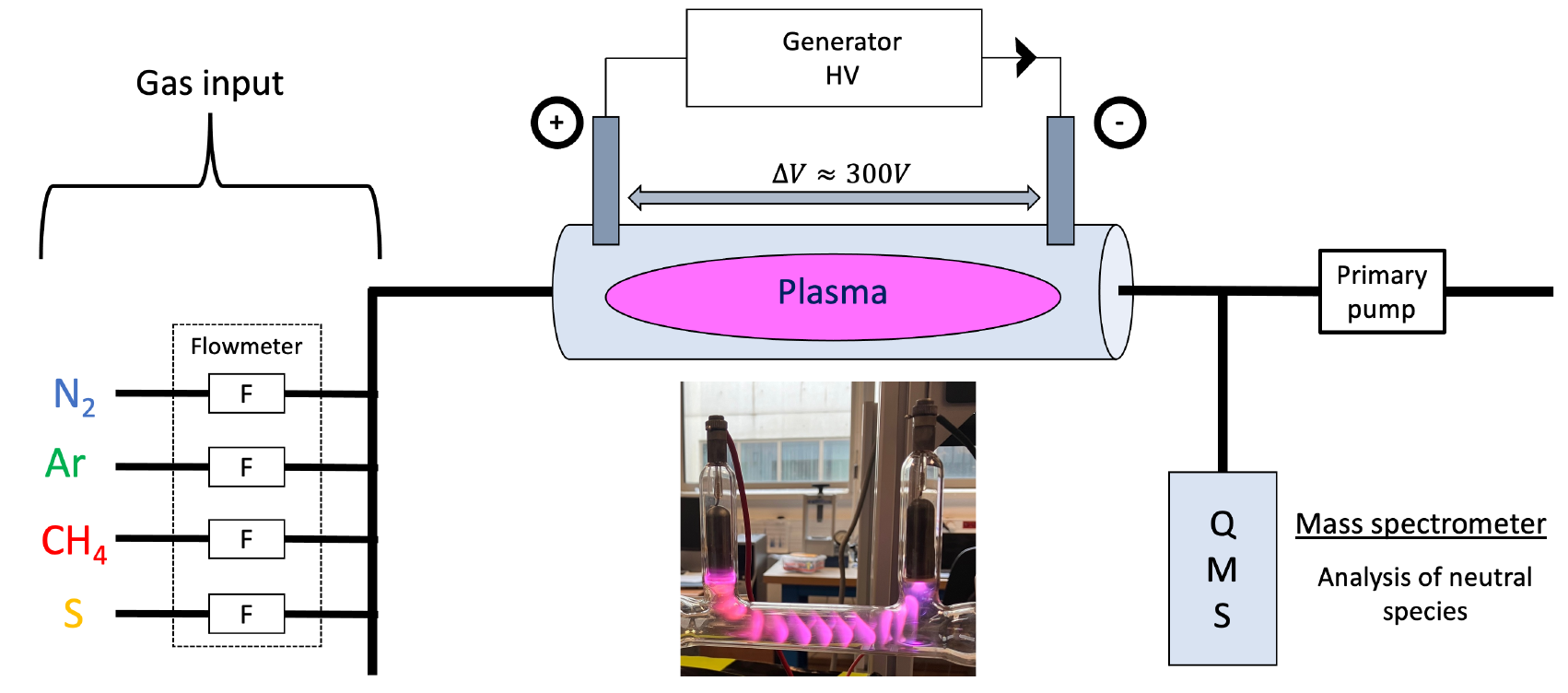}
\caption{Diagram and photograph of the VULCAIN DC plasma reactor.}
\label{fig1}
\end{figure}

To lead this study, we have designed a DC (Direct Current) cold plasma reactor called VULCAIN (acronym for Volcanic sULfur CAvity reactor for hIgh atmospheres of exoplaNets). The term cold plasma designates here a weakly ionized medium where the electron density is negligible in comparison to the molecular density of neutral stable species. This implies that elastic collisions between molecules and energetic electrons, which contribute to heat the gas, are quite unlikely. Cold plasmas are currently used as experimental analogs to reproduce the upper atmospheric chemistry of planetary bodies and are an efficient source of organic hazes \citep{cable_titan_2012}.
As shown in figure \ref{fig1}, VULCAIN is composed of a glass tube (22 mm inner diameter, 100 mm long) with two electrodes. The reactor is sealed with Teflon gaskets (Neyco glass-to-metal transitions). The electrodes are connected to a high voltage generator (JFA electronique ALBRDS 3000V-100mA) which is used to trigger and maintain the plasma discharge. Upstream of the plasma zone, mass flowmeters (MKS-Serie GE50A-100, MKS-Serie GM50A) control the input of gases (the “S” in the fourth gas line figure \ref{fig1} refers to sulfur gases, which are either SO$_2$ or H$_2$S, depending on conditions). The reactor works with a continuous input of gas which passes through the plasma zone and the gas flow is maintained by a primary pump located at the outlet of the reactor. Just before the pump, a Quadrupole Mass Spectrometer (QMS) (Pfeiffer QME 220) is inserted to detect the volatiles produced in the plasma. A secondary turbomolecular pump is also connected to the reactor (not shown in figure \ref{fig1}) to prevent atmospheric contamination and build-up of O$_2$/H$_2$O when no experiments are carried out. The residual vacuum under secondary pumping is typically 5×10$^{-7}$mbar. 

\vspace{-0.3cm}
For all of the mixtures studied in this work, the plasma parameters are identical. A total gas flow of 25 sccm (Standard Cubic Centimeter per Minute) ensures a pressure of 1 mbar inside the reactor. Finally, the intensity through the electrodes is set at 20 mA. 
\par The methodology applied to retrieve the energetic distribution of the electrons used to initiate the gas-phase chemistry inside the discharge is described in appendix A. This estimation is based on electric field measurement in the positive column of the discharge and previous results in pure N$_2$ plasmas \citep{cernogora_population_1981,loureiro_coupled_1986}. The addition of a moderate amount of gases relative to N$_2$ does not significantly influence the plasma properties \citep{chatain_characterization_2023}.
\vspace{-0.5cm}
\subsubsection{Experimental conditions: Gas mixture composition }
\begin{figure}[ht!]
\centering
\includegraphics[width=\textwidth/2]{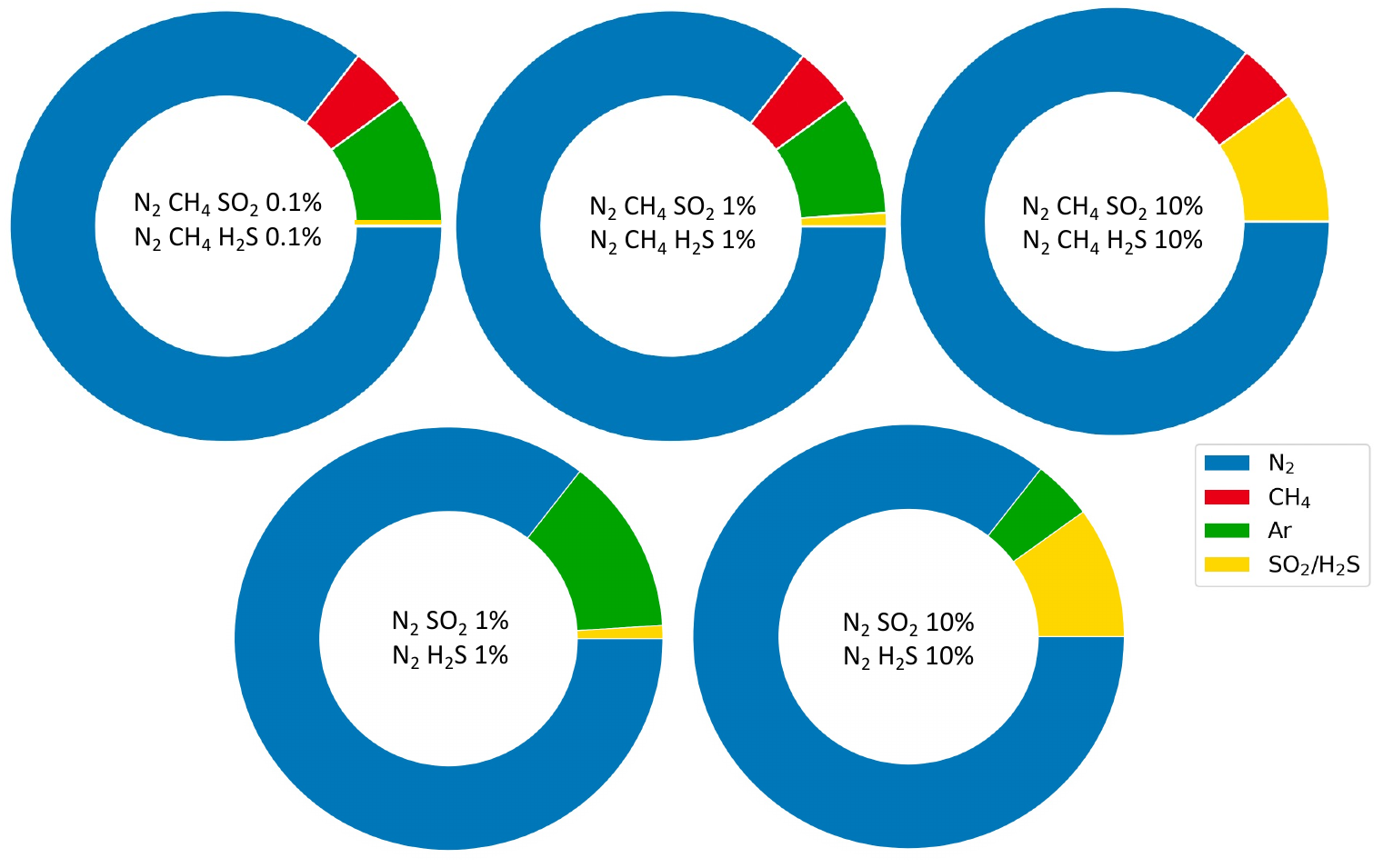}
\caption{Pie diagrams representing the initial composition of the main sulfur mixtures explored in this study described in Table 1.}
\label{fig2}
\end{figure}

To study sulfur chemistry in N$_2$ temperate atmospheres, we considered N$_2$-dominated mixtures composed of two different sulfur volatile inputs: H$_2$S or SO$_2$. Different orders of magnitude have been investigated for the initial proportion of sulfur gases in the discharge: 0.1 \%, 1 \%, and 10 \%. 0.1 \% sulfur reagent represents here an experimental limit below which sulfur volatiles produced in the plasma cannot be detected by \textit{in-situ} mass spectrometry. Moreover, to explore the possible coupling between carbon and sulfur volatiles, mixtures with an additional input of CH$_4$ have also been considered. In these mixtures including methane and sulfur gases, the ratio between N$_2$ and CH$_4$ is fixed and only the sulfur abundance varies. 
Finally, to modify the initial abundance of sulfur volatiles while keeping all other experimental parameters constant, namely the pressure, the total flowrate, and the partial N$_2$ and CH$_4$ flowrates, a variable input of Argon (Noble unreactive gas) was introduced into the mixture. Argon gradually replaces sulfur gases when their proportions decrease, so that the sum of argon and sulfur remains constant at 10 \% in abundance. 
The Table \ref{table:1} below summarizes the different mixtures performed and figure \ref{fig2} gives a graphic representation of these mixtures. The mixture names are composed as follows:\\
-	Argon is never mentioned as it does not influence the chemistry, so only the reactive species (N$_2$, CH$_4$, SO$_2$, or H$_2$S) are included. \\
-	The figure given after the list of reactive compounds corresponds to the percentage of sulfur, the percentages of N$_2$ and CH$_4$ are not included as they are identical in all the set of experiments.

\begin{table}[ht!]  
\centering    
\caption{Synthesis of the different mixtures considered}  
\label{table:1} 
\resizebox{\textwidth/2}{!}{%
\begin{tabular}{c c c c c c}   
\hline\hline              
Mixtures & N$_2$[\%] & CH$_4$[\%] & Ar[\%] & SO$_2$[\%] & H$_2$S[\%] \\         
\hline                      
   N$_2$-SO$_2$-10\% & 85.5 & 0 & 4.5 & 10 & 0  \\   
   N$_2$-SO$_2$-1\%& 85.5 & 0 & 13.5 & 1& 0 \\
   N$_2$-H$_2$S-10\%& 85.5 & 0 & 4.5 & 0& 10 \\
   N$_2$-H$_2$S-1\%& 85.5 & 0 & 13.5 & 0& 1 \\
   N$_2$-CH$_4$-SO$_2$-10\%& 85.5 & 4.5& 0 & 10& 0 \\
   N$_2$-CH$_4$-SO$_2$-1\%& 85.5 & 4.5& 9 & 1& 0 \\
   N$_2$-CH$_4$-SO$_2$-0.1\%& 85.5 & 4.5& 9.9 & 0.1& 0 \\
   N$_2$-CH$_4$-H$_2$S-10\%& 85.5 & 4.5& 0 & 0& 10 \\
   N$_2$-CH$_4$-H$_2$S-1\%& 85.5 & 4.5& 9 & 0& 1 \\
   N$_2$-CH$_4$-H$_2$S-0.1\%& 85.5 & 4.5& 9.9 & 0& 0.1 \\
\hline 
\end{tabular}
}
\end{table}

\par To interpret the results of the ten main mixtures of Table \ref{table:1}/Figure \ref{fig2} above and confirm the structure of specific volatiles, secondary mixtures composed of simpler combinations of the different reactants introduced were additionally considered. All secondary mixtures made for this study are specified in Table \ref{table:2} below, the experiment code names are built the same way as in Table \ref{table:1}. 

\begin{table}[ht!]  
\centering 
\caption{Synthesis of all the secondary mixtures tested} 
\label{table:2} 
\begin{threeparttable}
\resizebox{\textwidth/2}{!}{%
\begin{tabular}{c c c c c c c}   
\hline\hline              
Mixtures & N$_2$[\%] & CH$_4$[\%] & Ar[\%] & SO$_2$[\%] & H$_2$S[\%] & He[\%]\\      
\hline                      
   SO$_2$-10\%&0&0&90&10&0&0\\   
   SO$_2$-1\%& 0 & 0 & 99 & 1& 0 & 0 \\
   CH$_4$-SO$_2$-10\%& 0 & 4.5& 85.5 & 10& 0 & 0\\
   CH$_4$-SO$_2$-1\%& 0 & 4.5& 94.5 & 1& 0 & 0 \\
   CH$_4$-H$_2$S-10\%& 0 & 4.5& 16.5 & 0& 10 & 69\tnote{1} \\
   CH$_4$-H$_2$S-1\%& 0 & 4.5& 40 & 0& 1 & 54.5\tnote{1} \\
   Ref CH$_4$& 0 & 4.5& 15.5 & 0& 0 & 80\tnote{1} \\
   Ref N$_2$-CH$_4$& 85.5 & 4.5& 10 & 0& 0 & 0 \\
\hline
\end{tabular}
}
\begin{tablenotes}
\setlength{\parskip}{0pt}
\begin{minipage}{0.47\textwidth}
\scriptsize
\item[1] A controlled quantity of He was added to some mixtures to help to stabilize and maintain the plasma in the appropriated zone between the two electrodes. The He/Ar ratio chosen in this case minimizes the quantity of added He, as He discharges produce more energetic electrons\citep{zheng-de_electron_2002} causing a bias in the energy source.
\end{minipage}
\end{tablenotes}
\end{threeparttable} 
\end{table}
\par N$_2$, CH$_4$, Ar and He come from pure gas bottles with the following references: N$_2$ (Air liquide Alphagaz 2, Purity>99,9999 \%), Ar (Air liquide Alphagaz 2, Purity>99,9999 \%), CH$_4$ (Air liquide, N45, Purity >99,995 \%), He (Air liquide Alphagaz 2, Purity>99,9999 \%). Sulfur gases have two origins depending on the sulfur abundance of the mixture. For more concentrated mixtures (1 and 10 \%), sulfur comes from pure gas bottles: H$_2$S (Air liquide, N-23, Purity>99,3 \%) and SO$_2$ (Air liquide, N-30, Purity>99,9 \%). While for the more diluted conditions (0.1 \%), sulfur comes from pre-mixed bottles N$_2$-H$_2$S (Air liquide Mélange de gaz N$_2$-H$_2$S 1000 ppm (±100), purity unknown) and N$_2$-SO$_2$ (Air liquide Mélange de gaz N$_2$-SO$_2$ 1000 ppm (±100), purity unknown). 
\subsection{Analysis method: Mass spectrometry measurements}
\subsubsection{Measurement protocol}
Measurements are performed in the range between 1 and 100 mass units. Before each analysis, two blanks are performed: \\
-	one of the ionization chamber of the mass spectrometer alone. \\
-	one of the reactor at the same pressure of 1mbar filled with the dominant gas N$_2$, Ar or He depending on the conditions. 
\par These two blanks assess the residual contamination level of sulfur volatiles (SO$_2$ and H$_2$S) inside the reactor and the ionization chamber in order to check that these contributions are negligible in comparison to the sulfur content introduced for the experiment. 
For each condition, the data are composed of two measurements: first a reference spectrum is performed on the initial mixture without plasma discharge, and then a second spectrum is recorded at least five minutes after plasma ignition once the steady-state has been reached. Approximately 100 mass spectra are acquired for each measurement. In these different spectra, intensity is normalized using the mass of the main dominant gas which is either N$_2$, or Ar or He depending on the conditions. The mean intensities estimated respectively with and without plasma are then compared to identify the masses which show significant variations (superior to 2 times the standard deviation). The intensity of the mixture measured without plasma is taken as a reference. No increase in fluorinated fragments at m/z 19 was observed, confirming the absence of any detectable contamination induced by F$^+$ ions from Teflon gaskets. At least three replicates of the same experiment have been done for each mixture. Only the mean intensity calculated using all the different replicates are shown. 

\subsubsection{Sulfur isotopic ratio}
For a given sulfur species with a mass $m$, its isotopic ratio representing the proportion between $^{32}$S and $^{34}$S isotopes can be estimated thanks to the mean intensities I$_{m+2}$ and I$_{m}$ measured for the masses $m$ and $m+2$ respectively.  
\vspace{-0.2cm}
\begin{equation*}
    \frac{I_{\mathrm{m+2}}}{I_\mathrm{m}} = \frac{^{\mathrm{34}}S}{^\mathrm{32}S} \
\end{equation*}
\par At mass $m+2$, the contribution of isotopes of other elements (mainly $^{18}$O) can be neglected, due to the high natural abundance of $^{34}$S. We checked that the isotopic signatures found in SO$_2$ and H$_2$S bottles were consistent with the known sulfur natural abundance for $^{34}$S isotope \citep{thode_sulphur_1961}. 

\subsubsection{Sulfur volatiles quantification by deconvolution}
To quantify the amount of each sulfur volatile detected, the mass spectra for each N$_2$-CH$_4$-SO$_2$ and N$_2$-CH$_4$-H$_2$S mixtures have been deconvolved. The low mass fragments in the fragmentation pattern of each sulfur specie are indeed mixed with fragments of other volatiles and the deconvolution is needed to estimate the different contributions. Since the retrieved abundances depends on the volatile database chosen as an input, the overall approach must therefore be considered as semi quantitative providing only orders of magnitude for the targeted volatile mixing ratios.
\par This deconvolution is done using a mass spectra deconvolution code based on a Monte-Carlo approach which has been described previously \citep{gautier_decomposition_2020}. In brief, three phenomena are taken into account to estimate the abundances of sulfur volatiles: the mass-dependent transmission through the quadrupole, the ionization of species in the source which depends on their chemical properties, and the mass dependent diffusion through the micrometer hole used to adapt the pressure between the experiment and the ionization chamber.  The transmission response of the instrument with mass has been estimated using a calibration bottle of noble gases (He, Ne, Ar, Kr, Xe) (Messer). Ionization is corrected with total ionization cross section data at 70 eV taken from the NIST database and diffusion coefficients are calculated supposing a molecular regime which is a realistic assumption at the pressure of the mass spectrometer ($\approx5\times10^{-6}$ mbar) and the size of the micrometer hole ($50$ $\mu$$m$). Before the deconvolution, the instrumental noise is removed for each fragment, excepting those associated with the initial reactants, by subtracting the normalized OFF spectrum from the normalized ON spectrum.
Quantification performed on the initial mixtures has been done to validate the coherence of the deconvolution method. The results as well as the setting adopted for the deconvolution are detailed in appendix B. The appendix C shows the transmission curve as a function of mass, while the appendix D details the spectra resulting from the deconvolution and provides insights on the species selected.
\par NHSO and S$_2$O abundances could not be estimated by deconvolution with this Monte-Carlo approach due to the lack of data regarding their total ionization cross sections and their fragmentation patterns. That is why N$_2$-H$_2$S and N$_2$-SO$_2$ mixtures have not been considered for the deconvolution as no sulfur volatile products or only non-quantifiable sulfur volatile products have been observed under these conditions. 

\subsection{Molecule detectability: infrared absorption cross sections}
To further discuss the detectability in infrared spectroscopy of the sulfur products identified in our experiments, their absorption cross sections have been compared with those of numerous common volatiles. Absorption cross sections represent the absorptive contribution of a species for a given concentration and under defined pressure and temperature conditions. The relative comparison of these cross-sections generally allows the identification of volatile compounds likely to hide the spectral signatures of the sulfur species considered in the corresponding spectral range. Additionally, this comparison provides insights into the relative abundances required to mask these sulfur signatures. The paragraph below details the sources and the calculation methods used to estimate these cross sections. 

\subsubsection{NHSO detectability}
NHSO cross section has been estimated using Beer-Lambert law expressed as follows:
 \begin{equation*}
      A=log(\frac{I_0}{I_t}) = \frac{\sigma\times n \times l}{ln(10)} \Rightarrow 
      \sigma=\frac{A\times ln(10)}{n\times l}
   \end{equation*}
where A is the absorbance, $I_0$ the source intensity, $I_t$ the transmitted intensity, $\sigma$ the absorption cross section (cm$^2$), $n$ the molecular concentration (cm$^{-3}$), and $l$ the optical path length (cm). 
\par The NHSO cross section is derived from previous high-resolution absorbance measurements: \citep{joo_determination_1996}  for $\nu$$_2$ band, \citep{joo_analysis_1996} for $\nu$$_3$  band, and \citep{puskar_high-resolution_2006} for $\nu$$_4$  and $\nu$$_6$  bands. 
To assess its detectability, NHSO absorption cross section was compared with those of the following volatile species considered as relevant targets for temperate exoplanetary atmospheres: H$_2$S, SO$_2$, CH$_4$, CO, CO$_2$, H$_2$CO, CH$_3$OH, CH$_3$SH, DMS, OCS, CS$_2$, CS, H$_2$O, NH$_3$, NO, N$_2$O, NO$_2$, HCN, C$_2$H$_2$, C$_2$H$_4$, C$_2$H$_6$, C$_2$N$_2$, C$_4$H$_2$, HC$_3$N, CH$_3$CN. Halogen, and phosphorous species fall outside the scope of this study and were deliberately excluded. Additionally, more complex organic compounds were not considered in the absence of spectroscopic line-by-line parameters. However, the contributions of all these excluded compounds can reasonably be neglected, without affecting the conclusions of this study.
\par The cross sections of all the species considered, excepted CH$_3$SH and DMS, were calculated with Hitran line by line parameters, based on a Voigt profile which takes into account both Dopler and N$_2$ pressure broadening. A cut-off of 500 half-width is applied for each band, calculations are performed with a resolution of 0.001 cm$^{-1}$ inferior to Lorentzian and Dopler half-width. All cross sections were calculated at a pressure of 0.5 mbar and a temperature of 298 K to be consistent with NHSO data. For CH$_3$SH and DMS which are not included in Hitran line by line list, data were taken from Hitran cross section database and are given at 1 bar and 298 K. 
In absence of supplementary cross section measurements at other pressures and temperatures for NHSO, the influence of these two parameters on NHSO detectability could not be further explored in this study. 
\subsubsection{OCS and CS$_2$ detectability }
For CS$_2$ and OCS, cross sections were calculated as previously detailed for NHSO (resolution, cut-off, profile…) and compared with the following set of species: H$_2$S, SO$_2$, CH$_4$, CO, CO$_2$, H$_2$CO, CH$_3$OH, CH$_3$SH, DMS, CS, H$_2$O, NH$_3$, NO, N$_2$O, NO$_2$, HCN, C$_2$H$_2$, C$_2$H$_4$, C$_2$H$_6$, C$_2$N$_2$, C$_4$H$_2$, HC$_3$N, CH$_3$CN. NHSO was not included in this list in absence of spectroscopic data in the spectral ranges considered for OCS and CS$_2$ absorption bands. 
For the most relevant absorption bands of OCS and CS$_2$ (specifically the CS$_2$ $\nu$$_3$ band and the OCS $\nu$$_3$ band), absorption cross-sections were calculated at various pressure and temperature values to deeper investigate the influence of these two parameters on OCS/CS$_2$ detectability. The parameter space (Pressure, Temperature) was explored in greater detail for values relevant to temperate exoplanets: $(P,T)$ $\in$ $[0.001;100$ mbar]$\times$$[150;400$ K]. These calculations, and the methodological approach are detailed in Appendix E. 
\subsubsection{CH$_3$SH detectability}
CH$_3$SH absorption cross section is directly taken from the Hitran database and originates from previous measurements conducted in N$_2$ at 1 bar and 298 K \citep{sharpe_gas-phase_2004}. In absence of additional spectroscopic data, CH$_3$SH cross section is compared only at 1 bar and 298 K with the same species previously considered for OCS/CS$_2$.

\vspace{-0.3cm}
\section{Results}

\begin{figure*}[t]
\sidecaption
\includegraphics[width=12cm]{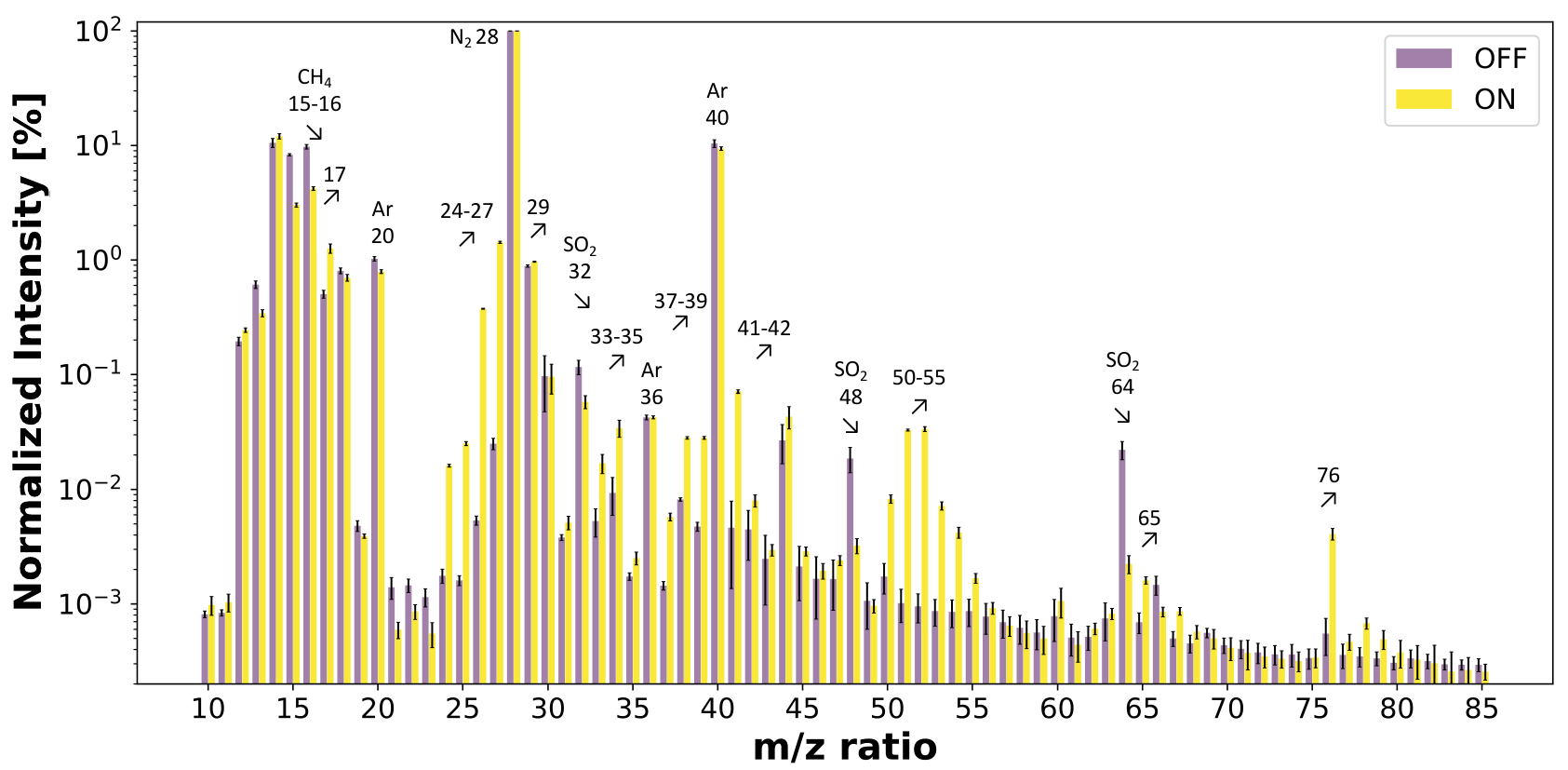}
\caption{Comparison of normalized mass spectra obtained before (purple) and after (yellow) plasma ignition for the N$_2$-CH$_4$-SO$_2$-0.1\% mixture without additional noise and transmission corrections. The arrows indicate the sense of variation of the corresponding masses.} 
\label{fig3}
\end{figure*}

Figure \ref{fig3} shows an example of the raw mass spectrometry data processed in this study. Comparison of the spectrum of the initial mixture (OFF) with the spectrum obtained after plasma ignition (ON) reveals numerous differences induced by the chemistry occurring in the gas phase of the plasma discharge. First, the decrease in intensity observed in the plasma spectrum “ON” figure \ref{fig3} for the main fragments of two reactants CH$_4$ $\{15,16\}$ and SO$_2$ $\{48,64\}$, highlights the consumption of these two species within the discharge. Regarding the third reactant, N$_2$, its consumption is too low to be measured by mass spectrometry due to the high stability and the large amount of this molecule. The quantity of N$_2$ can thus be considered as constant between the initial mixture and the plasma mixture. For this reason, N$_2$ is a relevant reference for signal normalization and for the deconvolution (see Appendix B).
\par Moreover, the increase in intensity observed on figure \ref{fig3} for several masses in the “ON” spectrum attests to the formation of numerous volatiles in the plasma. These increases are observed for the following m/z ratios in decreasing order: 76, 55 to 50, 42, 41, 39 to 37, 35 to 33, 27 to 24, and 17. The aim of this work is to identify the different species responsible of these mass peaks and among them more particularly sulfur-bearing molecules. The mass spectrometry data processed in each of the mixtures is provided in Appendix F. 
\vspace{-0.2cm}
\subsection{Identification of sulfur volatiles}
To discriminate sulfur-free volatiles from sulfur-bearing volatiles in the primary mixtures investigated (see Table \ref{table:1} and figure \ref{fig2}), the signals obtained in the presence and absence of sulfur (Ref-N$_2$-CH$_4$) were compared for each mass. Masses for which a significant increase is observed after plasma ignition only in the presence of sulfur are considered to be due to sulfur-bearing volatiles or their fragments. In addition, the use of secondary mixtures (SO$_2$-1/10\%, CH$_4$-H$_2$S-1/10\%, CH$_4$-SO$_2$-1/10\% mixtures) provide more information on the chemical structure of each volatile, highlighting the minimum combination of elements required for its formation.
\vspace{-0.3cm}
\subsubsection{Identification using secondary mixtures}
\begin{figure*}[t]
\centering
\includegraphics[width=0.95\textwidth]{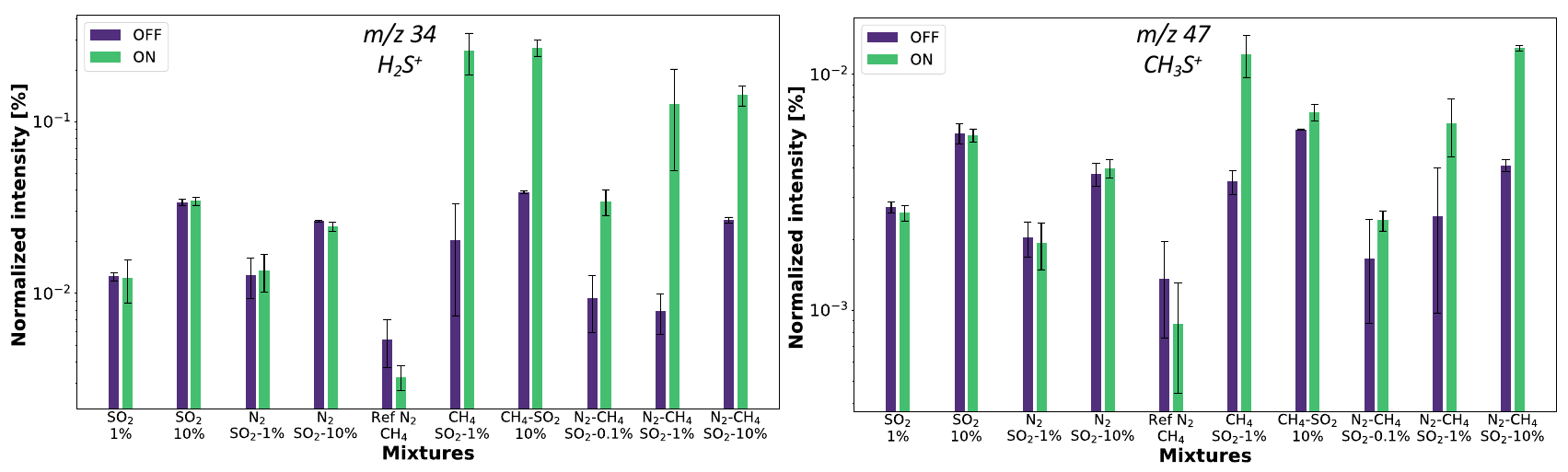}
\includegraphics[width=0.95\textwidth]{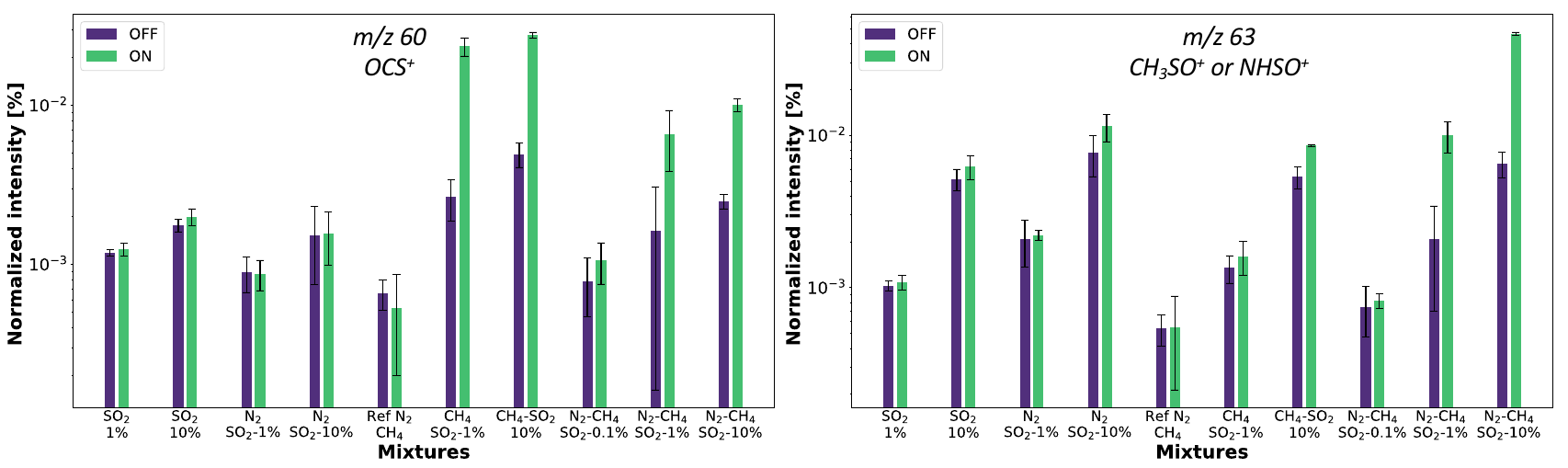}
\includegraphics[width=0.95\textwidth]{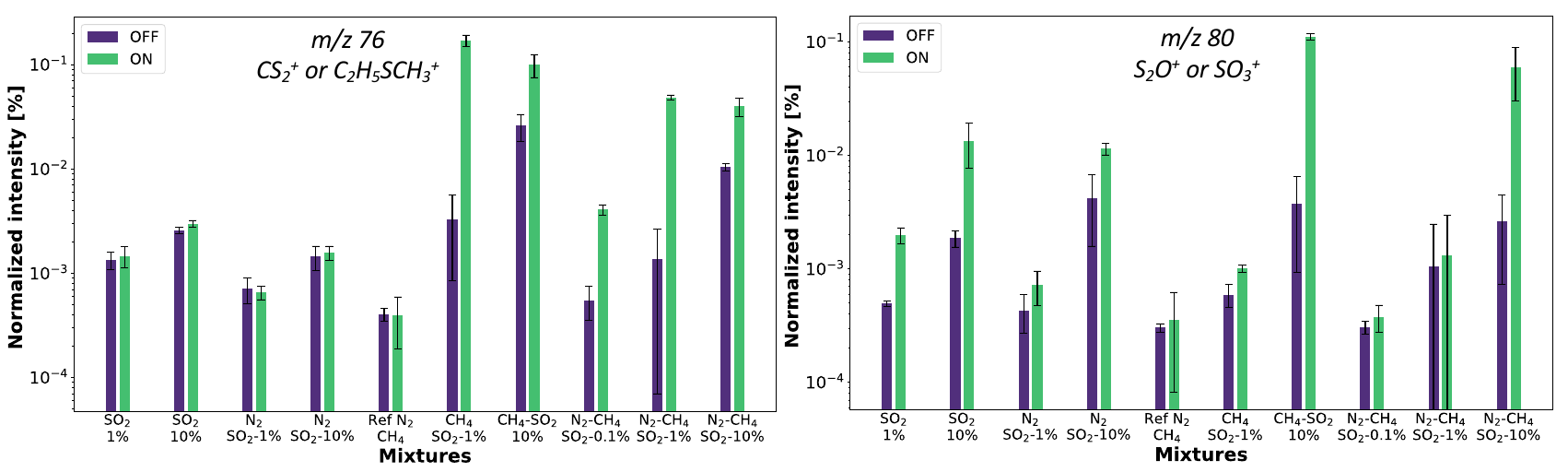}
\caption{Sulfur bearing fragments observed in SO$_2$ plasma discharges.}
\label{fig4}
\end{figure*}

{\hspace{0.6cm}\textit{i. SO$_2$ mixtures}\vspace{0.1cm}}\\
In N$_2$-CH$_4$-SO$_2$ mixtures, six mass fragments associated to sulfur volatile products were identified. They are highlighted figure \ref{fig4} by increasing order: we find	m/z 34, 47, 60, 63, 76, and 80. In N$_2$-SO$_2$ mixtures only the fragment at m/z 80 is detected. The presence of this fragment in SO$_2$-1\% and SO$_2$-10\% suggests that the associated volatile(s) is/are mainly formed by the reaction of SO$_2$ with itself. This is consistent with its/their detection(s) in the most concentrated N$_2$-SO$_2$-10\% and N$_2$-CH$_4$-SO$_2$-10\% mixtures only. Taking these remarks into account, two attributions are therefore possible for this/these volatile(s) with a nominal mass of 80: SO$_3$ or S$_2$O. The next paragraph will precise this m/z 80 attribution by using sulfur isotopes. 
For the other identified sulfur fragments, m/z 76, 63, 60, 47, and 34, they are all not observed under SO$_2$-1\%/SO$_2$-10\% and N$_2$-SO$_2$-1\%/N$_2$-SO$_2$-10\% conditions, which confirms that the corresponding volatiles contain either carbon or hydrogen in their structures. Thus, the following fragments can be proposed: 
-m/z 76 can be attributed to either CS$_2$ or C$_2$H$_5$SCH$_3$ (see next paragraph for discrimination), \\
-m/z 63 to CH$_3$SO (which is the main fragment of DMSO (CH$_3$SOCH$_3$) or the main fragment of CH$_3$SOH) or to NHSO (see next paragraph for discrimination), \\
-m/z 60 to OCS, \\
-m/z 47 to CH$_3$S the main fragment of CH$_3$SH, \\
-and m/z 34 to H$_2$S.\\

\begin{figure*}[t]
\centering
\includegraphics[width=0.95\textwidth]{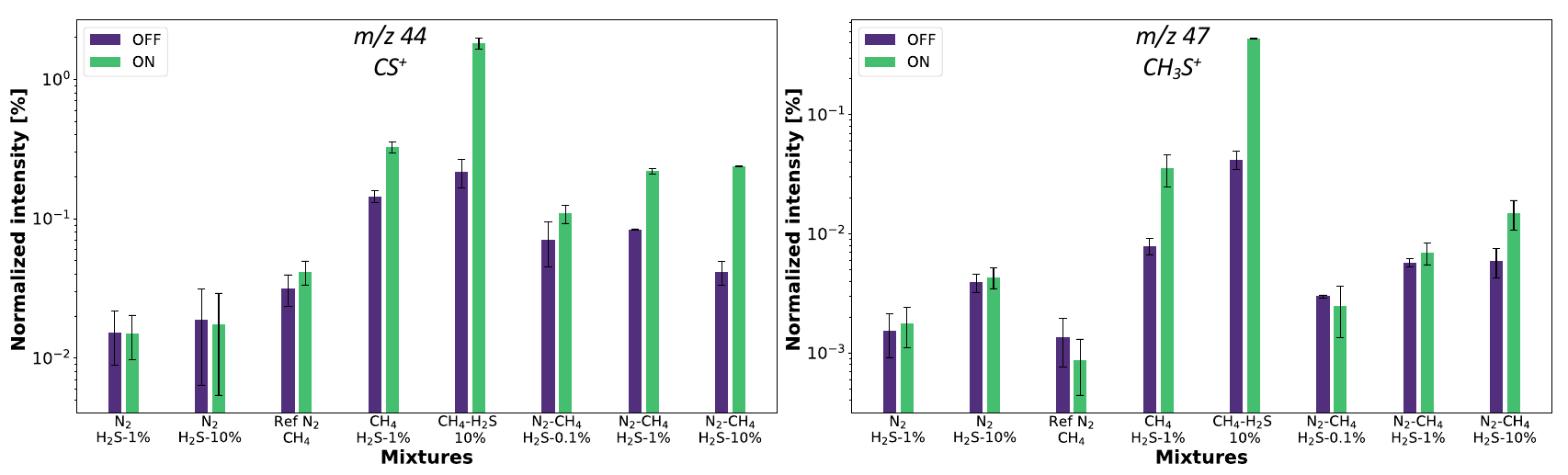}
\includegraphics[width=0.95\textwidth]{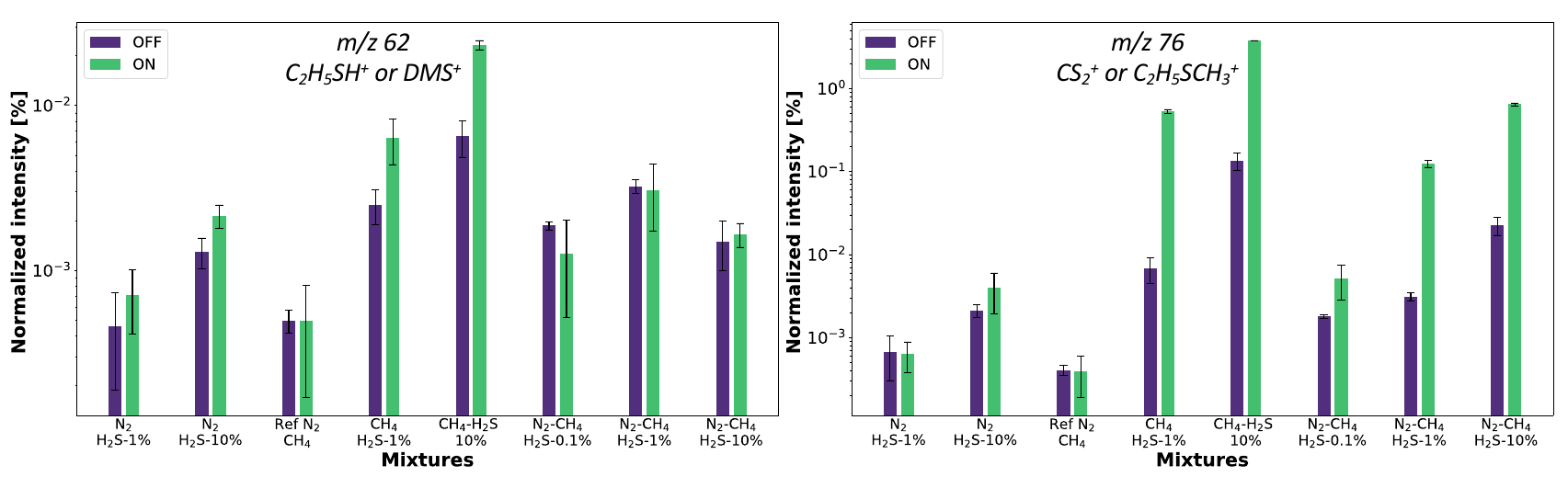}
\caption{Sulfur bearing fragments observed in H$_2$S plasma discharges.}
\label{fig5}
\end{figure*}

{\hspace{0.6cm}\textit{ii. H$_2$S mixtures}\vspace{0.1cm}}\\
In N$_2$-CH$_4$-H$_2$S mixtures, three fragments attributed to sulfur volatile products were identified as illustrated figure \ref{fig5}. These three masses are not observed in N$_2$-H$_2$S mixtures indicating that the corresponding volatiles are carbon-bearing volatiles. As suggested above for N$_2$-CH$_4$-SO$_2$ mixtures we can propose the following attributions for the fragments observed: \\
-	the fragment at m/z 76 could be due to either CS$_2$ or C$_2$H$_5$SCH$_3$.\\
-The fragment at m/z 47 reflects the presence of CH$_3$SH. As previously explained for SO$_2$ mixtures, m/z 47 is the main fragment in the CH$_3$SH fragmentation pattern. \\ 
- Finally, the mass peak at m/z 44 can be attributed to either CS or CS$_2$, CS$_2$ having a significant contribution at m/z 44 in its fragmentation pattern.
\par In addition, these different fragments are not systematically observed depending on the initial percentage of H$_2$S in the mixture. Fragment at m/z 76 is observed whatever the initial proportion of H$_2$S, while m/z 47 and m/z 44 are present only in the N$_2$-CH$_4$-H$_2$S-1\% and N$_2$-CH$_4$-H$_2$S-1\%/N$_2$-CH$_4$-H$_2$S-10\% gas mixtures respectively. 
\par Finally, we also looked for the presence of DMS (CH$_3$SCH$_3$) which main fragment is at m/z 62. It was not observed in any of the three N$_2$-CH$_4$-H$_2$S mixtures meaning that DMS is absent or below the detection limit for these experiments. Moreover, mass 62 was detected in CH$_4$-H$_2$S mixtures (see figure \ref{fig5}). However, this could be attributed to either DMS (CH$_3$SCH$_3$) or to ethanethiol (C$_2$H$_5$SH). The similarities between the two fragmentation patterns of these two function isomers make it difficult to discriminate them using mass spectrometry. 

\vspace{-0.3cm}
\subsubsection{Isotopic considerations}
The use of C and S isotopes has proven to be an effective way of resolving the degeneracy that exists in several of the previously formulated attributions (i.e., SO$_3$ vs S$_2$O, CS$_2$ vs C$_2$H$_5$SCH$_3$, and DMSO vs CH$_3$SOH vs NHSO). For the first two degeneracies we have used the natural abundance of the sulfur isotope $^{34}$S to determine the number of S elements in the corresponding volatile molecules. However, this approach using sulfur isotopes does not work for the last case where all possible attributions include only one sulfur. This is why for this specific case, we carried out complementary experiments with isotopically labeled methane $^{13}$CH$_4$ and CD$_4$ to discriminate the identification based on the number of constitutive C and H. 
\par
\begin{figure}[ht!]
\centering
\includegraphics[width=0.95\textwidth/2]{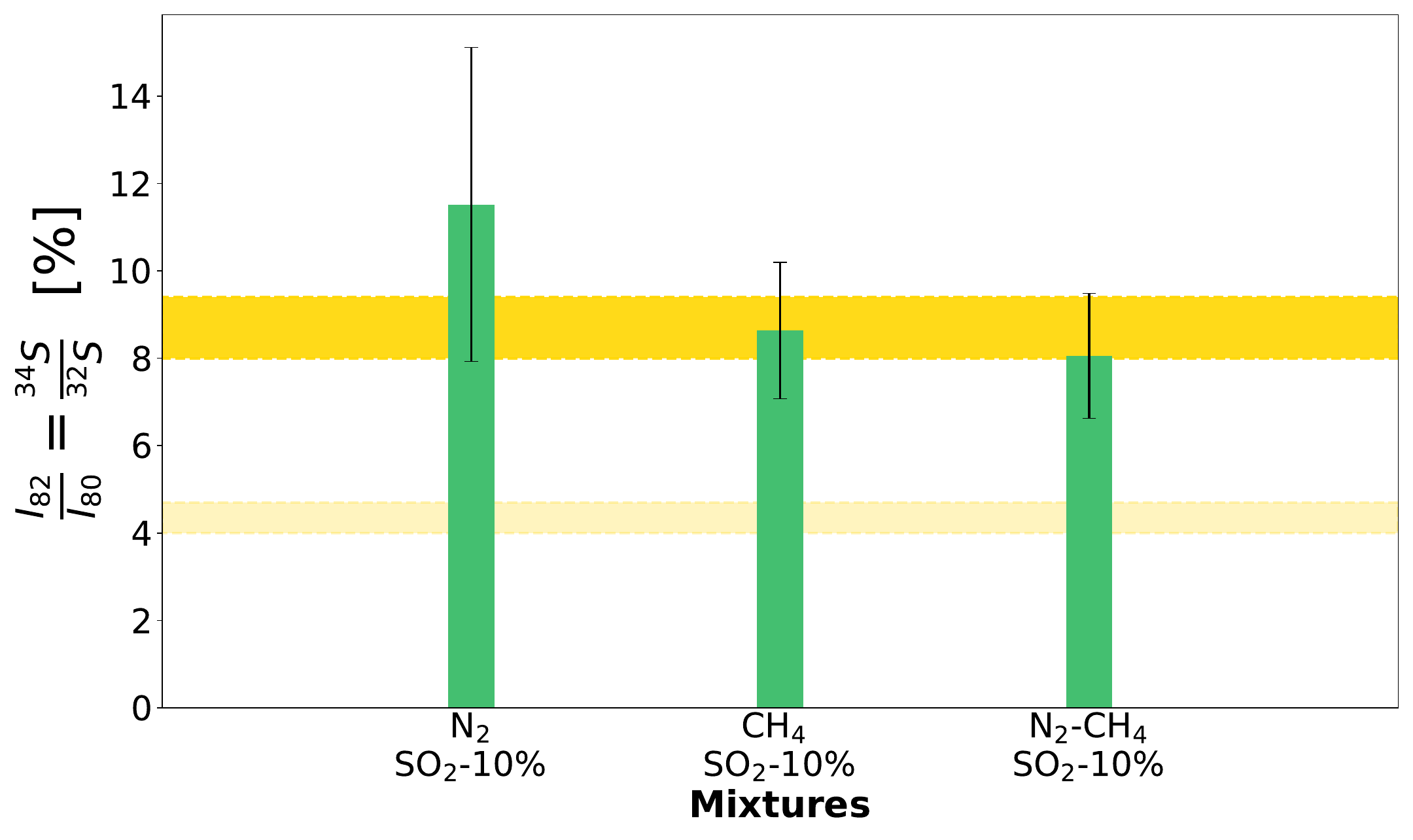}
\includegraphics[width=0.95\textwidth/2]{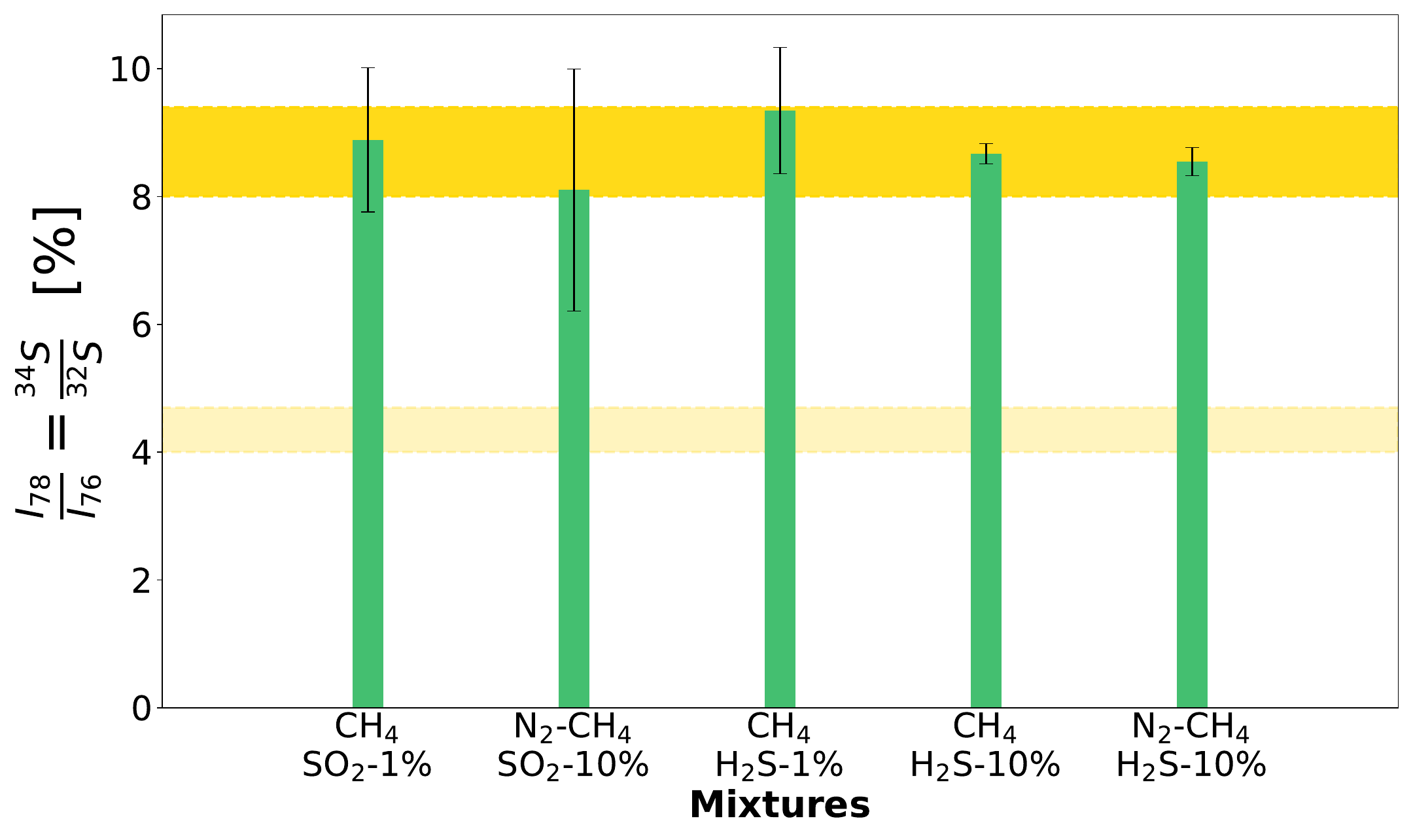}
\caption{Abundance of the $^{34}$S isotope (relative to $^{32}$S) measured for the sulfur volatile previously identified at mass m/z 80 (upper panel) and m/z 76 (lower panel). Only the mixtures with a signal-to-noise ratio superior to 5 at m/z 82 and 78 were included.}
\label{fig6}
\end{figure}

 Figure \ref{fig6} shows the $^{34}$S/$^{32}$S ratios obtained for the sulfur-containing volatiles observed at fragments 76 and 80. The natural value of this $^{34}$S/$^{32}$S ratio is contained in the range [4.0\%; 4.7\%] \citep{thode_sulphur_1961} (pale yellow area figure \ref{fig6}).  If the target molecule has only one sulfur atom, the $^{34}$S/$^{32}$S ratio must fall within this range, while if it contains two sulfur atoms, the $^{34}$S/$^{34}$S ratio must be in the range [8.0\%; 9.4\%] (bright yellow zone figure \ref{fig6}). In this case, the presence of two sulfur atoms in the molecule increases the statistical weight of the $^{34}$S isotopolog. These remarks thus show that, depending on the number of sulfur atoms, the $^{34}$S/$^{32}$S ratio takes on sufficiently distinct values to be discriminated by mass spectrometry.
\par For the two volatiles considered, the $^{34}$S/$^{32}$S ratios shown in figure \ref{fig6} are consistent with the presence of two sulfur atoms. These results suggest that the mass 80 is mainly due to S$_2$O, and the mass 76 to CS$_2$. This does not exclude a small contribution of SO$_3$ to mass 80. On the other hand, the contribution of C$_2$H$_5$SCH$_3$ to mass 76 seems unlikely.  Indeed, the non-detection of structurally simpler thioethers such as DMS (CH$_3$SCH$_3$) in N$_2$-CH$_4$-SO$_2$ and N$_2$-CH$_4$-H$_2$S mixtures show that these same conditions do not favor this type of functional group. \\

\vspace{-0.2cm}
\par
\begin{figure}[ht!]
\centering
\includegraphics[width=0.95\textwidth/2]{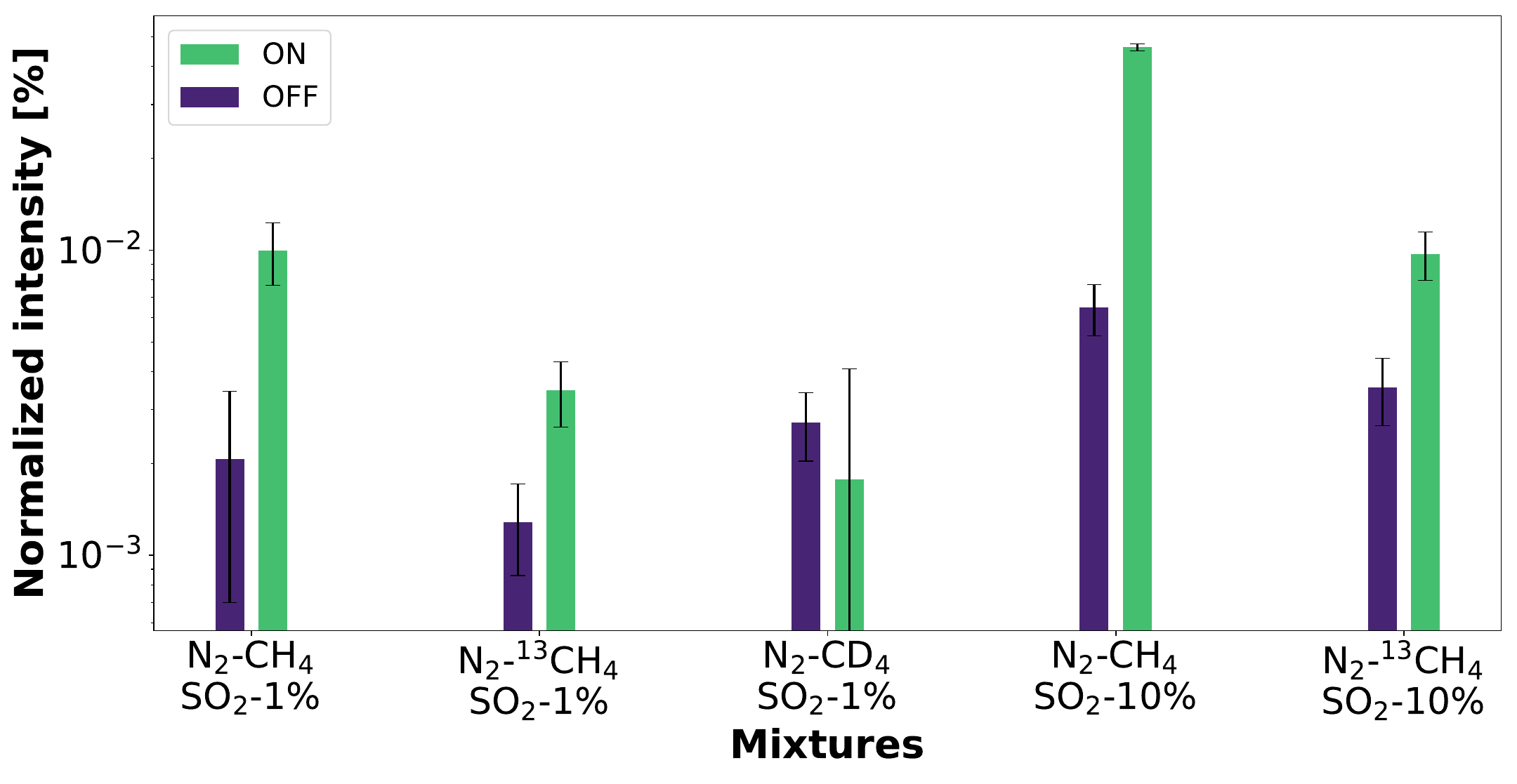}
\caption{Comparison of normalized intensities for the sulfur fragment at m/z 63 among the $^{13}$CH$_4$, $^{12}$CH$_4$, and CD$_4$ experiments.}
\label{fig7}
\end{figure}

 Figure \ref{fig7} compares the signals obtained at m/z 63 in discharges performed with $^{12}$CH$_4$ and isotopically marked methane $^{13}$CH$_4$ and CD$_4$. In the $^{13}$CH$_4$ mixtures, an increase is observed after plasma ignition at the same m/z 63 ratio as in the case of $^{12}$CH$_4$ experiments. This suggests that a significant contribution at m/z 63 is due to a specie which does not contain carbon atoms. The absence of increase in intensity at m/z 61 and 62 in the $^{12}$CH$_4$ experiments excludes a significant contribution of carbon-bearing species at m/z 63 in the $^{13}$CH$_4$ experiments. Considering the associated formula of this non-carbon-bearing species N$_x$H$_y$O$_w$S$_z$, with \((x,y,w,z)\in \mathbb{N}^4\) , we have z=1 because sulfur is a heavy elements and m(S$_2$)=64>63. Moreover, the CD$_4$ experiment highlights a disappearance of the peak at m/z 63. This is consistent with a mass shift of one mass unit and the convolution between SO$_2$ main fragment at m/z 64 and the searched specie marked with one deuterium atom. This imposes that y=1. Thus, given these two constraints on y and z, the only formula leading to a total mass of 63 and a stable structure is NHSO. 
However, Figure \ref{fig7} also highlights that the signal measured at m/z 63 in the presence of $^{13}$CH$_4$ is significantly lower than the one in $^{12}$CH$_4$ experiments. This suggests that additional sulfur oxidized organic specie(s) contribute to m/z 63, such as DMSO or methane sulfenic acid (CH$_3$SOH). Other main fragments of these molecules were targeted in CD$_4$ and $^{13}$CH$_4$ experiments (m/z 68 for CD$_3$SOD, m/z 80 for $^{13}$C$_2$H$_6$SO in N$_2$-CH$_4$-SO$_2$-1\% mixture where the S$_2$O contribution remains negligible at this m/z ratio, and m/z 84 for C$_2$D$_6$SO); however no increase was observed for the considered fragments. This result does not allow us to identify the nature of the sulfur organic specie(s) involved at m/z 63.

\subsubsection{Conclusion}
\begin{figure}[ht!]
\centering
\includegraphics[width=0.45\textwidth]{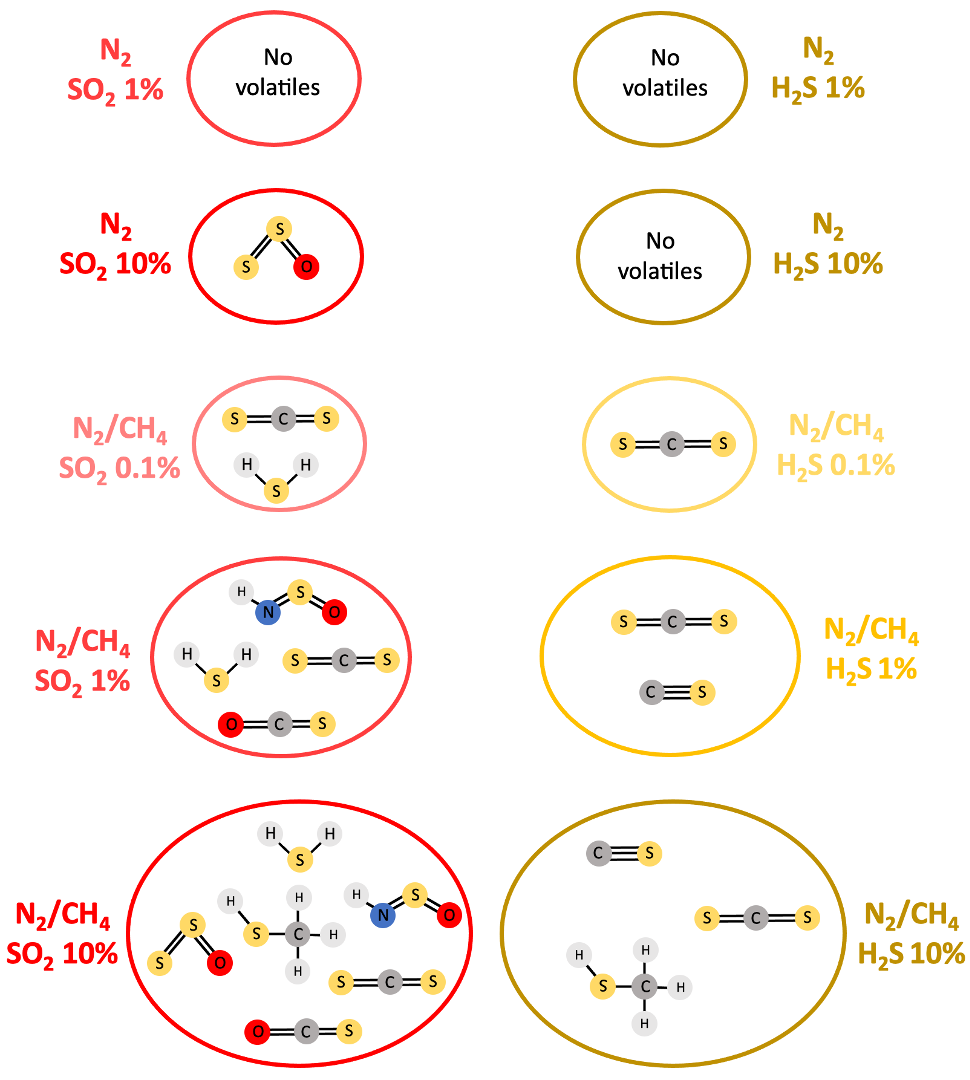}
\caption{Synthesis of the sulfur volatile products identified in the main mixtures investigated N$_2$-SO$_2$, N$_2$-H$_2$S, N$_2$-CH$_4$-H$_2$S, and N$_2$-CH$_4$-SO$_2$.}
\label{fig8}
\end{figure}

\vspace{-0.2cm}
Figure \ref{fig8} shows the various sulfur-containing products observed by mass spectrometry in the main mixtures explored : N$_2$-SO$_2$, N$_2$-H$_2$S, N$_2$-CH$_4$-SO$_2$, and N$_2$-CH$_4$-H$_2$S. 
Under the experimental conditions investigated here, the coupling between CH$_4$ and sulfur-bearing reactants produces a certain diversity of products containing carbon and sulfur elements. The reaction between H$_2$S and CH$_4$ mainly forms CS$_2$, CH$_3$SH, and CS, while in the presence  of SO$_2$, methane reactivity leads to the production CS$_2$, OCS, and CH$_3$SH. In the particular case of N$_2$-CH$_4$-SO$_2$ mixtures, it is difficult to confirm the presence of CS, as CO$_2$ is also observed in these conditions at the same m/z 44 ratio. 
Moreover, these first qualitative results highlight an inefficient coupling between nitrogen and sulfur elements in the gas phase. No volatiles combining nitrogen and sulfur, such as NOS, NS or NS$_2$, have been observed in both N$_2$-H$_2$S and N$_2$-SO$_2$ mixtures. 
\par Finally, the detection of NHSO and H$_2$S in N$_2$-CH$_4$-SO$_2$ mixtures suggests the existence of reduction pathways from SO$_2$ to H$_2$S and formation reactions of NHSO. Elemental processes leading to NHSO are currently missing in photochemical databases. Our detection highlights the need to implement reactions pathways involving SO$_2$ in the presence  of nitrogen and hydrogen in databases. 
\par Several volatiles illustrated in figure \ref{fig8}, including OCS, CS$_2$, and CH$_3$SH, have previously been identified in experimental studies investigating H$_2$S chemistry within CO$_2$-dominated Super-Earth atmospheres \citep{he_sulfur-driven_2020} and Archean early-earth environment \citep{reed_abiotic_2024}. These findings confirm that the formation of these molecules is not exclusive to the conditions explored in this work and can also be initiated by CO$_2$ and H$_2$S. 
However, regarding the other sulfur products observed figure \ref{fig8}, namely S$_2$O, NHSO, and H$_2$S, they have not been observed in previous experiments simulating sulfur chemistry in planetary/exoplanetary contexts. These molecules thus highlight new reactions pathways specific to SO$_2$ chemistry which remains poorly explored so far for exoplanetary atmospheres. 

\begin{figure*}[htbp]
  \centering
  \begin{subfigure}[b]{0.62865\textwidth}
    \centering \fbox{\includegraphics[width=0.955\linewidth]{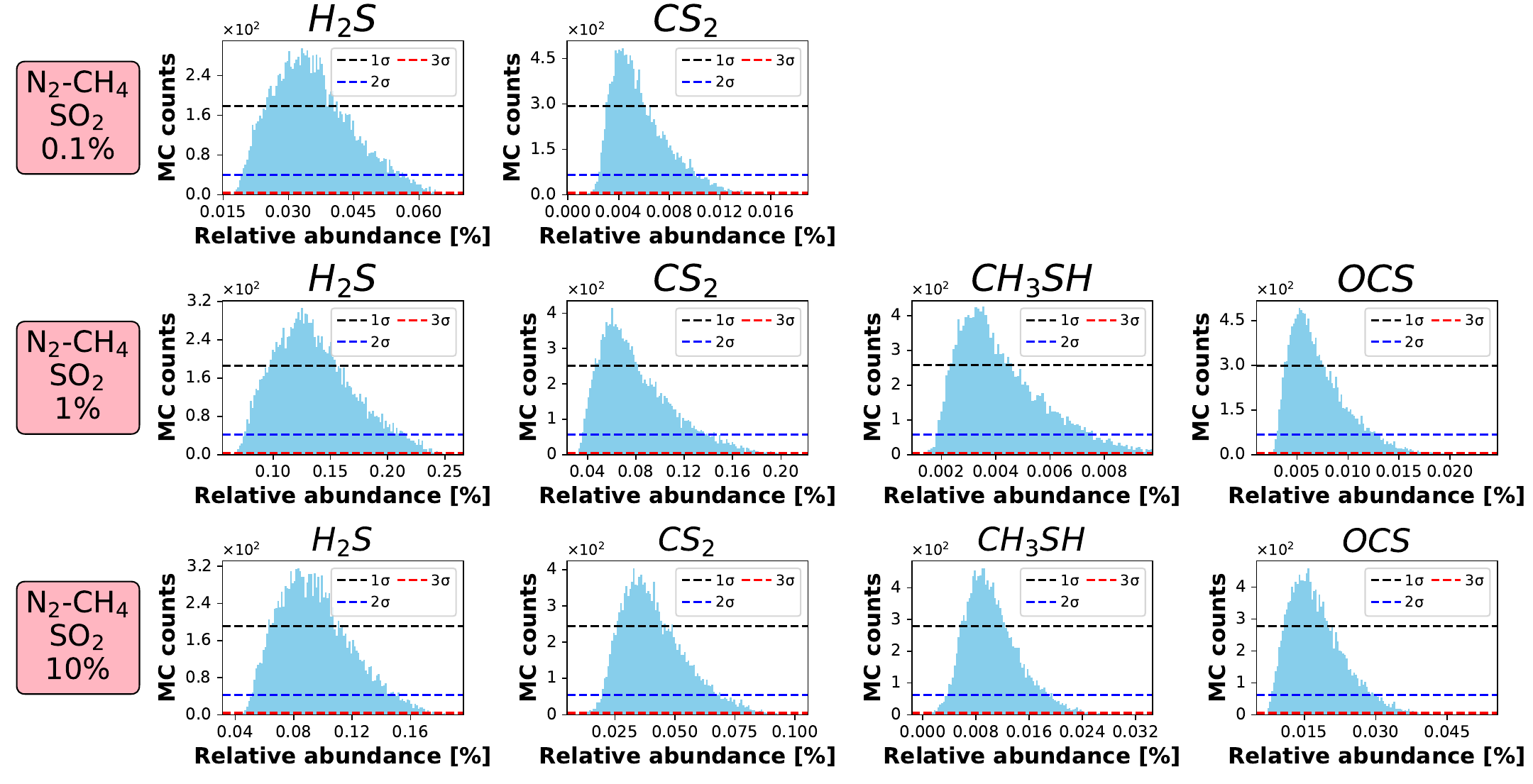}}
    \caption{N$_2$-CH$_4$-SO$_2$ mixtures}
    \label{fig:so2}
  \end{subfigure}
  \hfill
  \begin{subfigure}[b]{0.3564\textwidth}
    \centering
    \fbox{\includegraphics[width=0.958\linewidth]{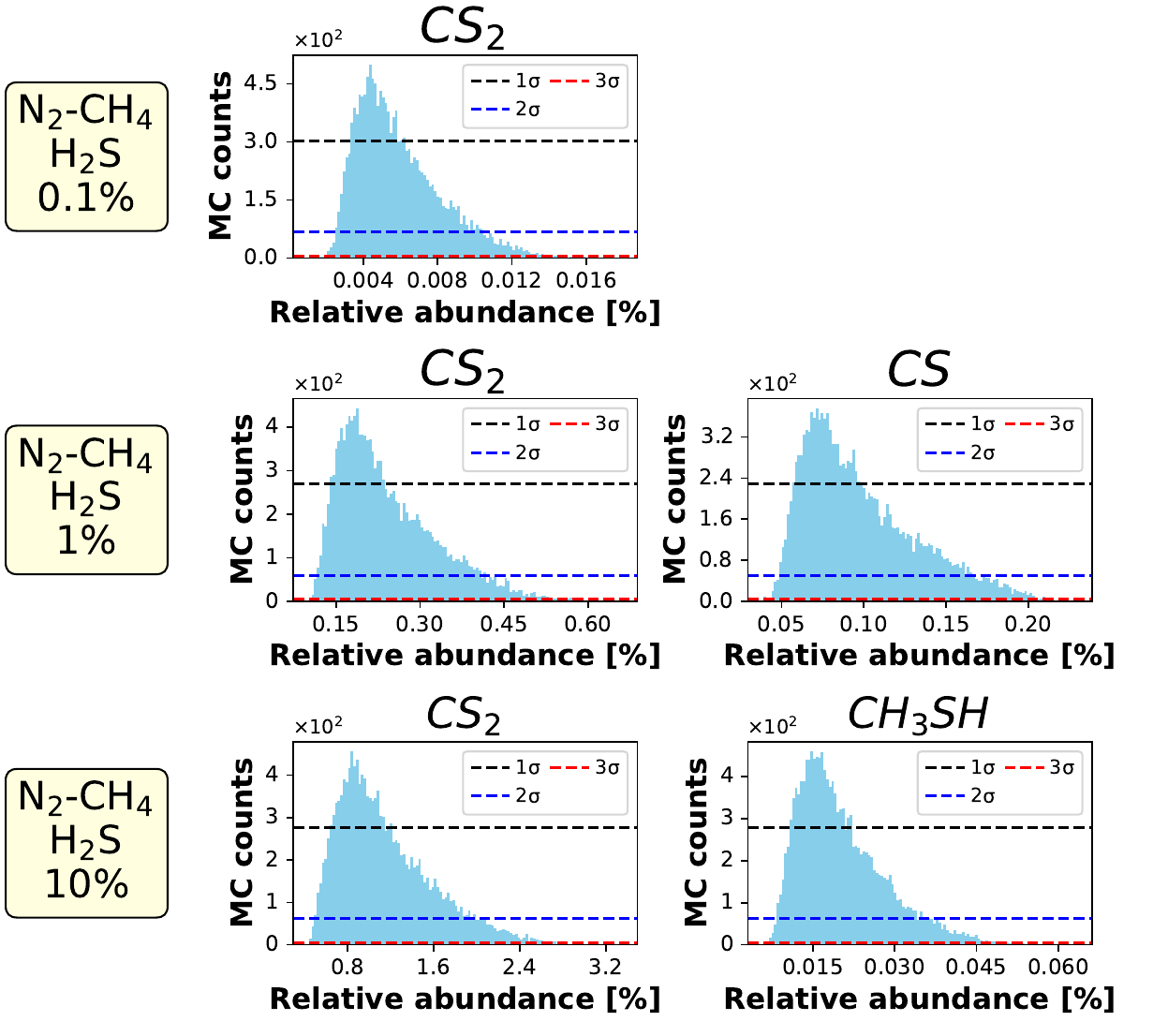}}
    \caption{N$_2$-CH$_4$-H$_2$S mixtures}
    \label{fig:h2s}
  \end{subfigure}
  \caption{Abundances retrieved by deconvolution for the sulfur quantifiable volatile products observed in N$_2$-CH$_4$-H$_2$S and N$_2$-CH$_4$-SO$_2$ mixtures}
  \label{fig9}
\end{figure*}

\begin{figure*}[t]
\sidecaption
\includegraphics[width=12cm]{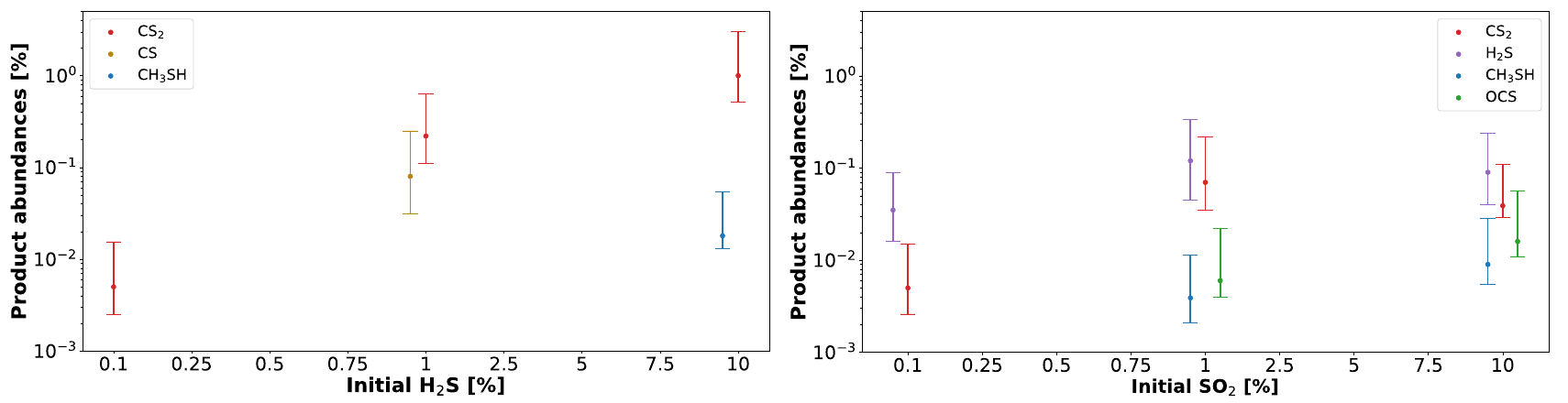}
\caption{Abundances retrieved for the different sulfur products quantified as a function of sulfur initial percentage for N$_2$-CH$_4$-H$_2$S mixtures (left panel) and N$_2$-CH$_4$-SO$_2$ mixtures (right panel). The central value corresponds to the median abundance value of the distribution observed in figure \ref{fig9}, and the uncertainties are given at 2$\sigma$.}
\label{fig10}
\end{figure*}

\subsection{Quantification of sulfur-bearing volatiles }
Figure \ref{fig9} displays the abundances of sulfur-bearing gaseous products estimated by mass spectra deconvolution in the different N$_2$-CH$_4$-SO$_2$ and N$_2$-CH$_4$-H$_2$S mixtures. 
\par In N$_2$-CH$_4$-SO$_2$ mixtures, H$_2$S and CS$_2$ are the two major sulfur products formed with an initial SO$_2$ percentage of 0.1 \% and 1 \%. In N$_2$-CH$_4$-SO$_2$-0.1\% mixture, H$_2$S and CS$_2$ are the only species detected and quantified. While at 1 \% of initial SO$_2$, the abundances retrieved with a 3$\sigma$ reliability for both H$_2$S and CS$_2$ are at least two times higher than the other quantified products CH$_3$SH and OCS. These results likely highlight a selectivity in favor of CS$_2$ relatively to other carbon and sulfur bearing volatiles under both experimental conditions. Moreover, the important H$_2$S abundance suggests the preponderance of SO$_2$ reduction pathways in the overall SO$_2$ reactivity. Finally, for the third mixture with an initial SO$_2$ percentage of 10 \%, the order between the species abundances is less defined. CS$_2$ exceeds CH$_3$SH abundance with a reliability of 2$\sigma$, and exceeds OCS only at 1$\sigma$. Therefore, CS$_2$ cannot be identified in this case with certainty as the main carbon and sulfur bearing volatile, but is formed at least in comparable amounts as CH$_3$SH and OCS. For H$_2$S, its estimated abundance is still superior to OCS and CH$_3$SH and higher than CS$_2$ at 1$\sigma$. H$_2$S thus remains either the most or the second-most abundant sulfur-bearing volatile products at 10 \% of initial SO$_2$. This shows the persistence of SO$_2$ reduction pathways even in the presence  of significant oxygen concentrations. 
\par In N$_2$-CH$_4$-H$_2$S mixtures, CS$_2$ represents a major volatile product in the sulfur budget whatever the initial H$_2$S percentage considered. CS$_2$ is the only volatile quantified with an initial H$_2$S concentration of 0.1 \%. At 1 \% of H$_2$S, CS$_2$ exceeds CS abundance with a 1$\sigma$ reliability, and is present at least in similar amounts as CS. Finally, CS$_2$ abundance is one order of magnitude higher than CH$_3$SH in N$_2$-CH$_4$-H$_2$S-10\% mixture. These results highlight the selectivity induced by H$_2$S and CH$_4$ in favor of CS$_2$ over CH$_3$SH with reasonable confidence.
\par Figure \ref{fig10} compares the evolution of the quantified abundances for the different volatile products in function of the initial sulfur percentage introduced. It notably highlights a decrease in the abundance of CS$_2$ retrieved in SO$_2$-mixtures, compared to H$_2$S-mixtures. For an initial SO$_2$ concentration of 10 \%, the estimated CS$_2$ abundance is lower than CS$_2$ abundance obtained with 10 \% H$_2$S. For an initial sulfur content of 1\%, the median CS$_2$ abundance value follows this same trend between N$_2$-CH$_4$-SO$_2$-1\% and N$_2$-CH$_4$-H$_2$S-1\% mixtures. However, in this case, the uncertainties on the abundances due to the deconvolution are too high to directly conclude to a CS$_2$ decrease in the presence  of SO$_2$. Nevertheless, the evolution of the CS$_2$ main fragment intensity (at m/z 76) clearly reveals this variation between the two conditions. The uncertainties on the measured intensities are lower because they integrate only measurement uncertainties and not deconvolution uncertainties which are considerable. This comparative approach between the main fragment intensity of a given species is more precise than the quantitative approach based on the abundance to highlight the variation observed between two conditions. 
\par Using this method, it can be demonstrated that CS$_2$ abundance increases with the initial H$_2$S percentage, whereas it decreases over the range [1-10 \%] in SO$_2$-based mixtures. These opposing trends highlight the inhibition induced by SO$_2$ and oxygen on CS$_2$ formation. The presence of significant oxygen concentrations seems to promote competing reactions pathways leading to the production of other sulfur and oxygen-bearing species such as S$_2$O, NHSO, or OCS. This hypothesis is supported by the anti-correlation observed between the intensity of the CS$_2$ main fragment at m/z 76 and the intensities of the characteristic fragments of the previous species (m/z 60 for OCS, m/z 63 for NHSO, m/z 80 for S$_2$O). 
\section{Discussion}
\subsection{Volatile formation processes}
As shown in Figure 8, SO$_2$ and H$_2$S reactivity in N$_2$ plasma discharges leads to the detection of numerous volatile products containing sulfur under the experimental conditions explored. This section discusses the formation pathways of these volatiles based on the basic steps existing in the literature to describe sulfur gas phase chemistry and highlights some of the limitations of this modeling. These reactions include both neutral and ionic processes. They are extracted from the four following databases commonly used in atmospheric sciences: KIDA (KInetic Database for Astrochemistry) \citep{wakelam_kinetic_2012}, UMIST (University of Manchester Institute of Science and Technology) \citep{millar_umist_2024}, NIST (National Institute of Standards and Technology) \citep{manion_j_a_nist_2015}  and the 1D-photochemical model VULCAN \citep{tsai_vulcan_2017}. So far, these are the most relevant sources for describing temperate sulfur chemistry even if they have not been specifically built for this purpose. A more robust and validated photochemical network is needed to properly describe sulfur chemistry and its coupling with other volatile elements C, H, N, O, in temperate atmospheres. The processes highlighted in this section and in appendix G constitute an interesting starting set of reactions to potentially develop such a network. To date, validated networks only exist for hot exoplanetary environments at temperatures above 500 K \citep{veillet_development_2026}. 

\subsubsection{Sulfur and Nitrogen bearing volatiles}
{\hspace{0.6cm}\textit{i. NHSO}\vspace{0.1cm}}\\
The detection of NHSO in N$_2$-CH$_4$-SO$_2$ discharges is particularly interesting as this volatile is absent from the photochemical reaction databases considered (NIST, UMIST, and KIDA). However, HNSO is known experimentally to be produced by the reaction between NH$_3$ and SO$_2$ \citep{hata_reactions_1964}. Moreover, a recent study reported the detection of this species in a molecular cloud within the interstellar medium using rotational spectroscopy \citep{sanz-novo_discovery_2024}. These remarks suggest that formation processes for NHSO likely exist both in the gas phase and on the surface of ice grains. 
\par In our experiment, NHSO may result from the reactions between NH radicals and SO or NH and SO$_2$ in the gas phase. As shown by the equations below, NH radicals can be produced by dissociation of N$_2$ and capture of hydrogen atoms or by dissociation of NH$_3$ which is present in significant quantities as evidenced by the increase observed at m/z ratio 17. Regarding SO, this intermediate species is directly formed by the dissociation of SO$_2$.
\begin{center}
$
\begin{array}{l}
    \text{N$_2$}\longrightarrow 2 \: \text{N} \\[5pt]
    \text{N} + \text{H$_2$} \rightarrow \text{NH} + \text{H} \\[5pt]
    \text{NH$_3$}\rightarrow \text{NH} + 2 \:\text{H} \\[5pt]
    \text{NH} + \text{SO} \rightarrow \text{NHSO} \\[5pt]
    \text{NH} + \text{SO$_2$} \rightarrow \text{NHSO} + \text{O}
\end{array}
$
\end{center}

\par In this mechanism, kinetic constants associated with the NHSO formation processes are missing from the current databases. No mention of the reaction between NH and SO$_2$ has been found in the literature. However, this pathway represents a potentially efficient route for NHSO production. Regarding the coupling between NH and SO, both experimental \citep{blitz_experimental_2010,hughes_experimental_2002} and modeling \citep{hassani_kinetics_2020,yu-juan_structures_2003} studies have demonstrated the reactivity of NH radicals in the presence of SO, but the products resulting from this reaction have not been yet characterized, as suggested by the following equation ($\text{NH} + \text{SO} \rightarrow \text{Products}$).\\

{\vspace{0.1cm}\hspace{0.6cm}\textit{ii. Non-detection of NO$_x$S$_y$: N$_2$S, NS, NOS, and NS$_2$}\vspace{0.1cm}}\\
In mixtures involving N$_2$ and sulfur reactants (H$_2$S or SO$_2$), the presence of nitrogen and sulfur bearing volatiles, such as NS, N$_2$S, NOS and NS$_2$, has also been investigated. Sulfur being in the same column of the periodic table than oxygen, such species are the sulfur equivalents of NO$_x$.
\par However, no intensity increase at the corresponding nominal masses of these different volatiles (46 for NS, 60 for N$_2$S, 62 for NOS, and 78 for NS$_2$) has been observed in both N$_2$-H$_2$S and N$_2$-SO$_2$ mixtures. This suggests that the corresponding species are either absent or in trace amounts below the detection limit of the mass spectrometer. Consequently, the direct coupling between nitrogen and sulfur atoms in gas phase does not appear as a major reaction pathway under the experimental conditions explored. 

\subsubsection{Carbon and sulfur-bearing products}

{\hspace{0.6cm}\textit{i. CS$_2$}\vspace{0.1cm}}\\
As detailed in appendix G,  only neutral pathways currently exist in  databases to describe the formation of CS$_2$. This involves two main processes: the trimolecular recombination between CS and S, and the reaction with OCS and CS (denoted (1) and (2) in appendix G).
This scheme gives clues to understand the lower CS$_2$ abundance retrieved when SO$_2$ is used as initial reactant. First, several key species in the CS$_2$ formation pathway, such as CS, CH, or CH$_2$ react with the two main SO$_2$ dissociation products oxygen and sulfur monoxide. These additional reactions favor the consumption of these intermediate species and therefore limit the efficiency of the subsequent reactions leading to CS$_2$. In addition to these processes impacting the formation of CS$_2$, the presence of SO$_2$ and oxygen also activate CS$_2$ sink reactions such as $\text{CS$_2$} + \text{OH} \rightarrow \text{OCS} + \text{H}$, $\text{CS$_2$} + \text{O} \rightarrow \text{CS} + \text{SO}$, or $\text{CS$_2$} + \text{O} \rightarrow \text{OCS} + \text{S}$.
\par Nevertheless, the mechanism exposed in appendix G also presents some limitations, highlighting the current gaps in building a
robust description of CS$_2$ photochemical sources. One of the CS$_2$ pathways recently proposed to account for the high CS$_2$ abundances retrieved in the atmosphere of several exoplanets \citep{dai_notitle_2026,triantafillides_notitle_2026,yang_photochemical_2026}, still needs to be investigated more deeply. This concerns one of the key intermediates of this new route: CH$_2$S. The abundances estimated with this kinetic scheme for CH$_2$S are significant, meaning that this species should, in principle, be a stable and detectable product. However, such compound has never been observed experimentally in this work as well as in previous studies investigating carbon and sulfur gas-phase reactivity \citep{he_sulfur-driven_2020,raulin_formation_1975,reed_abiotic_2024}. 
\par In addition to this inconsistency between model and experiments, uncertainties remain on the value adopted for some kinetic constants such as the reaction $(1)$ between CS and S. The $k_\infty$ constant in the Lindemann formalism used to parameterize three body reactions is derived from an extrapolation at lower temperature of an expression validated within the 1900-2800 K range \citep{glarborg_temperature_2015}. The authors themselves caution against the reliability of extrapolation to lower temperatures. \\

{\hspace{0.6cm}\textit{ii. CH$_3$SH}\vspace{0.1cm}}\\
As illustrated in appendix G, the formation of CH$_3$SH through the coupling of CH$_4$ with H$_2$S or CH$_4$ with SO$_2$ involves the same key process which is the neutral recombination between CH$_3$ and SH radicals (reaction $(3)$). This trimolecular reaction is, so far, the only efficient source of CH$_3$SH in the current description of sulfur chemistry. This remark highlights the significant lack of knowledge regarding this species and, more generally, coupling reactions between carbon and sulfur. 
\par Major uncertainties still remain on the value adopted for the kinetic constant associated with reaction $(3)$ forming CH$_3$SH. More precisely, the $k_0$ parameter used in the Lindemann formalism for trimolecular constant has been estimated in the case of reaction ($2$) based on the value known for the recombination reaction of methanol (CH$_3$OH) \citep{humpfer_reaction_1994} which may not be directly transferable. In addition to this lack of constraints existing on kinetic constant, other sources may exist for CH$_3$SH, as suggested by the reactivity observed for S($^1$D) in the presence of CH$_4$ \citep{black_rate_1985}. 
\par Thus, a more in-depth characterization of CH$_3$SH formation pathways, combined with precise estimates of the associated rate constants, is currently needed to provide a more robust and rigorous description of sulfur chemistry. Such considerations are essential for better assessing the photochemical sources of CH$_3$SH and further discussing its potential bio-signature status \citep{catling_exoplanet_2018,domagal-goldman_using_2011,seager_biosignature_2013}.
\par Although incomplete, the reaction scheme presented in the appendix nevertheless accounts for certain experimental observations, such as the lower abundance of CH$_3$SH relative to CS$_2$. This result is consistent with the kinetic constants associated with the main formation processes of CH$_3$SH and CS$_2$ proposed in this scheme (reactions ($1$) and ($2$)). At 1 mbar and room temperature, the trimolecular CH$_3$SH formation constant is several orders of magnitude lower than CS$_2$, highlighting a less efficient final recombination process for CH$_3$SH in agreement with observations.

{\hspace{0.6cm}\textit{iii. OCS}\vspace{0.1cm}}\\
As shown in appendix G, three main neutral processes lead to the formation of OCS with CH$_4$ and SO$_2$ as initial reactants: the trimolecular recombination reaction between CO and S atom ($4$), the reaction between CO and SH ($5$), and finally the oxidation of CS by SO ($6$). Beyond these neutral reactions, an additional ionic pathway leading to OCS is found in current databases \citep{turner_searches_1990,woon_quantum_2009}. This route involves the intermediate OCSH$^+$ formed by the reaction between $\text{CH}_3^+$ and SO. 
\par The processes detailed above provide insights to discuss the differences observed in OCS abundance between this work and a previous experimental study which explored sulfur reactivity in N$_2$/CO$_2$/CH$_4$/H$_2$S mixtures \citep{reed_abiotic_2024}. The percentage of OCS retrieved relatively to the sulfur initially introduced is significantly higher in Reed’s experiment than in the present study. This remark suggests major differences in the OCS formation pathways between the two experiments. In Reed’s study, the initial reactants H$_2$S and CO$_2$ seem to promote more efficiently OCS formation than the CH$_4$ and SO$_2$ used in this work. This observation can be rationalized by the reaction mechanisms outlined earlier. Several OCS formation pathways involve CO, as well as S and SH radicals, which are direct dissociation products of CO$_2$ and H$_2$S. In contrast, as detailed in the appendix G, the formation of OCS from CH$_4$ and SO$_2$ requires in this case a greater number of elementary steps and intermediates.
\par Such differences observed in the experimental results of these two studies could finally suggest a different selectivity for sulfur photochemistry in temperate environments depending on the oxidation state of carbon volatiles. The findings of Reed et al. highlight that oxidized conditions with substantial CO$_2$ concentrations favor OCS as the primary carbon and sulfur bearing photochemical product, whereas our results suggest that more reduced conditions, dominated by CH$_4$, produce more efficiently CS$_2$.
\vspace{-0.4cm}
\subsubsection{Other sulfur products }
\vspace{-0.1cm}
{\hspace{0.6cm}\textit{i. H$_2$S}\vspace{0.1cm}}\\
As detailed in appendix G, both neutral and ionic processes exist in the literature to form H$_2$S with SO$_2$ and CH$_4$ as initial reactants. The neutral pathway involves a progressive hydrogenation of S atoms produced by SO$_2$ dissociation. Moreover, in the presence  of important sulfur concentrations, HS$_2$ might be a key H$_2$S precursor: its reactions with SH and H may constitute in this case efficient sources of H$_2$S.  This second pathway could explain the significant abundance retrieved for H$_2$S with an initial SO$_2$ percentage of 10 \% despite the important oxygen amount introduced.
Regarding the ionic pathway, this process is also based on a two-step hydrogenation of S$^+$ ions.
\par The neutral and ionic pathways highlighted above as well as the significant abundance retrieved for H$_2$S in N$_2$-CH$_4$-SO$_2$ mixture figure \ref{fig9} suggest that SO$_2$ reduction in the presence  of hydrogenated species is an efficient source of H$_2$S. This scheme is the reversed process of the H$_2$S photochemical oxidation proposed to explain the SO$_2$ band observed in the transit spectra of several H$_2$-dominated atmospheres like the warm Neptune WASP-107 b \citep{dyrek_so2_2024} or the hot-Jupiter WASP-39 b \citep{tsai_photochemically_2023}. This remark highlights the strong impact of atmospheric conditions, mainly the composition and temperature, on sulfur chemistry and its products.\\
{\hspace*{0.6cm}\textit{ii. S$_2$O}\vspace*{0.1cm}}\\
S$_2$O has been observed specifically with an initial SO$_2$ percentage of 10 \% (i.e., in N$_2$-SO$_2$-10\% and N$_2$-CH$_4$-SO$_2$-10\% mixtures), which corresponds to the highest concentration explored in this study. This result is consistent with previous analyses confirming the detection of S$_2$O in concentrated SO$_2$ or SO$_2$-S$_x$ plasma discharges \citep{meschi_disulfur_1956,schenk_uber_1937,schenk_new_1965}.
\par In the literature five main pathways are proposed to explain the formation of S$_2$O in gas phase \citep{steudel_sulfur-rich_2003} as shown by the first set of equations in the left column below. S$_2$O can be produced either by a disproportionation involving sulfur monoxide and its dimer (SO)$_2$ ($7$), by the reaction between (SO)$_2$ and S$_2$ ($8$), by the capture of a sulfur atom by sulfur monoxide (($9$) and ($10$)), or lastly by an oxidation of S$_2$ ($11$). These five reactions mainly involve four intermediates, SO, (SO)$_2$, S, and S$_2$, which result from SO$_2$ dissociation as detailed by the second set of equation on the right column.
\begin{table}[ht!]
\label{table:8}    
\centering       
\begin{tabular}{l l}     
$(\text{SO})_2 + \text{SO}\rightarrow \text{S}_2\text{O} + \text{SO}_2$\:($7$)&
$\text{SO}_2 \longrightarrow \text{SO}+ \text{O}$ \\[3pt]
$(\text{SO})_2 + \text{S}_2\rightarrow 2\:\text{S}_2\text{O}$\:($8$) &
$\text{SO} + \text{SO} + \text{M} \rightarrow (\text{SO})_2 + \text{M}$ \\[3pt]
$\text{SO} + \text{S} + \text{M} \rightarrow \text{S}_2\text{O} + \text{M} $ \:($9$)&
$\text{SO} + \text{O} \longrightarrow \text{S} + \text{O}_2$\\[3pt]
$\text{SO} + \text{S}_2\longrightarrow \text{S}_2\text{O} + \text{S}$ \:($10$)&
$\text{SO} + \text{SO} \rightarrow \text{SO}_2 + \text{S}$ \\[3pt]
$\text{S}_2 + \text{O} \rightarrow \text{S}_2\text{O}$ \:($11$) &
$\text{SO} \longrightarrow \text{S} + \text{O}$ \\[3pt]
&$\text{S} + \text{S}  \longrightarrow \text{S}_2$
\end{tabular}
\end{table}
\par This mechanism thus helps to understand the efficiency of S$_2$O formation at high SO$_2$ concentrations, as this product is exclusively generated by reactions between SO$_2$ dissociation intermediates.  The rate of reactions (7) to (11) are influenced by SO abundance which is strongly correlated to the initial SO$_2$ percentage through SO$_2$ dissociation.
\par In the literature, kinetic rate constants exist only for the SO dimerization and S$_2$O formation through disproportionation (reaction (7)) \citep{herron_rate_1980}. No kinetic rate constants have been determined so far for the other processes (reactions (8) to (11)) which have been proposed to account of the more efficient S$_2$O gas phase formation observed in the presence  of sulfur vapor S$_x$ \citep{schenk_new_1965,steudel_sulfur-rich_2003}. These remarks highlight the lack of data and elemental processes describing SO$_2$ reactivity with its own dissociation intermediates. Kinetic constants are needed to assess the relative contributions of the five pathways proposed above in the formation of S$_2$O. Moreover, (SO)$_2$ sink reactions are incomplete, but necessary to properly describe the reactivity of this key intermediate.
\par Thus, these elements point out the existing lacks in the current description of SO$_2$ chemistry. Such missing processes are important to properly represent sulfur photochemistry in a wide diversity of exoplanetary contexts.

\subsection{Detectability of the sulfur products identified}

This part discusses the possibility for the different sulfur products previously detected to be observed by space or ground-based telescopes using their spectral signatures in the infrared range. This discussion aims at highlighting specific band for each compound which could be targeted to identify these sulfur species in temperate exoplanetary atmospheres. This is done by comparing the absorption cross sections of the targeted sulfur volatiles with an extended set of 25 volatiles considered as relevant for exoplanetary atmospheres (see Methods). This comparison defines more precise conditions in term of atmospheric composition to observe unambiguous signatures for the considered sulfur volatile species. 

\subsubsection{CS$_2$}
In the mid-infrared, CS$_2$ exhibits two fundamental absorption bands observable with space- or ground-based telescopes: the $\nu_3$ band (CS$_2$ asymmetric stretching) around 6.5$ \mu$m, and the $\nu_1+\nu_3$ band near 4.6 $\mu$m \citep{agar_high_1962}.
\par
\begin{figure}[ht!]
\centering
\includegraphics[width=0.45\textwidth]{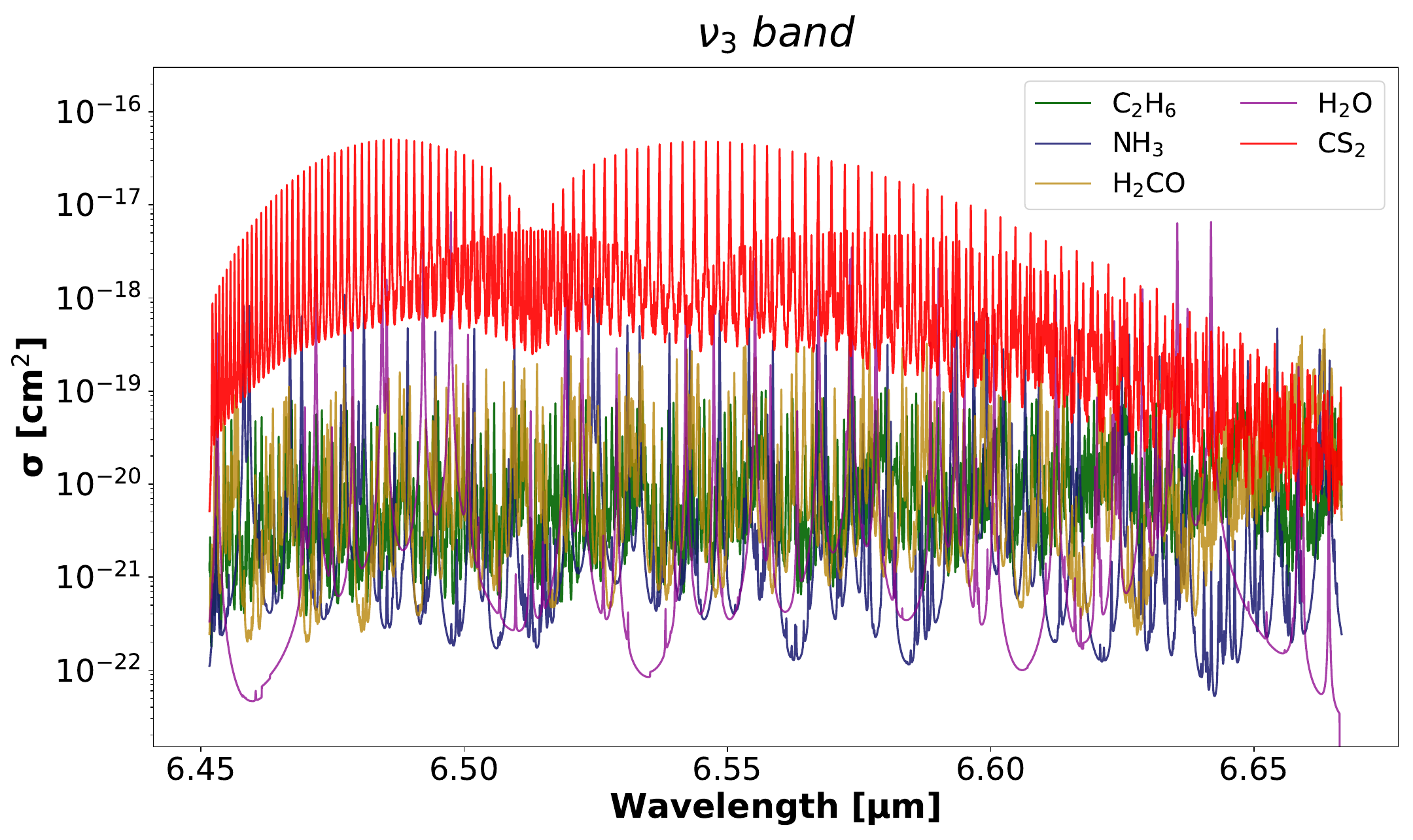}
\includegraphics[width=0.45\textwidth]{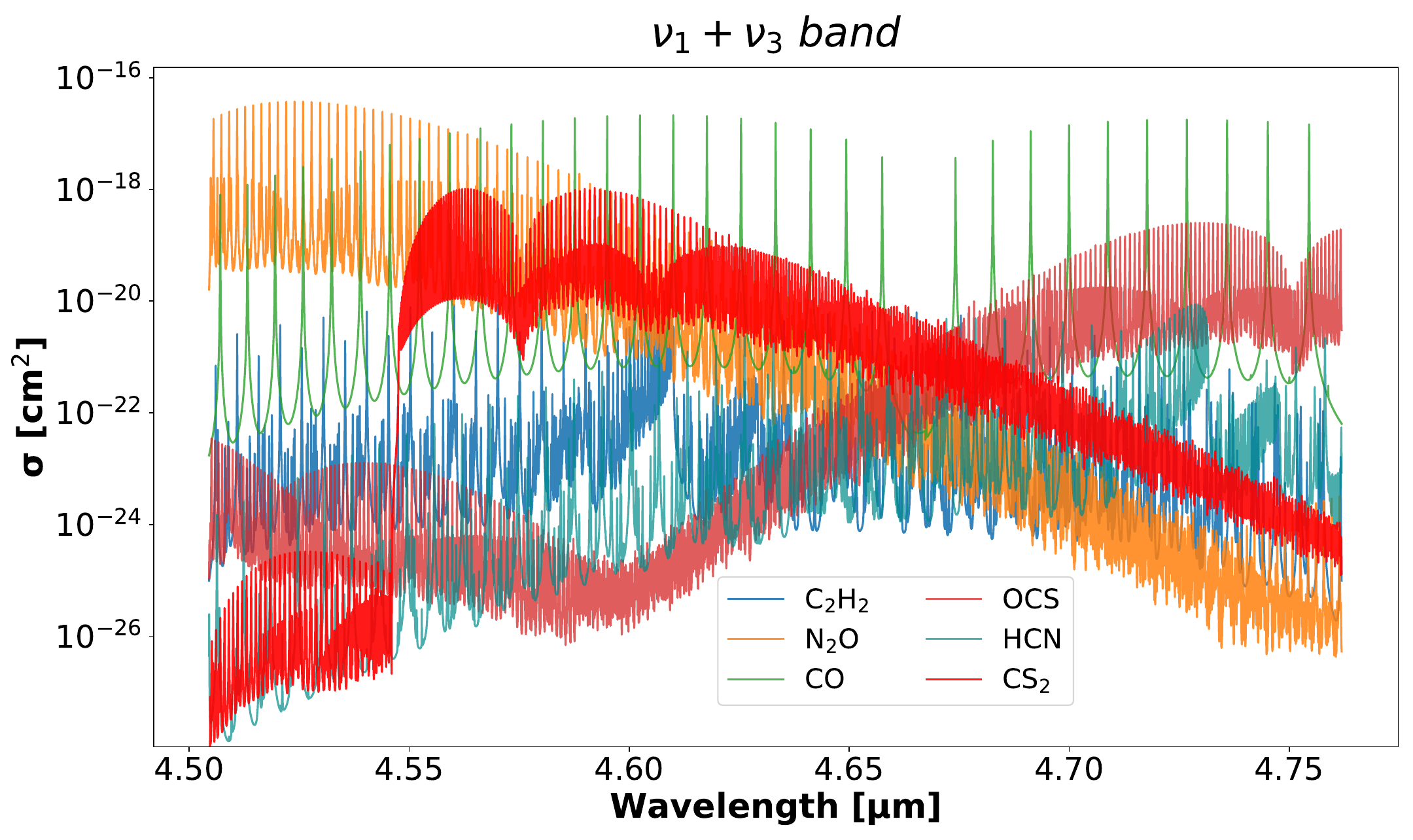}
\caption{CS$_2$ absorption cross sections compared with the strongest volatile cross sections in the spectral regions of the fundamental $\nu_3$ (upper panel) and $\nu_1+\nu_3$ (lower panel) CS$_2$ bands. For clarity, only species with an average contribution higher than one-thousandth the average contribution of CS$_2$ have been included.}
\label{fig11}
\end{figure}

 Focusing first on the most intense $\nu_3$ band, the upper panel figure \ref{fig11} revealed that CS$_2$ absorption features can be distinguished from other volatile signatures, except in humid atmospheres rich in water vapor. CS$_2$ absorption cross sections exceed those of the 25 benchmark volatiles by at least one order of magnitude. Moreover, given the low resolution of the spectral bins used by space telescopes such as JWST (typically between 0.1 and 0.5 $\mu$m), the average contribution of CS$_2$ cross sections over this extended spectral range is two orders of magnitude higher than H$_2$O, and three than C$_2$H$_2$, NH$_3$, and H$_2$CO, which are the four species with the most intense cross sections in the considered range. These ratios indicate the minimum relative abundance versus CS$_2$ required for these four latter species to mask the $\nu_3$ band. For water vapor, a concentration hundred times higher than CS$_2$ is needed to overcome CS$_2$ absorption. This remark suggests that humid atmospheres, rich in water vapor, could potentially hide the CS$_2$ $\nu_3$ band. However, for the three other volatiles (i.e., C$_2$H$_6$, NH$_3$, and H$_2$CO) their absorption contribution could reasonably be neglected compared to CS$_2$. To ensure that these conclusions were not specific to the chosen conditions figure \ref{fig11} (i.e., $p=100$ mbar and $T=250$ K), additional (pressure, temperature) pairs were considered over a larger parametric space ($[0.001;100]$ mbar and $[150;400]$ K see methods and Appendix E). The results remain consistent with those exposed figure \ref{fig11}, CS$_2$ having signatures significantly higher than other volatiles under all the conditions explored. 

For the second $\nu_1+\nu_3$ band, the detection of CS$_2$ appears to be more challenging in this spectral range. As shown in the lower panel figure \ref{fig11}, the $\nu_1+\nu_3$ absorption cross sections are between one and two orders of magnitude inferior to the $\nu_3$ band. Moreover, several volatiles such as CO, N$_2$O, or to a lesser extent OCS and C$_2$H$_2$, can potentially hide CS$_2$ features. However, in the absence of these species, the $\nu_1+\nu_3$ band can be used as an indicator of CS$_2$. The $\nu_1+\nu_3$ band could be particularly relevant for H$_2$-dominated atmospheres where the oxygen-bearing previously mentioned could be absent or at least minor. In the JWST NIRSpec spectrum of the temperate sub-Neptune TOI-270 d, the band observed near 4.6$\mu$m is interpreted as an absorption feature induced by the $\nu_1+\nu_3$ band of CS$_2$ \citep{felix_competing_2025,holmberg_possible_2024}. Nevertheless, the existence of other absorbing species on this spectral range points out the potential degeneracies inherent to the JWST NIRSpec data of TOI-270 d. More particularly, the contributions of CO and CS$_2$ seem to be difficult to discriminate in this specific range, leading to several possible interpretations of the spectrum. The observation of TOI-270 d at higher wavelength with the MIRI instrument could be a way to remove these degeneracies. The intense $\nu_3$ band of CS$_2$ around 6.5$\mu$m could be targeted to confirm with more reliability the presence of CS$_2$ in TOI-270 d atmosphere.
\par Thus, the previous remarks highlight CS$_2$ as a promising target to probe active sulfur chemistry in N$_2$-dominated and temperate exoplanetary atmospheres. The spectral properties of CS$_2$, more particularly its intense $\nu_3$ band, offers a real opportunity to detect unambiguous CS$_2$ signatures with MIRI instrument of the JWST or even in the context of future missions. 

\begin{figure}[t]
\centering
\includegraphics[width=0.45\textwidth]{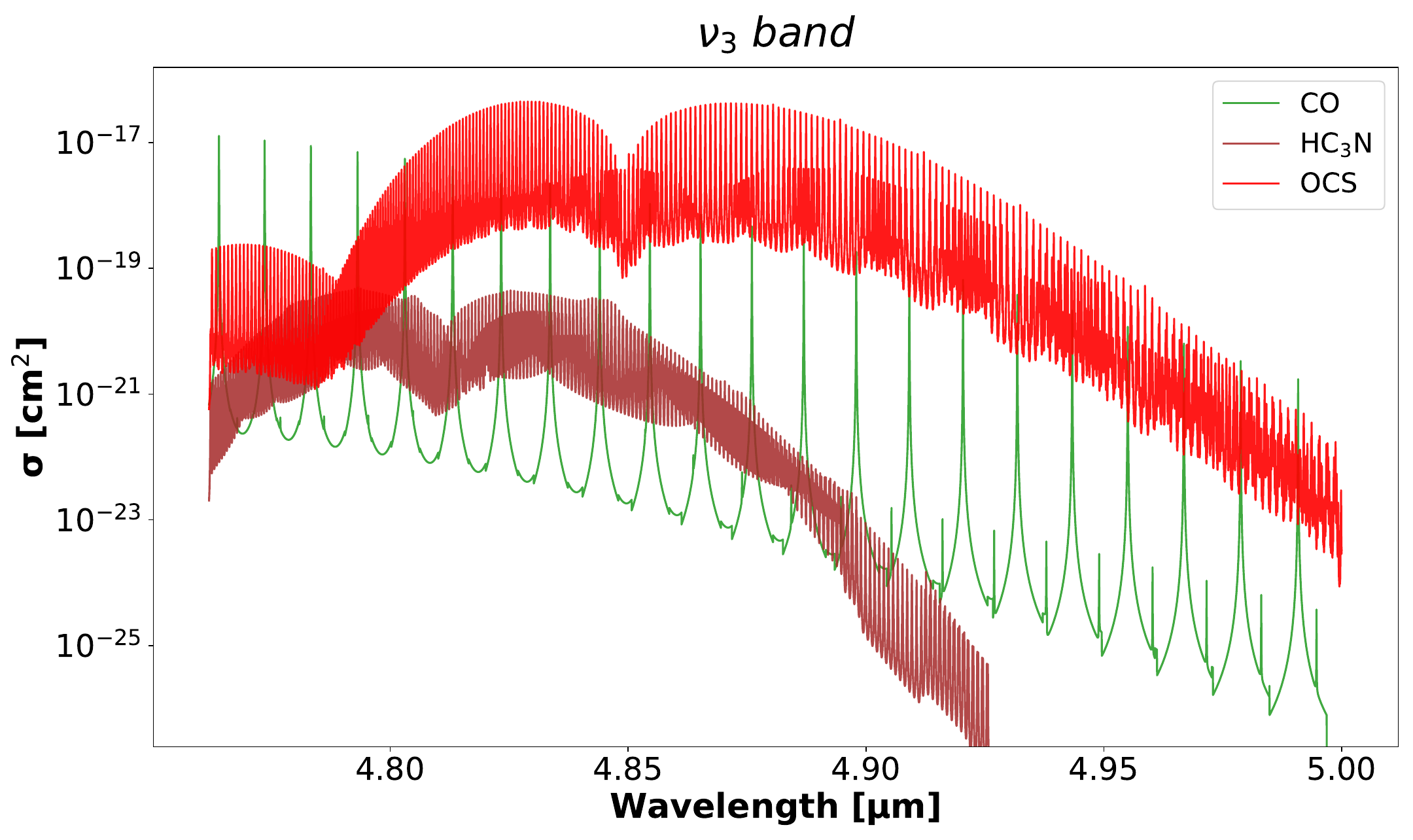}
\includegraphics[width=0.45\textwidth]{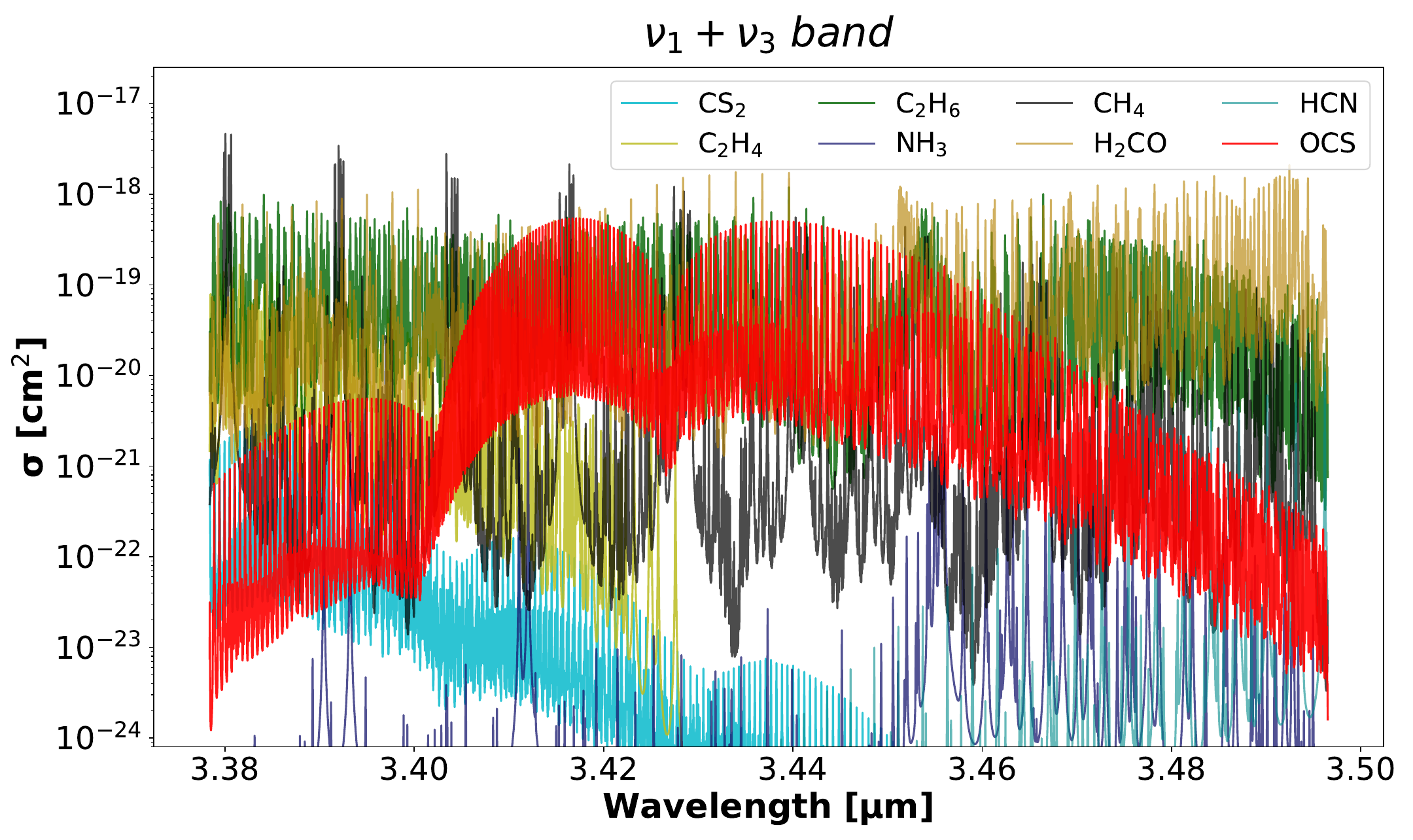}
\includegraphics[width=0.45\textwidth]{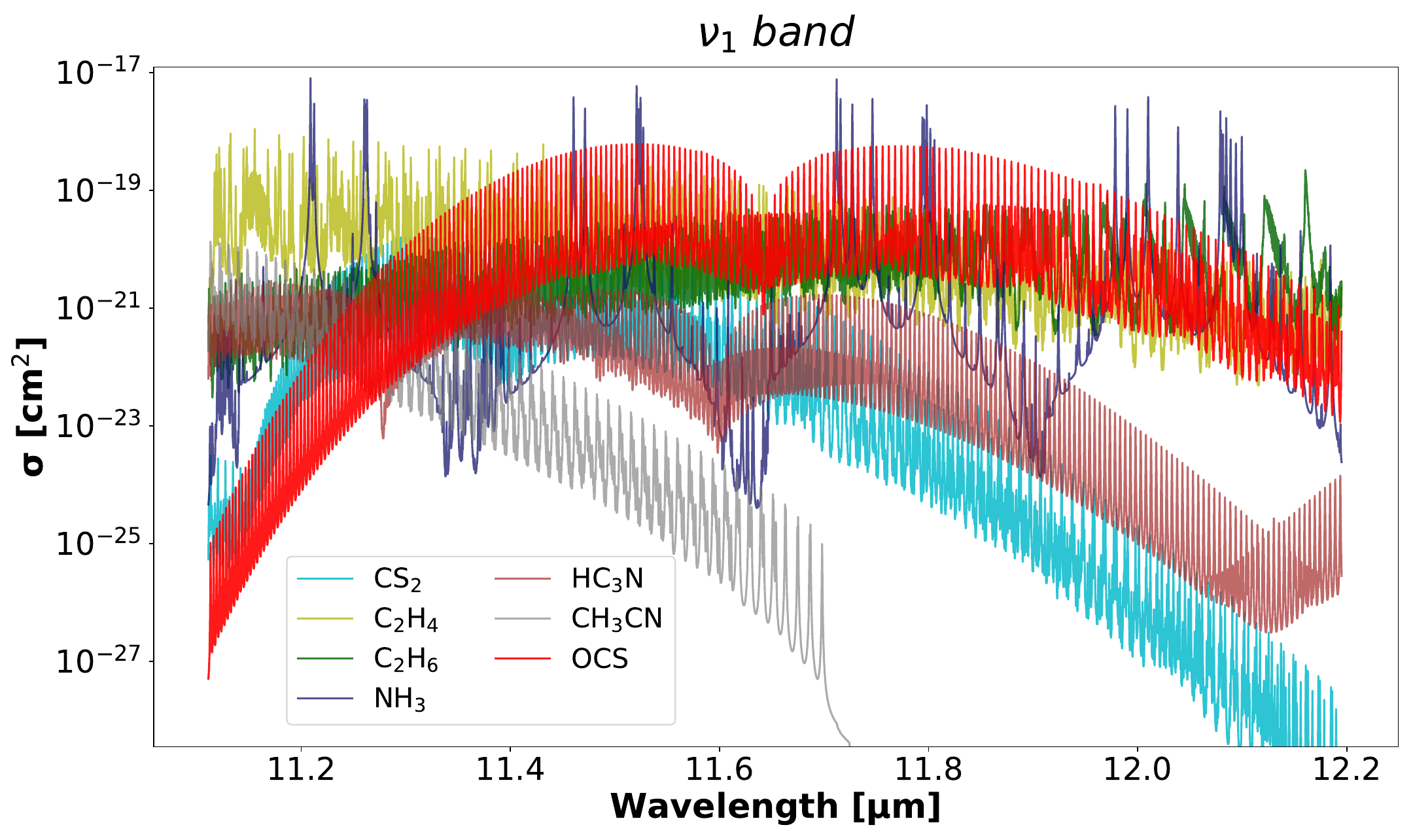}
\caption{OCS absorption cross sections for the three main absorption bands, compared to the strongest volatile cross sections in the associated spectral ranges.  As for Figure 11, the species with an average contribution inferior to one thousandth have not been shown.}
\label{fig12}
\end{figure}

\subsubsection{OCS}
In the mid-infrared OCS has three main fundamental absorption bands which are in decreasing order of intensity \citep{bocharov_vibrational_2008}: the $\nu_3$ band (CO stretch) around 4.5 $\mu$m, the $\nu_1+\nu_3$ band near 3.4 $\mu$m, and the $\nu_1$ band (CS stretch) around 11.7 $\mu$m. 
\par The $\nu_3$ band constitutes an interesting target to reveal potential OCS signatures for a temperate N$_2$-dominated atmosphere in the absence of important CO concentrations. As shown on the first panel figure \ref{fig12}, OCS cross sections for the $\nu_3$ band are at least one order of magnitude higher than the most absorbing species in this spectral range, HC$_3$N and CO. Considering the average cross sections on the whole $\nu_3$ range [4.75;5 $\mu$m], the OCS contribution exceeds CO and HC$_3$N by a factor 100 and 1000 respectively. These proportions between OCS and CO as well as OCS and HC$_3$N cross sections are not strongly dependent on pressure and temperature in the ranges explored for these two parameters (see Appendix E). Therefore, these remarks suggest that the absorption induced by HC$_3$N could reasonably be neglected compared to OCS, and highlight the important CO concentrations required to mask OCS $\nu_3$ band. 
\par The detection of OCS using the $\nu_1+\nu_3$ and  $\nu_1$ bands is more conditional. As revealed by the second and third panels of figure \ref{fig12}, numerous species, mainly C2 hydrocarbons (C$_2$H$_6$, C$_2$H$_4$), ammonia and methane have higher cross sections than OCS over these two spectral ranges. The observation of OCS absorption signatures therefore requires sufficiently low abundances of these latter hydrogenated species. Nevertheless, in such conditions, the $\nu_1+\nu_3$ band can be interesting from a methodological point of view: the proximity in wavelength between the $\nu_1$ and $\nu_1+\nu_3$ bands offers the opportunity to use two different spectral windows on the same dataset and achieve a more reliable detection of OCS. 

\subsubsection{NHSO}

The detection of NHSO highlights the existence of chemical pathways which could be active in several exoplanetary environments. Our results first suggest that SO$_2$ chemistry in N$_2$-rich secondary volcanic atmospheres could be an efficient source of this molecule. Moreover, according to the formation pathway previously described, SO$_2$ reactivity in temperate H$_2$-dominated atmospheres containing nitrogen species could also favor the formation of this molecule. NHSO therefore appears as a potential interesting target to search for in H$_2$ or N$_2$ temperate exoplanetary atmospheres. 
\par In the middle infra-red, NHSO has mainly four characteristic bands potentially observable by space or ground-based telescopes \citep{joo_determination_1996}: $\nu_1$  in the range [7.7;8.3 $\mu$m]  associated to S-O stretch, $\nu_3$  in the range [8.6;9.8 $\mu$m] associated to N-S stretch, $\nu_4$  in the range [10;11.5 $\mu$m]  associated to H-N-S bend, and $\nu_6$  in the range [11.5;15 $\mu$m] associated to out-of-plane vibrations. 
\par
\begin{figure}[t]
\centering
\includegraphics[width=0.45\textwidth]{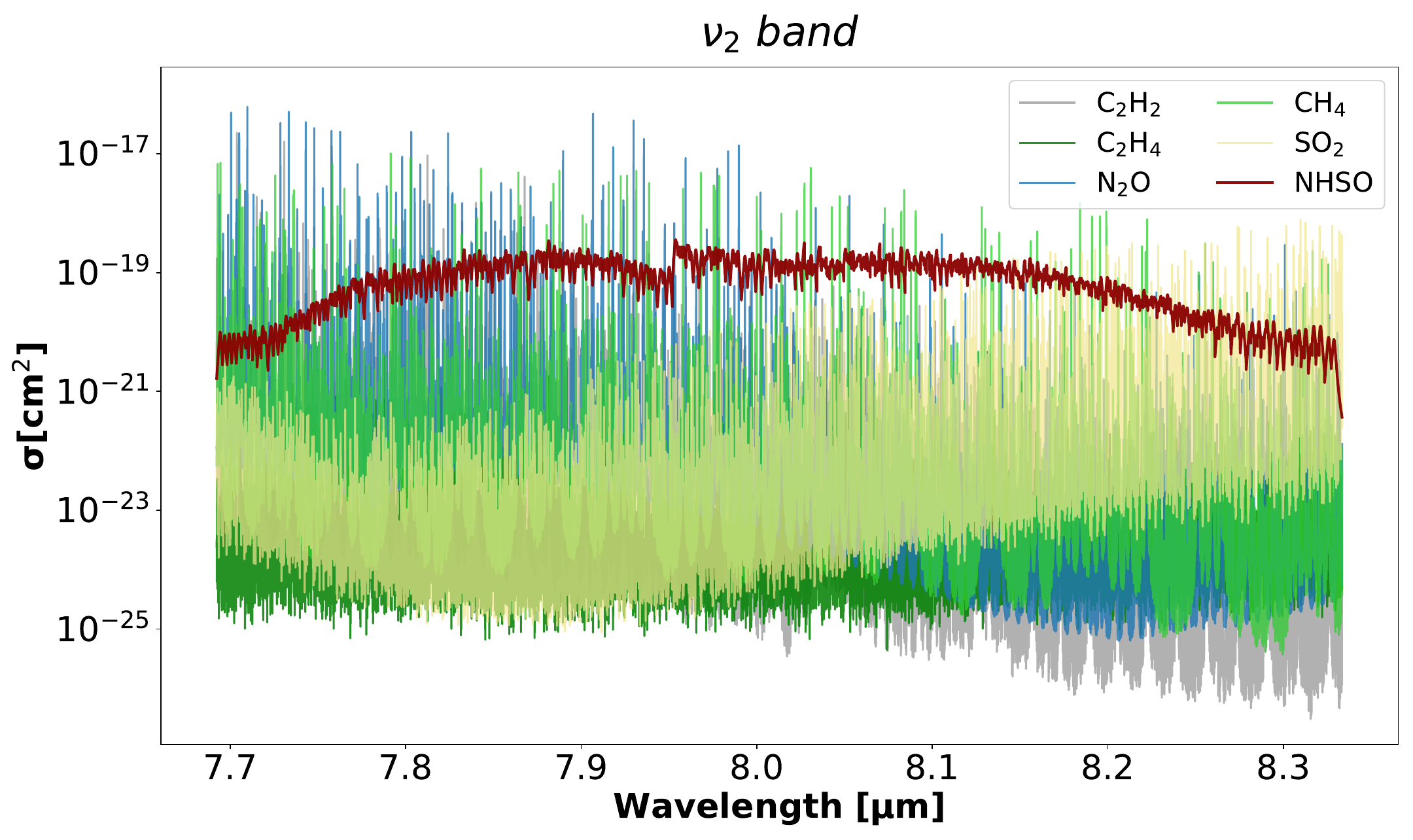}
\includegraphics[width=0.45\textwidth]{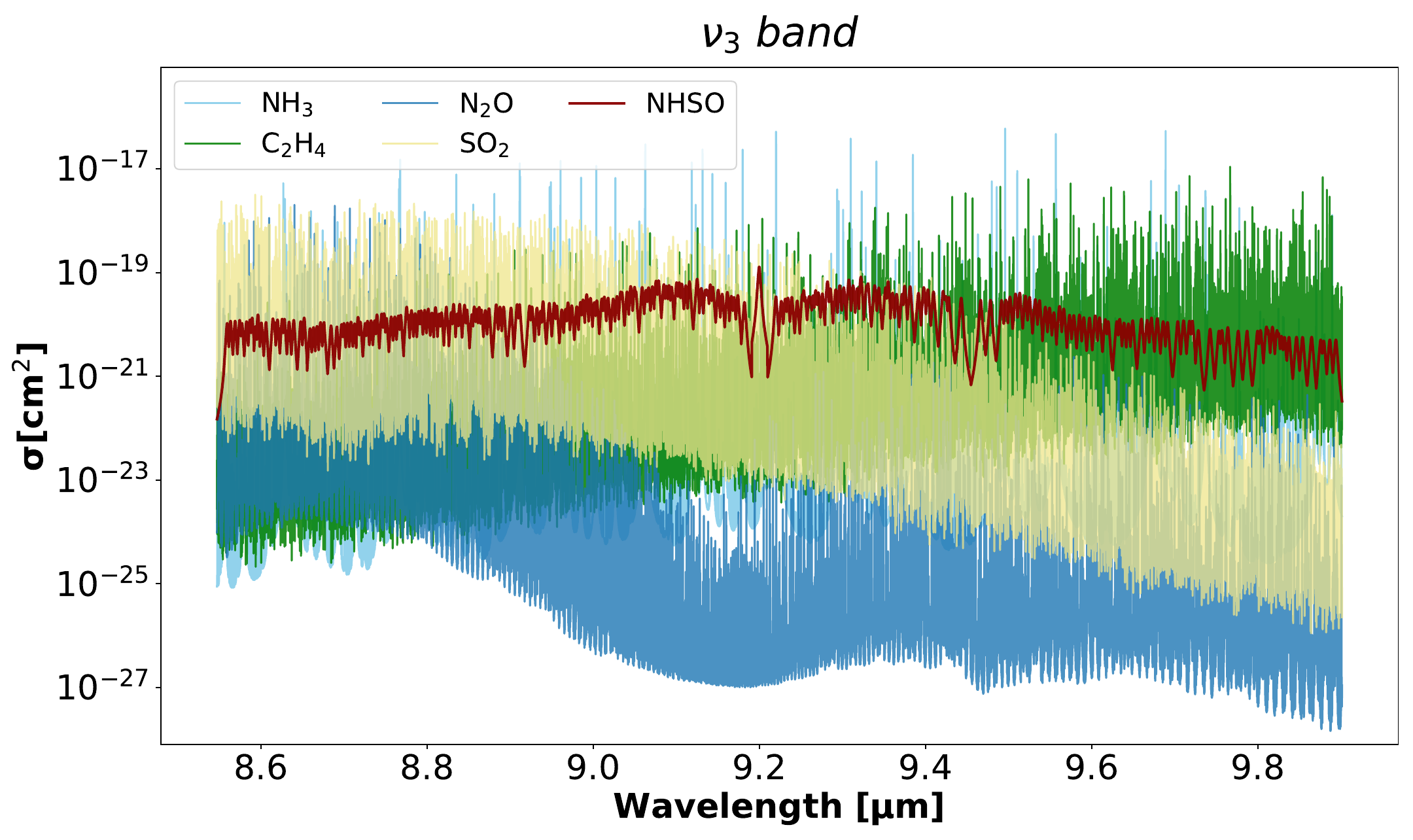}
\includegraphics[width=0.45\textwidth]{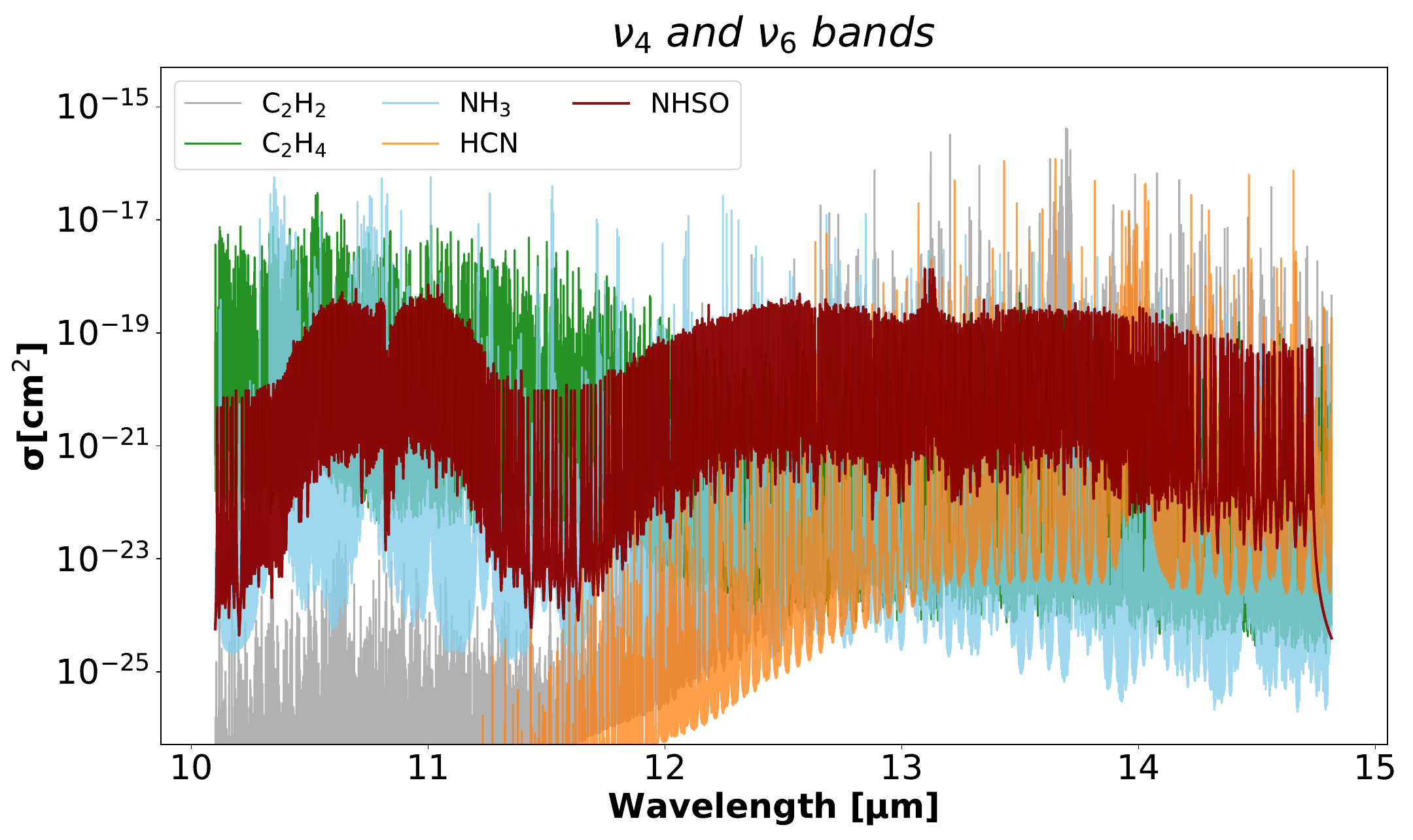}
\caption{NHSO absorption cross sections compared with the strongest volatile absorption cross sections in the spectral range of for four observable NHSO bands.}
\label{fig13}
\end{figure}

 As illustrated in figure \ref{fig13}, several volatile molecules exhibit spectral features that overlap with NHSO signatures, thereby complicating the unambiguous detection of NHSO at first order. 
\par The $\nu_2$ band of NHSO is obscured by spectral signatures of SO$_2$, N$_2$O, CH$_4$, C$_2$H$_2$ and C$_2$H$_4$. Distinguishing NHSO features from those of SO$_2$ and N$_2$O seems particularly challenging, as both species are expected to be found in atmospheres containing NHSO. SO$_2$ is a known precursor of NHSO with NH$_3$, while N$_2$O is a potential competitive product of NHSO efficiently formed in the presence of N$_2$ and SO$_2$. Regarding CH$_4$, this species is not necessarily expected to co-exist with NHSO, but it has already been detected volatiles in several exoplanetary atmospheres \citep{roy_diversity_2026,madhusudhan_carbon-bearing_2023}.
\par The $\nu_3$ band is mainly masked by two NHSO precursors, NH$_3$ and SO$_2$, as well as N$_2$O features. As previously explained, these three species are inseparable from NHSO.
\par Finally, cross sections of $\nu_4$ and $\nu_6$ bands are inferior to those of NH$_3$, and typical products of CH$_4$ chemistry in N$_2$ atmospheres namely HCN, C$_2$H$_4$, C$_2$H$_2$. 

Moreover, taking into account the potential differences in abundance between NHSO and the molecules previously mentioned in order of magnitude, the conclusions regarding the difficulties caused by NHSO detection can be even amplified. 
\par Thus, the previous remarks \textit{a priori} exclude the possibility to detect NHSO using low spectral resolution data such as those provided by JWST. However, high resolution detection of NHSO, targeting specific NHSO transition lines remains feasible, although it requires prior spectral characterization of both line positions and broadening bands associated to the targeted exoplanetary atmosphere. 

\subsubsection{CH$_3$SH}

Three main absorption bands of CH$_3$SH have been considered as potential observation windows: the band centered around 3.5 $\mu$m associated with C-H stretching, the band at 7 $\mu$m due to C-H bending, and the band around 10 $\mu$m linked to in-plane and out-of-plane deformations of C-H bonds. 
\par
\begin{figure}[t]
\centering
\includegraphics[width=0.45\textwidth]{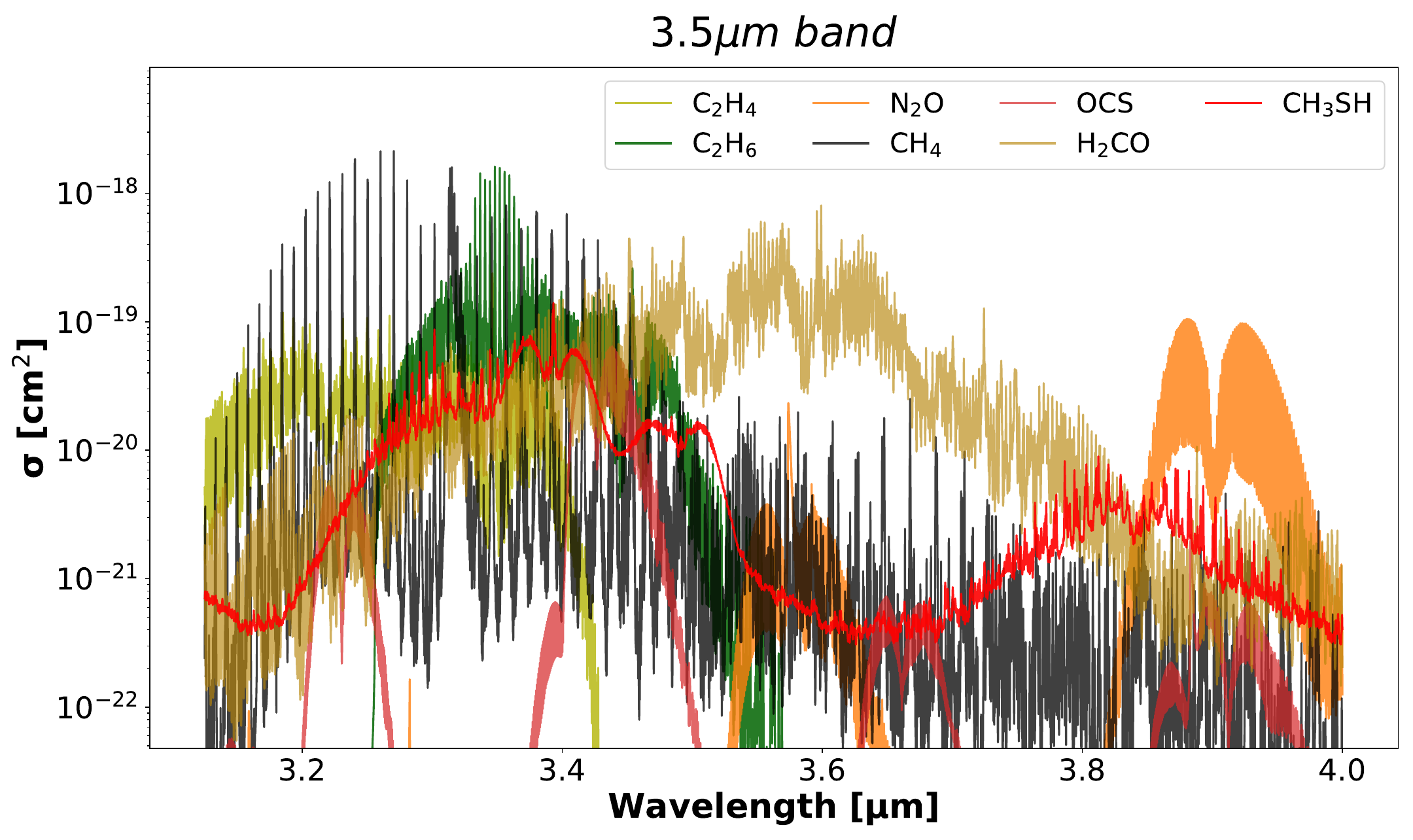}
\includegraphics[width=0.45\textwidth]{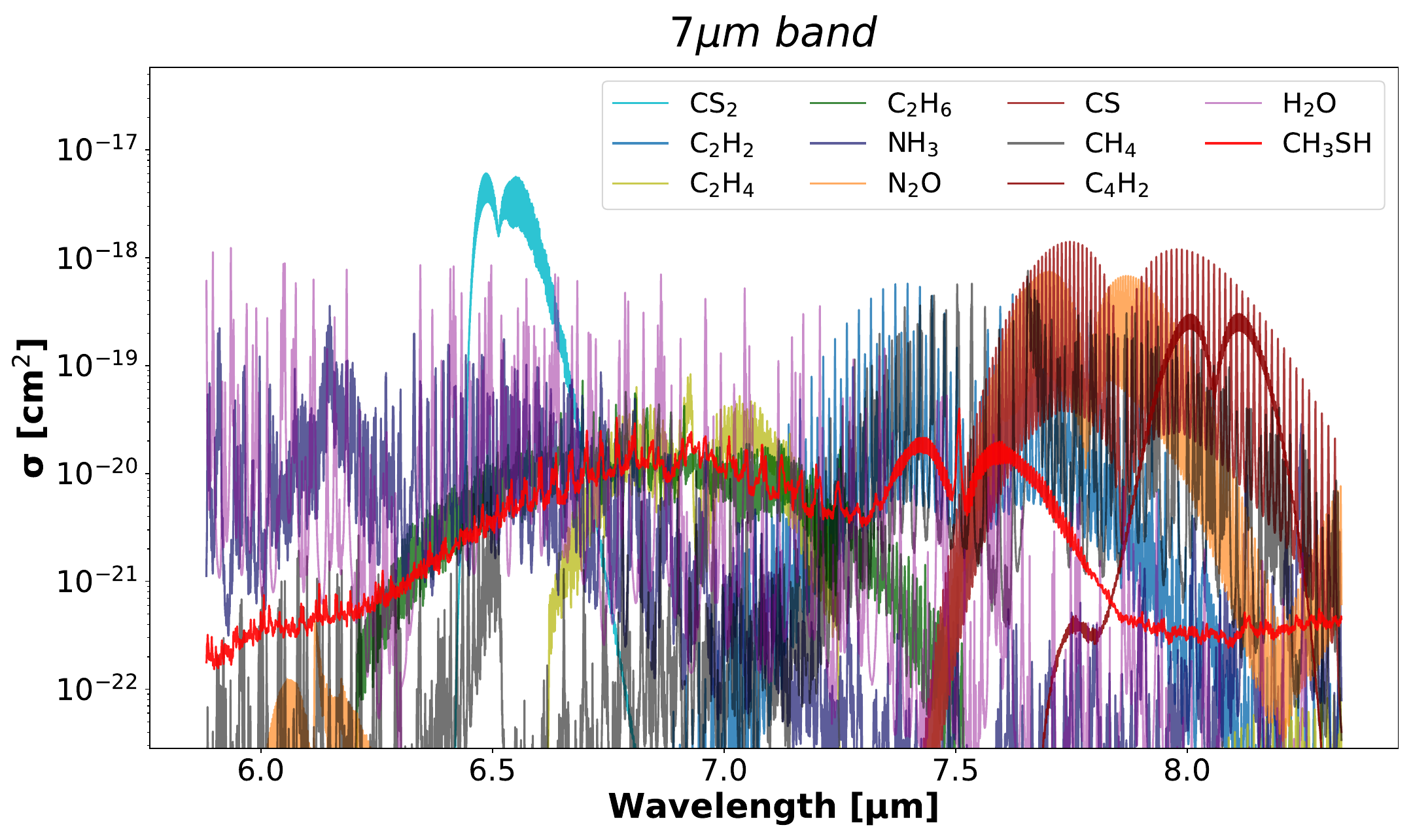}
\includegraphics[width=0.45\textwidth]{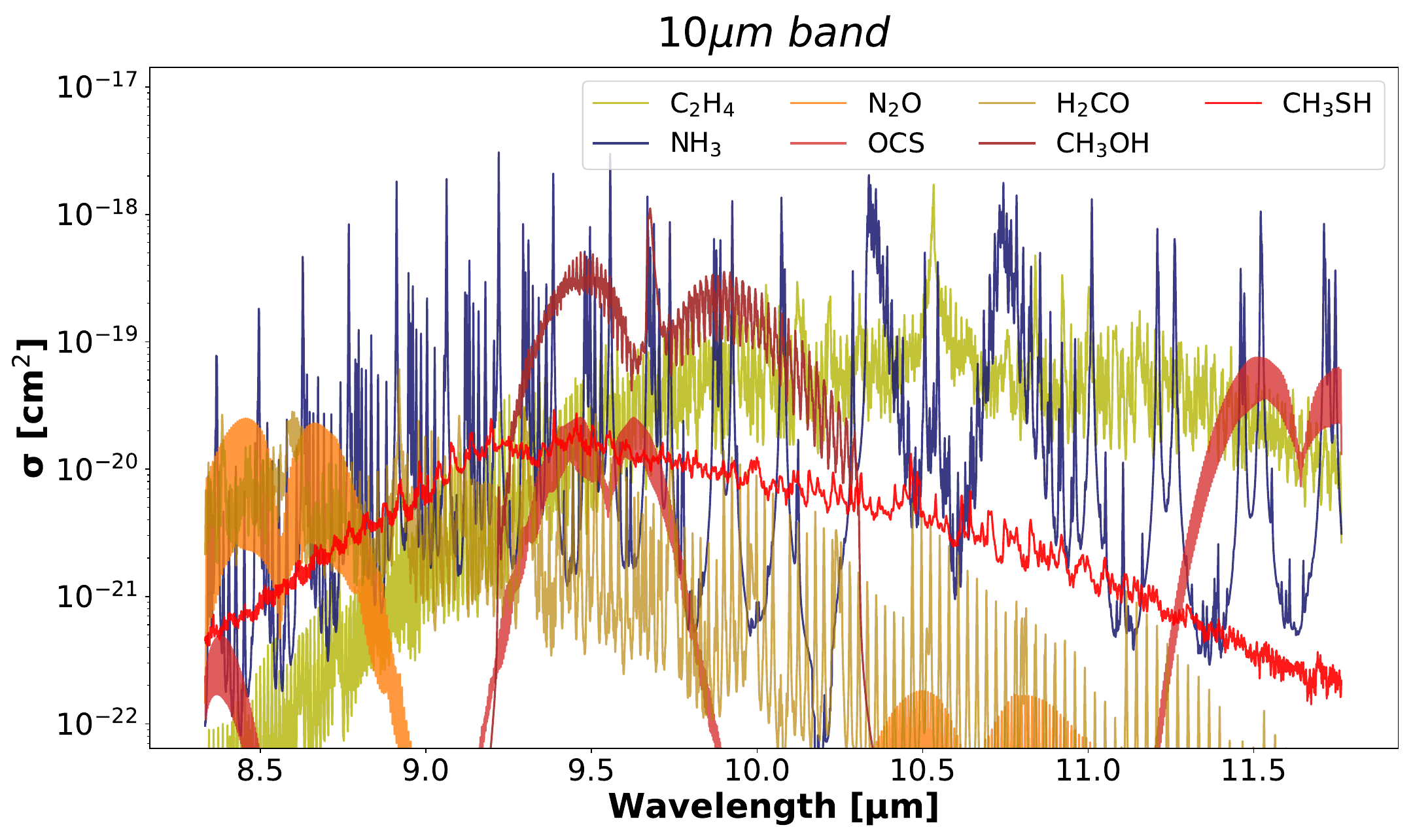}
\caption{Absorption cross sections of the three main bands of CH$_3$SH compared with those of the strongest volatile cross sections in the corresponding spectral range. For clarity, only species whose average contribution exceeds one tenth that of CH$_3$SH have been shown.} 
\label{fig14}
\end{figure}

 Figure \ref{fig14} below highlights numerous volatile compounds likely to overlap with the CH$_3$SH bands under consideration. The targeted bands are not specific to CH$_3$SH and their associated cross-sections are not sufficiently intense to significantly exceed those of other volatiles. Thus, the detection of unambiguous CH$_3$SH spectral signatures requires the absence of many volatile species.
\par The primary CH$_3$SH band around 3.5 $\mu$m is obscured by numerous volatiles including methane, C2 hydrocarbons (C$_2$H$_6$ and C$_2$H$_4$), and H$_2$CO. The cross-sections of the first band around 3.4 $\mu$m  are one order of magnitude lower than those of methane and ethane. Only the second band near 3.8 $\mu$m exhibits cross-sections slightly higher than methane and appears to be the least unfavorable target for observing CH$_3$SH signatures. 
\par The other two bands are similarly masked by multiple species. The presence of intense signatures from C2 hydrocarbons, sulfur-bearing volatiles such as CS$_2$ and CS, and water vapor significantly compromises the detection of the 7 $\mu$m band. For the 10 $\mu$m band, only ammonia, methanol, and C$_2$H$_4$ have stronger contributions over the considered interval. The overlap with C2 hydrocarbons and methane signatures is therefore less pronounced than with the two other bands. However, the low absolute intensity of this band (<10$^{-20}$ cm$^2$) and its important spectral width present substantial obstacles to its detection. 
\par Assuming that CH$_3$SH is formed through photochemical processes, the simultaneous presence of methane and hydrocarbons strongly constrains the detection of its main absorption bands. Methanethiol thus constitutes a sulfur photochemical product difficult to target via infrared spectroscopy. 

\subsubsection{S$_2$O}
In the absence of measured infrared absorption cross-sections for S$_2$O, the detectability of this molecule cannot be discussed as rigorously as previously done for OCS, CS$_2$, NHSO and CH$_3$SH. Nevertheless, remote detection of S$_2$O by infrared spectroscopy appears challenging at first glance given its known spectral properties. 
\par To our knowledge, only one study has characterized the gas-phase IR spectrum of S$_2$O \citep{jones_infrared_1950}, and the corresponding data are no longer accessible due to the age of the publication. This study highlights a band primarily observable in the mid-infrared region centered around 8.5 $\mu$m. However, this spectral range also contains a major absorption band of SO$_2$ (the $\nu_1$ band associated to S=O stretching mode). As our results have shown, SO$_2$ is a precursor of S$_2$O, and important SO$_2$ concentrations are required to form S$_2$O. That is why, this spectral overlap between S$_2$O and its photochemical precursor SO$_2$ severely compromises the detection of S$_2$O at low resolution. Only higher-resolution characterization might potentially resolve these degeneracies, but this would require high-resolution spectra of both SO$_2$ and S$_2$O under conditions representative of the targeted exoplanetary atmosphere.

\section{Conclusions}
This study experimentally explored the gas-phase reactivity of sulfur in temperate N$_2$-dominated exoplanetary atmospheres using plasma discharges. Several mixtures were considered to specifically target the coupling between nitrogen (N$_2$) and sulfur (in the form of H$_2$S or SO$_2$) as well as the more complex reactivity involving nitrogen, sulfur and carbon (CH$_4$). The volatiles produced by this gas-phase chemistry were identified using \textit{in situ} mass spectrometry. 
\par Four carbon and sulfur-bearing species were observed: CS$_2$, OCS, CH$_3$SH, and CS. Quantification of the abundance through deconvolution of the spectra containing these four organo-sulfur volatiles revealed that CS$_2$ was the main product among these four species under nearly all the conditions examined. Moreover, products specific to SO$_2$ chemistry were observed such as H$_2$S, NHSO, or S$_2$O. The detection of NHSO and S$_2$O highlights the existence of previously unconsidered and poorly understood reaction pathways, while the observation of H$_2$S suggests efficient SO$_2$ reduction processes. 
\par Formation pathways were proposed for each volatile based on elementary reactions from photochemical databases. These processes constitute an initial set of reactions potentially relevant to establish a coherent photochemical network describing sulfur chemistry under temperate conditions. The development of this network could prove essential for the future characterization of temperate exoplanetary atmospheres. 
\par Finally, the detection via transit spectroscopy of the sulfur volatile products previously observed was addressed by comparing their mid-infrared spectral signatures with those of numerous other volatiles. Based on this approach, the detection of several molecules such as NHSO, S$_2$O, or CH$_3$SH, appears challenging due to the significant overlaps with the signatures of certain precursors of these species. In contrast, OCS and CS$_2$ represent promising targets potentially detectable at low resolution. More precisely, the ranges centered around 6.5$\mu$m (CS$_2$ $\nu_3$ band) and 4.85$\mu$m (OCS $\nu_3$ band) are the most favorable spectral windows for detecting unambiguous signatures of these two species.

\begin{acknowledgements}
This project has received funding from the European Research Council (ERC) under the ERC OxyPlanets projects (grant agreement No. 101053033). 
We would like to thank Pr. G.Cernogora for its contribution to the development of VULCAIN reactor. We also thank T.Gautier which give us access to the MS deconvolution code.
\end{acknowledgements}

\bibliography{aa60617-26}
\bibliographystyle{bibtex/aa.bst}
\begin{appendix}
\section{\centering Electronic distribution in the plasma discharge}
\begin{figure}[ht!]
\centering
\includegraphics[width=6cm]{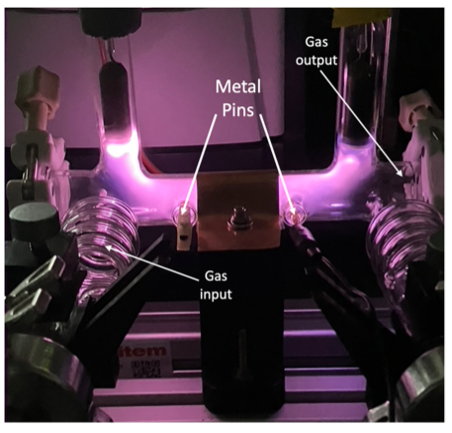}
\caption{Copy of the Vulcan reactor with the same dimensions (internal diameter, inter-electrode distance) specifically designed for electric field measurements.}
\label{figS-1}
\end{figure}

Electric field measurements were performed in a reactor with dimensions identical to those of VULCAIN, shown in figure \ref{figS-1}. This reactor has metal pins that measure the voltage in the positive column of the discharge. Since the electric field $E$ is uniform in the positive column, it is estimated thanks to the voltage between the two metal pins, $\Delta V$, and the distance between the two pins $l=50$ mm : $E=\frac{\Delta V}{l}$. \\

\begin{figure*}[ht!]
\sidecaption
\includegraphics[width=9cm]{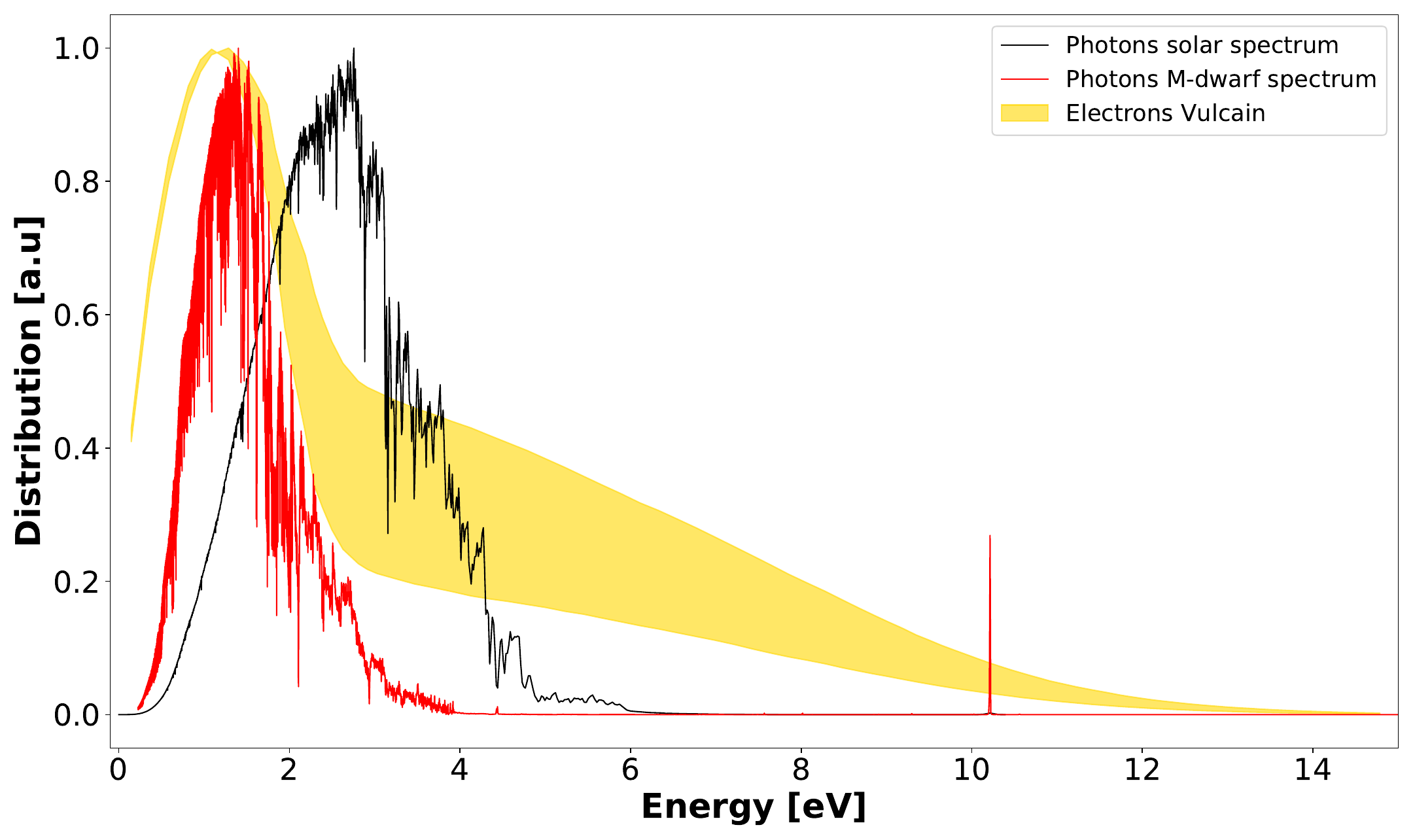}
\caption{Normalized solar and red dwarf GJ-436 spectra compared to Vulcain electron distribution. The envelope of the Vulcan distribution consists of two distributions from (Loureiro et al 1986) calculated for two different vibrational temperatures of N$_2$ molecules: $T_{vib}=4000$ K forms the lower envelope and $T_{vib}=6000$ K the upper envelope.}
\label{figS-2}
 \end{figure*}

The determined value of the electric field $E$ is then used to estimate the $E/N$ ratio with $N$ the total density of neutral species (in cm$^{-3}$). $N$ is calculated based on the ideal gas law at reactor pressure $P_r=1$ mbar, and the gas temperature inside the plasma $T_g=400$ K: $N=\frac{P_r}{k_b \times T_g}$. Finally, the energetic distribution in the VULCAIN discharge has been estimated based on theoretical calculations performed in previous modeling studies for a pure N$_2$ plasma under similar pressure conditions (1 mbar), in a cylindric reactor of similar dimensions, and for an $E/N$ ratio of 10$^{15}$ V.cm$^2$, comparable to the value measured on VULCAIN. The distribution is shown in yellow figure \ref{figS-2} and is compared with a solar and M-dwarf energetic distribution. 

\section{\centering Deconvolution validation method}
Figure \ref{figS-3} below shows the abundances retrieved for the six initial mixtures (plasma OFF) considered in the deconvolution. Species abundances are calculated relatively to N$_2$, and then converted to percentage by multiplying by N$_2$ mixing ratio which is equal to 85.5\% in each case. N$_2$ is the reference species for the deconvolution and its abundance could not be retrieved with this approach. This relative approach also works in case of measurements taken in plasma discharge (plasma ON) as N$_2$ dissociation is very low relatively to its initial abundance. The N$_2$ percentage in the discharge can therefore be considered equal to its initial value. The examples below show that the real value is not necessarily equal to the maximum of the distribution (see H$_2$S in N$_2$-CH$_4$-H$_2$S-0.1\% or CH$_4$ in N$_2$-CH$_4$-SO$_2$-1\%) but still contained in $2\sigma$ confidence interval of the distribution.
\begin{figure*}[ht!]
\centering
\includegraphics[width=17cm]{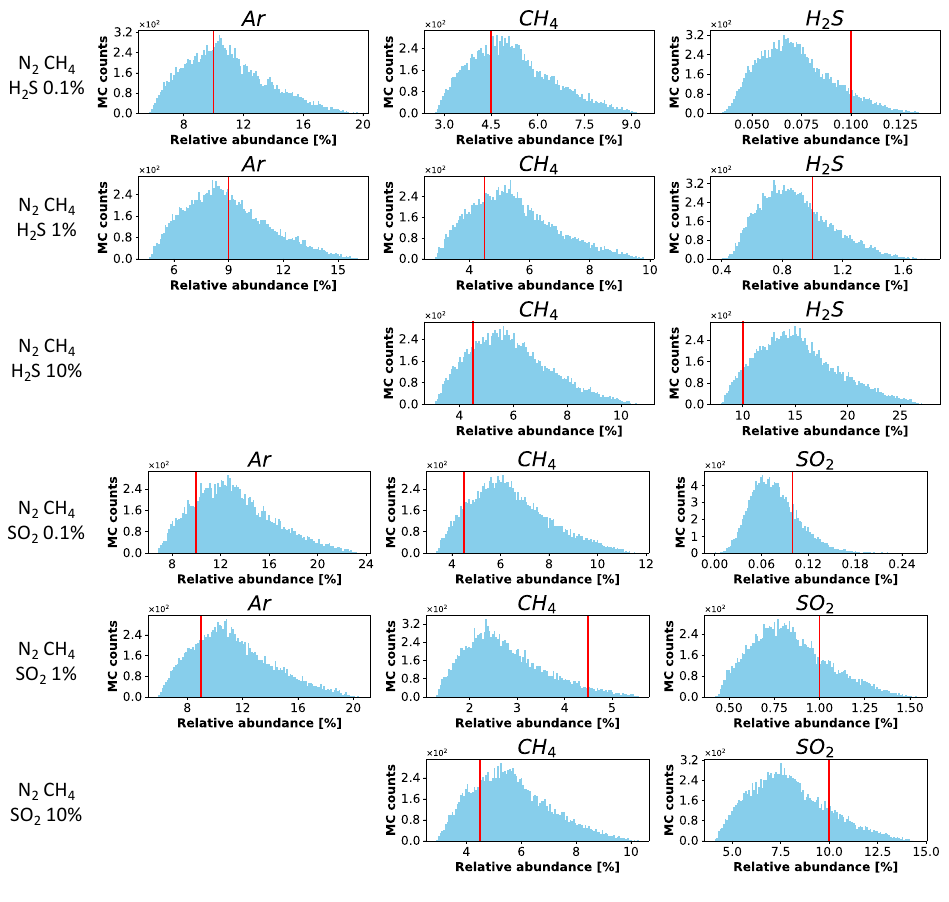}
\caption{Abundance distributions of the reactants retrieved by deconvolution compared with the real value represented by the red line.}
\label{figS-3}
\end{figure*}
\par The setting adopted for the deconvolution is the following: the number of Monte-Carlo draws is $3\times10^5$. On these draws, only 5\% of the best fits with the minimum residue are selected. The distributions are built on this restricted set of data. Two different inputs are needed for this Monte-Carlo approach: species fragmentation patterns and species sensitivities with their associated uncertainties. The sources of these fragmentation patterns and sensitivities as well as the values chosen for the uncertainties are detailed below. 

\subsection{Fragmentation patterns}

The fragmentation patterns of the reactants N$_2$, CH$_4$, Ar, SO$_2$, and H$_2$S have been measured experimentally by flushing these species into the QMS. For the other species considered in the deconvolution, CH$_3$SH, CS$_2$, OCS, and CS, their fragmentation patterns have been taken from the NIST database. Some fragmentation patterns of non-sulfur bearing volatiles included in the deconvolution are also taken from the calibration file of the INMS (Ion and Neutral Mass Spectrometer) spectrometer on board Cassini, which is considered more reliable than NIST. 
The relative error applied for the deconvolution is chosen depending on the reliability of the source or the method used to get fragmentation patterns. For reactants, the relative errors were estimated experimentally from the observed variations in their QMS fragmentation patterns. This results in the following relative errors: 15\% for N$_2$, CH$_4$, and Ar; 25\% for H$_2$S, and SO$_2$. For the products, the errors on the NIST and INMS fragmentation patterns could not be directly estimated. In the reference study describing the Monte-Carlo deconvolution method, a relative error of 30 \% is applied to NIST fragmentation patterns \citep{gautier_decomposition_2020}. In this work, since the species considered in the deconvolution, particularly sulfur species, are less common and well-referenced than those in Gautier’s article, we deliberately chose higher relative errors of 50\% for NIST species and 40\% for INMS species. 

\subsection{Sensitivites}
Sensitivities are the factors used to convert the total intensity measured for a given species into its corresponding abundance. They mainly relie on the total ionization cross section at 70ev which is taken from the NIST database. Uncertainties are applied to sensitivities to account for the errors in NIST ionization cross sections and transmission correction. A relative error of 30\% is assumed for reactants, while a higher value of 50\% is taken for sulfur products. The higher values for sulfur products are justified by their high masses, which are particularly sensitive to the transmission correction. Such values ensures sufficiently large distributions for sulfur products abundances, providing access to orders of magnitude rather than precise estimations of the different mixing ratios.
 
\section{\centering Noise and transmission corrections}

Before deconvolution, the noise contribution is removed by subtracting the normalized ON and OFF intensities for each fragment associated with products formed in the discharge. This correction cannot be applied to reactant fragments, as the normalized OFF intensity in this particular case is not representative of the noise level. However, the signals associated with the reactants are considered to be significantly higher than the noise level, which justifies why no noise correction is required. 
\par
\begin{figure}[]
\centering
\includegraphics[width=0.49\textwidth]{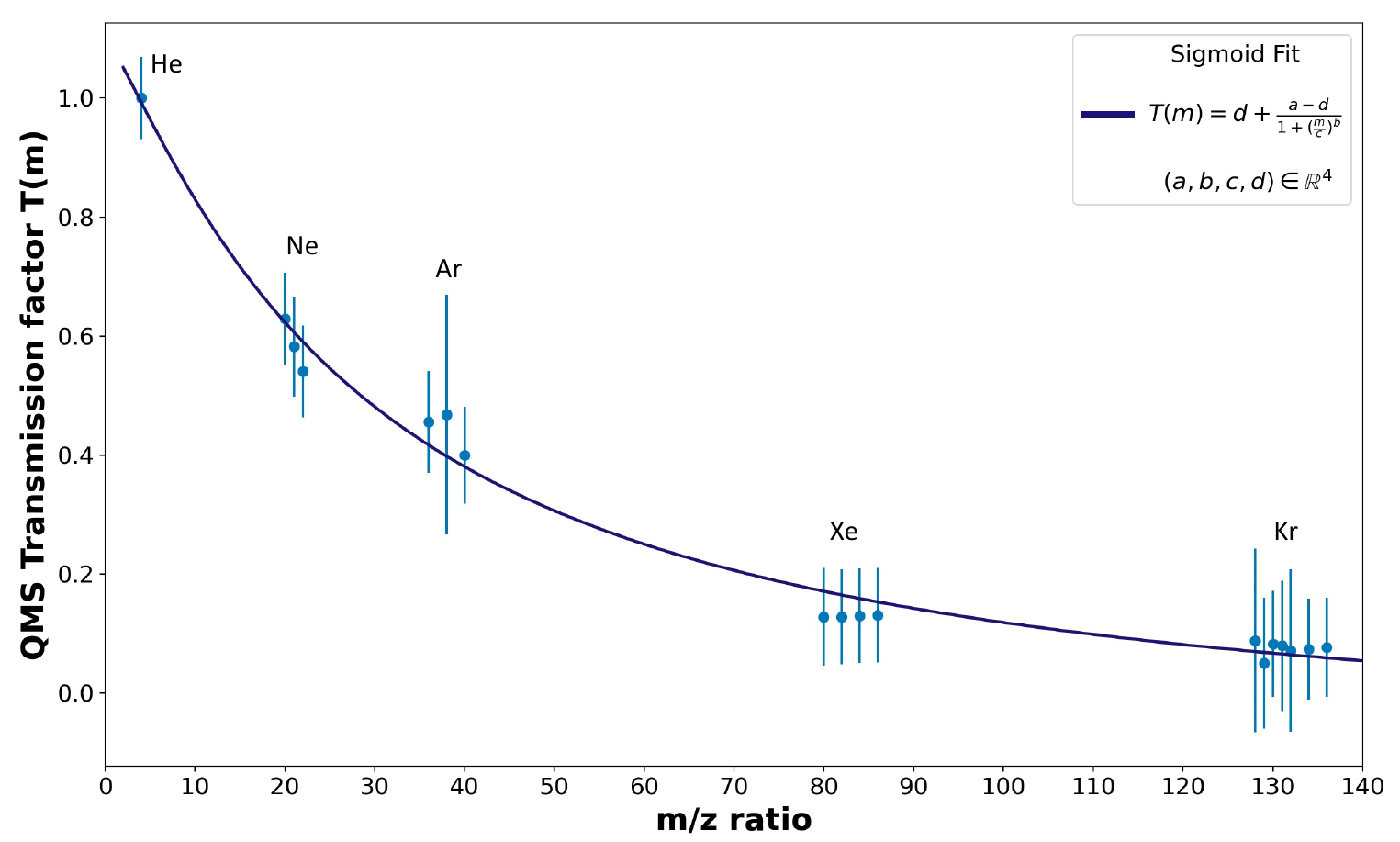}
\caption{Transmission response curve of the quadrupole as a function of mass obtained using a calibrated mixture of rare gases, He, Ne, Ar, Kr, and Xe.}
\label{figS-4}
\end{figure}

 Then, the intensities are corrected by the transmission factor which describes the signal loss at high mass over charge ratios. As sulfur species are heavy volatiles, this correction is essential to properly estimate their abundances and avoid underestimation due to these losses in the quadrupole. This transmission curve has been determined using a calibration bottle of rare gases He, Ne, Ar, Xe, and Kr in the following proportion 75\%, 15\%, 5\%, 3\%, and 2\%. For each isotope of each gas having a mass $m$, the transmission factor is defined as the ratio between the real abundance and the abundance retrieved from the intensity measured at mass $m$ and the corresponding sensitivity. The following sigmoid curve figure \ref{figS-4} is obtained to describe the intensity loss in function of the mass.

\section{\centering Deconvolution databases and deconvolution spectra}

This appendix details the choice of species used to deconvolve the MS data and show the resulting spectra. For each mixture we built the database based on the fragments observed experimentally, as detailed in tables \ref{tab:fragments:SO2} and \ref{tab:fragments:H2S}. 
\par For N$_2$-CH$_4$-SO$_2$ mixtures, the presence of oxygen significantly improves the possibilities for attribution, as the combination of C, H, N and O leads to numerous potential volatile products. Among the observed fragments, several are likely to originate from combined contributions of oxygen-bearing species and hydrocarbons or nitriles. Using mass spectrometry alone, it is difficult to distinguish the contributions of these different species with overlapping fragments. In particular, the fragment at m/z 44, which is significant in the fragmentation patterns of CS$_2$ and OCS, is likely to have contributions from both CO$_2$ and C$_3$H$_8$. In this case, the nature of the species included in the database could \textit{a priori} influence the retrieved abundances for OCS and CS$_2$.
\par For this reason, four different databases, plausible with the observed fragments, were considered for N$_2$-CH$_4$-SO$_2$ mixtures as detailed in Table \ref{table:database}. Two minimal databases were built including only reactants, sulfur products, and CO$_2$ (database n°2) or C$_3$H$_8$ (database n°3), to test the impact of the nature of the species contributing at m/z 44. Moreover, two extended databases (n°1 and n°4), comprising all the species potentially involved in the observed fragments, were also tested to explore combined contributions of species particularly at m/z 44, as well as the influence of other fragments on the deconvolution (such as m/z 17, 29, or 50-55). All these databases lead to similar results for the abundances of sulfur products such as those exposed \ref{fig9}, showing that the deconvolution is not highly dependent in this case on the nature of the species involved at certain fragments. 
This low sensitivity of the results likely originates from the absence of competing species for the major fragments of sulfur products (m/z 60 for OCS, and m/z 76 for CS$_2$) in the four databases tested.This absence is justified by the non-detection of these fragments in N$_2$-CH$_4$, and CH$_4$ reference mixtures, confirming the absence of significant contributions from non-sulfur-bearing volatiles at these masses. 
\par
\begin{figure*}[ht!]
\centering
\includegraphics[width=0.79\textwidth]{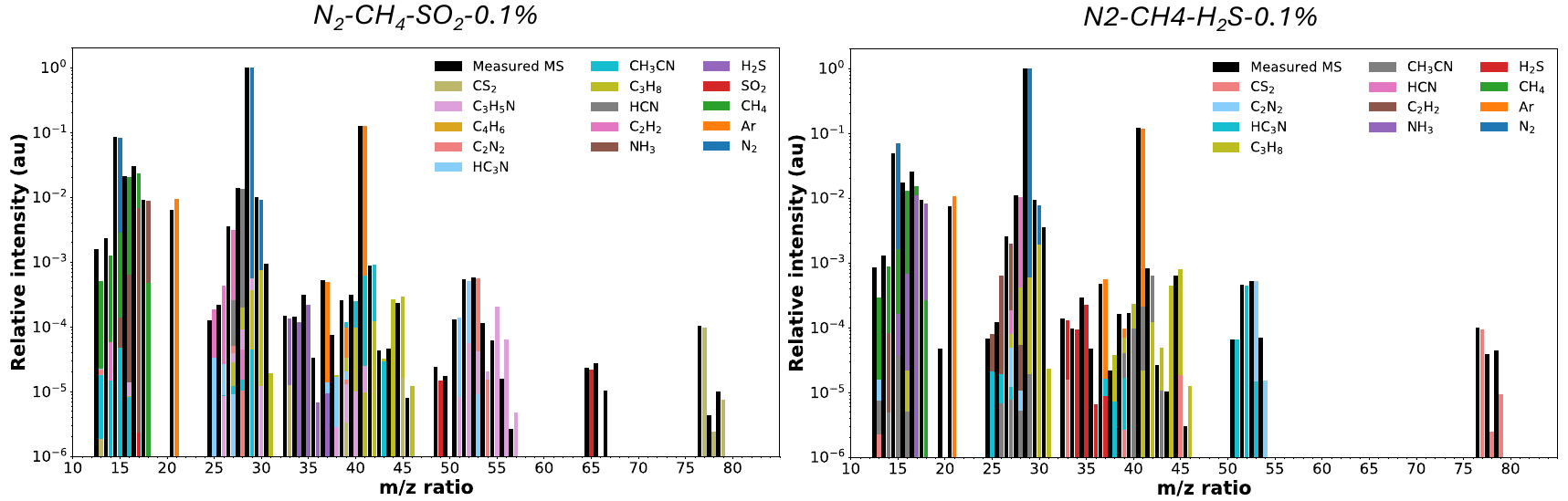}
\includegraphics[width=0.79\textwidth]{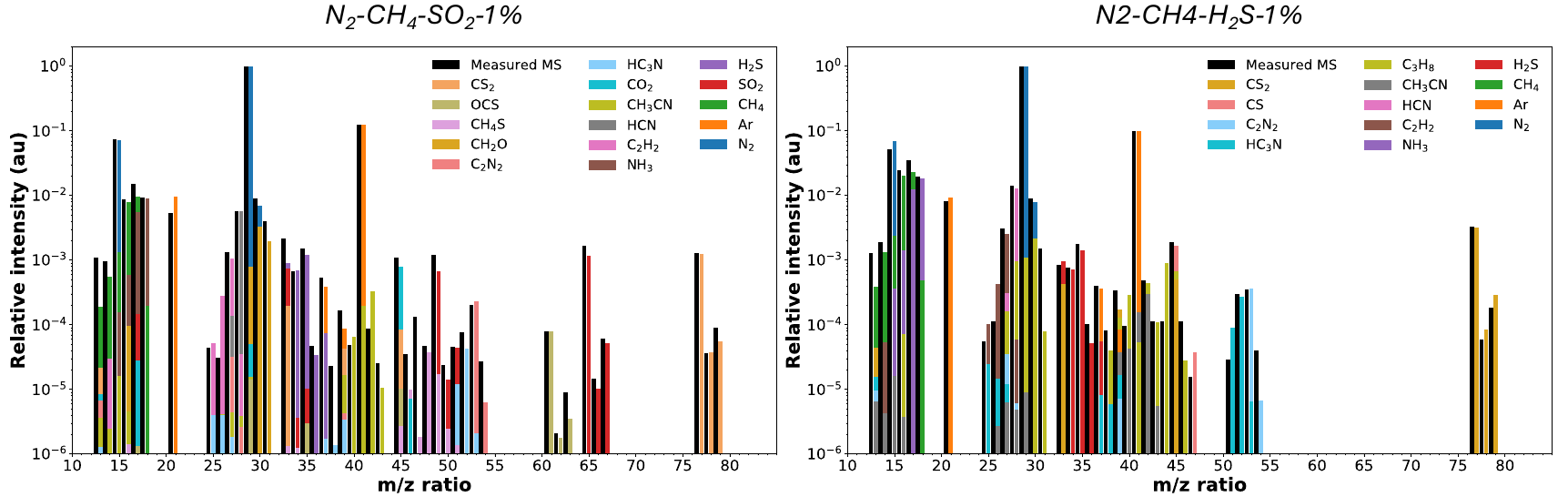}
\includegraphics[width=0.79\textwidth]{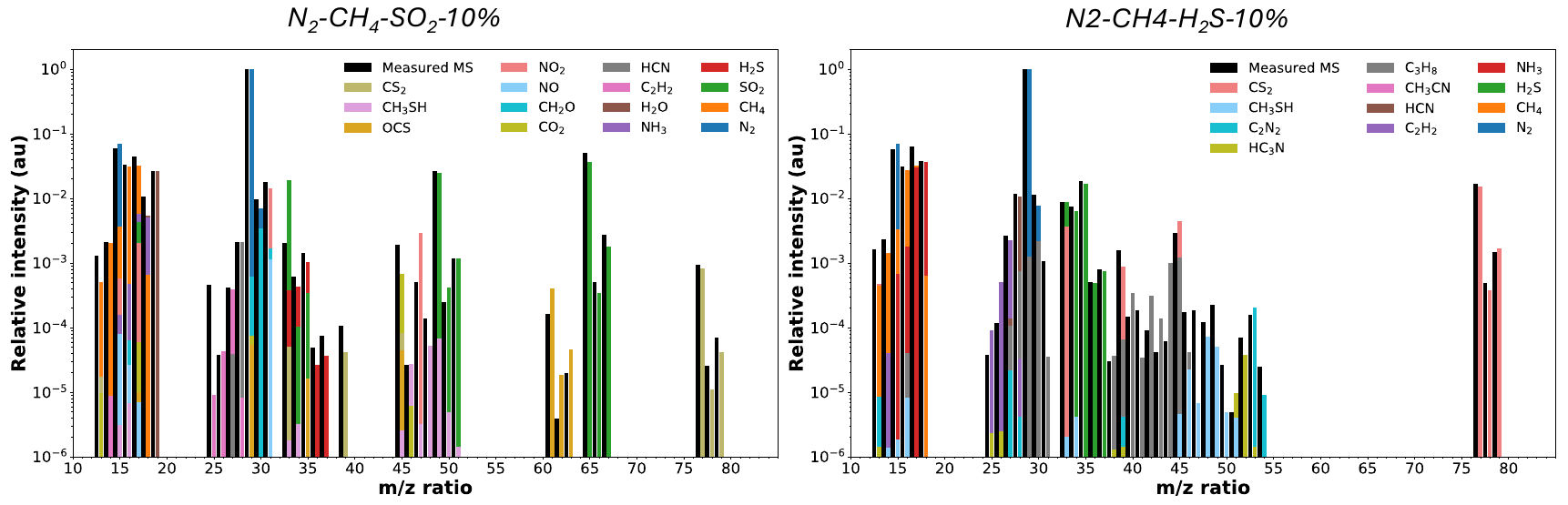}
\caption{Comparison between experimental spectrum corrected of transmission and noise (black) with synthetic spectrum retrieved from deconvolution (minimum residual) with the apparent contribution of each species (colors) for N$_2$-CH$_4$-SO$_2$ and N$_2$-CH$_4$-H$_2$S mixtures.}
\label{figS-5}
\end{figure*}

 For N$_2$-CH$_4$-H$_2$S mixtures, the combinations of possible products are more restricted in the absence of oxygen. That is why, only one database for each mixture was considered in this case, built according to the fragments observed in Table \ref{tab:fragments:H2S}. The figure \ref{figS-5} compares the deconvoluted and experimental spectra, for N$_2$-CH$_4$-SO$_2$, this deconvolution is performed in the case of database n°1.\\

\begin{table*}[ht!]
\centering
\caption{Observed fragments and possible contributors for N$_2$-CH$_4$-SO$_2$ mixtures}
\label{tab:fragments:SO2}
\resizebox{0.85\textwidth}{!}{%
\begin{tabular}{|>{\centering\arraybackslash}p{1.5cm}|>{\centering\arraybackslash}p{3.5cm}!{\vrule width 1.2pt}>{\centering\arraybackslash}p{1.5cm}|>{\centering\arraybackslash}p{3.5cm}!{\vrule width 1.2pt}>{\centering\arraybackslash}p{1.5cm}|>{\centering\arraybackslash}p{3.5cm}|}
\hline
\multicolumn{2}{|c|}{\textbf{N$_2$-CH$_4$-SO$_2$-0.1\%}} &
\multicolumn{2}{|c|}{\textbf{N$_2$-CH$_4$-SO$_2$-1\%}}&
\multicolumn{2}{|c|}{\textbf{N$_2$-CH$_4$-SO$_2$-10\%}} \\
\hline
\textit{\makecell{Observed \\ fragments}} & \textit{\makecell{Possible \\contributors}}  &
\textit{\makecell{Observed \\ fragments}} & \textit{\makecell{Possible \\contributors}} &
\textit{\makecell{Observed \\ fragments}} & \textit{\makecell{Possible \\contributors}} \\
\hline
{\footnotesize 17} & {\tiny NH$_3$} &
{\footnotesize 17}& {\tiny NH$_3$} &
{\footnotesize 17}& {\tiny NH$_3$} \\

{\footnotesize 24-27} & {\tiny HCN, C$_2$H$_2$, C$_2$N$_2$, HC$_3$N} &
{\footnotesize 24-27}  & {\tiny HCN, C$_2$H$_2$, HC$_3$N} &
{\footnotesize 18}  & \tiny {H$_2$O} \\

{\footnotesize 29}  & \tiny {CH$_2$O, C$_3$H$_8$} &
{\footnotesize 33-36}  & \tiny {H$_2$S} &
 {\footnotesize 26-27}& {\tiny HCN, C$_2$H$_2$, C$_2$N$_2$, HC$_3$N} \\

{\footnotesize 33-35} & \tiny {H$_2$S} &
{\footnotesize 37-39} & {\tiny C$_3$H$_8$, HC$_3$N, CH$_3$CN} &
{\footnotesize 29} & {\tiny CH$_2$O, C$_3$H$_8$} \\

{\footnotesize 37-39} & {\tiny C$_3$H$_8$, C$_4$H$_6$, HC$_3$N, CH$_3$CN} &
{\footnotesize 41-42}& {\tiny C$_3$H$_8$, CH$_3$CN}&
{\footnotesize 30} & {\tiny CH$_2$O, NO, NO$_2$} \\

{\footnotesize 41-42}& {\tiny C$_3$H$_8$, CH$_3$CN} &
{\footnotesize 44}& {\tiny CO$_2$, C$_3$H$_8$, CS$_2$} &
{\footnotesize 33-36} & {\tiny H$_2$S} \\

{\footnotesize 44} & {\tiny CO$_2$, C$_3$H$_8$, CS$_2$} &
{\footnotesize 47} & {\tiny CH$_3$SH} &
{\footnotesize 37-39} & {\tiny C$_3$H$_8$, HC$_3$N, CH$_3$CN} \\

{\footnotesize 50-55} & {\tiny HC$_3$N, C$_2$N$_2$, C$_4$H$_6$, C$_3$H$_5$N} &
{\footnotesize 51-53} & {\tiny HC$_3$N, C$_2$N$_2$} &
{\footnotesize 41-42} & {\tiny C$_3$H$_8$, CH$_3$CN} \\

{\footnotesize 76}& {\tiny CS$_2$} &
{\footnotesize 60} & {\tiny OCS} &
{\footnotesize 44} & {\tiny CO$_2$, C$_3$H$_8$, CS$_2$} \\

 & &
{\footnotesize 76, 78}& {\tiny CS$_2$} & 
{\footnotesize 46} & {\tiny NO$_2$, CO$_2$, CH$_3$SH} \\

 & &
 & &
{\footnotesize 47} & {\tiny CH$_3$SH} \\

 & &
 & &
{\footnotesize 51-52} & \tiny {HC$_3$N, C$_2$N$_2$} \\

 & &
 & &
{\footnotesize 60, 62} & {\tiny OCS} \\

 & &
 & &
{\footnotesize 76, 78} & {\tiny CS$_2$} \\
\hline
\end{tabular}
}
\end{table*}

\begin{table*}[ht!]
\centering
\caption{Observed fragments and possible contributors for N$_2$-CH$_4$-H$_2$S mixtures}
\label{tab:fragments:H2S}
\resizebox{0.85 \textwidth}{!}{%
\begin{tabular}{|>{\centering\arraybackslash}p{1.5cm}|>{\centering\arraybackslash}p{3.5cm}!{\vrule width 1.2pt}>{\centering\arraybackslash}p{1.5cm}|>{\centering\arraybackslash}p{3.5cm}!{\vrule width 1.2pt}>{\centering\arraybackslash}p{1.5cm}|>{\centering\arraybackslash}p{3.5cm}|}
\hline
\multicolumn{2}{|c|}{\textbf{N$_2$-CH$_4$-H$_2$S-0.1\%}} &
\multicolumn{2}{|c|}{\textbf{N$_2$-CH$_4$-H$_2$S-1\%}}&
\multicolumn{2}{|c|}{\textbf{N$_2$-CH$_4$-H$_2$S-10\%}} \\
\hline
\textit{\makecell{Observed \\ fragments}} & \textit{\makecell{Possible \\contributors}}  &
\textit{\makecell{Observed \\ fragments}} & \textit{\makecell{Possible \\contributors}} &
\textit{\makecell{Observed \\ fragments}} & \textit{\makecell{Possible \\contributors}} \\
\hline
{\footnotesize 24-27} & {\tiny HCN, C$_2$H$_2$, C$_2$N$_2$, HC$_3$N} &
{\footnotesize 24-27}& {\tiny HCN, C$_2$H$_2$, C$_2$N$_2$, HC$_3$N} &
{\footnotesize 17}& {\tiny NH$_3$} \\

{\footnotesize 29} & {\tiny C$_3$H$_8$} &
{\footnotesize 29} & {\tiny C$_3$H$_8$} &
{\footnotesize 24-27}& {\tiny HCN, C$_2$H$_2$, C$_2$N$_2$, HC$_3$N} \\

{\footnotesize 38, 39}  & \tiny {CH$_3$CN, C$_3$H$_8$, HC$_3$N, C$_4$H$_6$} &
{\footnotesize 38, 39}  & \tiny {CH$_3$CN, C$_3$H$_8$, HC$_3$N}&
{\footnotesize 29} & {\tiny C$_3$H$_8$} \\

{\footnotesize 41, 42} & \tiny {CH$_3$CN, C$_3$H$_8$} &
{\footnotesize 41} & \tiny {CH$_3$CN, C$_3$H$_8$} &
{\footnotesize 38, 39}  & \tiny {CH$_3$CN, C$_3$H$_8$, HC$_3$N}\\

{\footnotesize 44} & {\tiny C$_3$H$_8$, CS, CS$_2$} &
{\footnotesize 44}& {\tiny C$_3$H$_8$, CS, CS$_2$}&
{\footnotesize 41} & \tiny {CH$_3$CN, C$_3$H$_8$} \\

{\footnotesize 50-54}& {\tiny HC$_3$N, C$_2$N$_2$, C$_4$H$_6$} &
{\footnotesize 51-53}& {\tiny HC$_3$N, C$_2$N$_2$}  &
{\footnotesize 44}& {\tiny C$_3$H$_8$, CS, CS$_2$} \\

{\footnotesize 76} & {\tiny CS$_2$} &
{\footnotesize 76, 78} & {\tiny CS$_2$} &
{\footnotesize 45, 46}& {\tiny C$_3$H$_8$, CH$_3$SH} \\

 &  &
 &  &
{\footnotesize 47}& {\tiny CH$_3$SH}\\

&  &
 &  &
{\footnotesize 51-53}& {\tiny HC$_3$N, C$_2$N$_2$}\\

 &  &
 &  &
{\footnotesize 76, 78} & {\tiny CS$_2$}\\
\hline
\end{tabular}
}
\end{table*}

\begin{table*}[ht!]
\caption{Synthesis of the different databases tested for N$_2$-CH$_4$-SO$_2$ mixtures.}                
\label{table:database}    
\centering    
\resizebox{\textwidth}{!}{%
\begin{tabular}{@{}p{3cm}p{4cm}p{4cm}p{4cm}p{4cm}@{}}   \toprule            
Mixtures & database n°$1$ & database n°$2$  & database n°$3$  & database n°$4$   \\        
\cmidrule(lr){1-5}                    
   N$_2$-CH$_4$-SO$_2$-0.1\%&
   {\tiny N$_2$, CH$_4$, Ar, SO$_2$, H$_2$S,
   NH$_3$, C$_2$H$_2$, HCN, C$_3$H$_8$, HC$_3$N,
C$_2$N$_2$, C$_4$H$_6$, CS$_2$, C$_3$H$_5$N}
    &{\tiny N$_2$, CH$_4$, Ar, SO$_2$,
    H$_2$S, CO$_2$, CS$_2$}
&{\tiny N$_2$, CH$_4$, Ar, SO$_2$, 
H$_2$S, C$_2$H$_2$, HCN, C$_3$H$_8$, CS$_2$}&{\tiny N$_2$, CH$_4$, Ar, SO$_2$, H$_2$S, NH$_3$, C$_2$H$_2$, HCN, H$_2$CO, NO,
C$_3$H$_8$, CO$_2$, CH$_3$CN, HC$_3$N, 
C$_2$N$_2$, C$_4$H$_6$, CS$_2$, C$_3$H$_5$N}\\   
\cmidrule(lr){1-5}   
   N$_2$-CH$_4$-SO$_2$-1\%&
   {\tiny N$_2$, CH$_4$, Ar, SO$_2$, H$_2$S, NH$_3$, C$_2$H$_2$, HCN, CH$_3$CN, CH$_2$O, CO$_2$, HC$_3$N, C$_2$N$_2$, CH$_3$SH, OCS, CS$_2$}
    &{\tiny N$_2$, CH$_4$, Ar, SO$_2$, H$_2$S, CO$_2$, OCS, CH$_3$SH, CS$_2$}
&{\tiny N$_2$, CH$_4$, Ar, SO$_2$, H$_2$S, C$_2$H$_2$, HCN, C$_3$H$_8$, CH$_3$SH, OCS, CS$_2$}&   {\tiny N$_2$, CH$_4$, Ar, SO$_2$, H$_2$S, NH$_3$, C$_2$H$_2$, HCN, NO, C$_3$H$_8$, CH$_3$CN, CH$_2$O, CO$_2$, HC$_3$N, C$_2$N$_2$, CH$_3$SH, OCS, CS$_2$}\\ 
\cmidrule(lr){1-5}   
   N$_2$-CH$_4$-SO$_2$-10\%& 
   {\tiny N$_2$, CH$_4$, SO$_2$, H$_2$S, NH$_3$, H$_2$O, C$_2$H$_2$, HCN, CH$_2$O, NO, CO$_2$, NO$_2$, CH$_3$SH, OCS, CS$_2$}
    &{\tiny N$_2$, CH$_4$, SO$_2$, H$_2$S, CO$_2$, OCS, CH$_3$SH, CS$_2$}
&{\tiny N$_2$, CH$_4$, SO$_2$, H$_2$S, C$_2$H$_2$, HCN, C$_3$H$_8$, CH$_3$SH, OCS, CS$_2$}
&{\tiny N$_2$, CH$_4$, SO$_2$, H$_2$S, NH$_3$, H$_2$O, C$_2$H$_2$, HCN, CH$_2$O, NO, C$_3$H$_8$, CO$_2$, NO$_2$, CH$_3$SH, OCS, CS$_2$}\\
\bottomrule                               
\end{tabular}
}
\end{table*} 

\section{\centering Pressure and Temperature dependence\\
on CS$_2$/OCS $\nu_3$ bands detectability}

The (Pressure, Temperature) space has been discretized as follows. 26 different values of temperature are considered within the range $[150,400]$ K which corresponds to an increment of 10 K between each temperature. For pressure, 4 values are considered for each decade between $10^{-3}$ and $100$ mbar. This division of the $(P, T)$ space is therefore composed of a total of 546 couples ($26\times21$). 
\par
\begin{figure*}[ht!]
\centering
\includegraphics[width=0.78\textwidth]{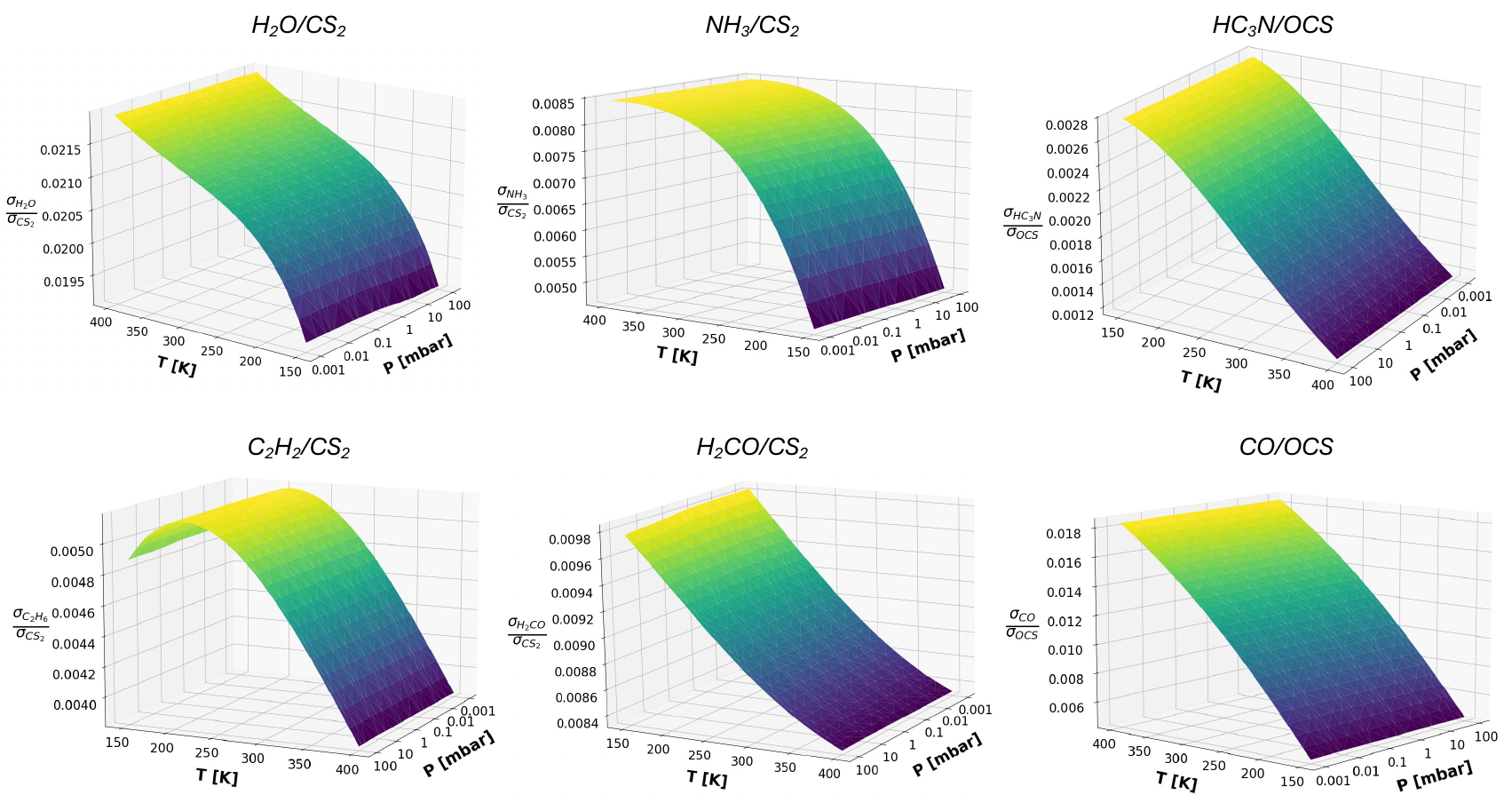}
\caption{Ratios between the average cross sections of the main absorbing volatiles over the CS$_2$ $\nu_3$ band spectral range and the average CS$_2$ cross section  as a function of pressure and temperature. Same figure for OCS $\nu_3$ band are shown in the third column.}
\label{figS-11}
\end{figure*}

 To study the evolution with temperature and pressure of the bands associated to the most absorbing species identified in the range of OCS and CS$_2$ $\nu_3$ band, we define the following ratio $r_i$ for the species $i$, expressed as: 
\vspace{0.4cm}
\\
\scalebox{1.2}{\centering $r_i (T,P)=\frac{
\int_{\nu_3} \sigma_i(\nu, T, P)\,d\nu}{\int_{\nu_3}\sigma_{ref}(\nu, T, P) \, d\nu}$}\\
\vspace{0.4cm}
\\with $\sigma_{ref}$ the cross section of the reference species OCS or CS$_2$, $\sigma_i$ the cross section of the species $i$. This quantity represents the ratio between average cross section of the species $i$ over the range of the reference species absorption band and the average cross section of the reference species (CS$_2$ or OCS) in this same interval.  In figure \ref{figS-11} are shown the evolution of this ratio as a function of pressure and temperature in a 3D space.\\
Figure \ref{figS-11} below show that for the four main absorbing species in the spectral range of the CS$_2$ $\nu_3$ band, H$_2$O, NH$_3$, C$_2$H$_6$ and H$_2$CO, their average cross section in this same interval remains several orders of magnitude below CS$_2$ average cross section whatever the value taken for pressure and temperature. The ratios between the average cross sections are similar in order of magnitude than in figure 11 for a temperature of $250$ K and a pressure of $100$ mbar. This suggests that the results highlighted figure 11 are not specific to the $(P,T)$ conditions chosen for this figure and can be extended to the integrality of the $(P,T)$ space explored here. 
\par The same conclusions can be drawn for the comparison of OCS cross sections with CO, and HC$_3$N cross sections. The two panels in the third column of figure \ref{figS-11} show that the contributions of HC$_3$N and CO over the OCS $\nu_3$ spectral range remain several orders of magnitude below the OCS cross sections as for the first panel of figure \ref{fig12}.

\newpage
\section{\centering Complete mass spectrometry dataset}
The mass spectra of the main mixtures explored in this study are shown figure \ref{figS-13} below. The SO$_2$ signatures at m/z 64 and 48 observed in H$_2$S mixtures (right column) are not likely due to a contamination source within the reactor but more plausibly to side reactions of S$_x$ in the ionization source. This interpretation is corroborated by the absence of SO$_2$ signals of this magnitude in the reactor blank (see section 2.2.1) as well as by the detection of intense SO$_2$ signals after flushing the source with oxygen-bearing volatiles such as CO$_2$ or O$_2$, which demonstrates the occurrence of S$_x$ post-oxidation reactions in the ionization chamber. The decrease observed in SO$_2$ signals after plasma ignition can then be explained by H$_2$S dissociation in the discharge limiting the S$_x$ formation in the source and, therefore, the subsequent S$_x$ side reactions.
\begin{figure*}[ht!]
\centering
\includegraphics[width=0.95\linewidth]{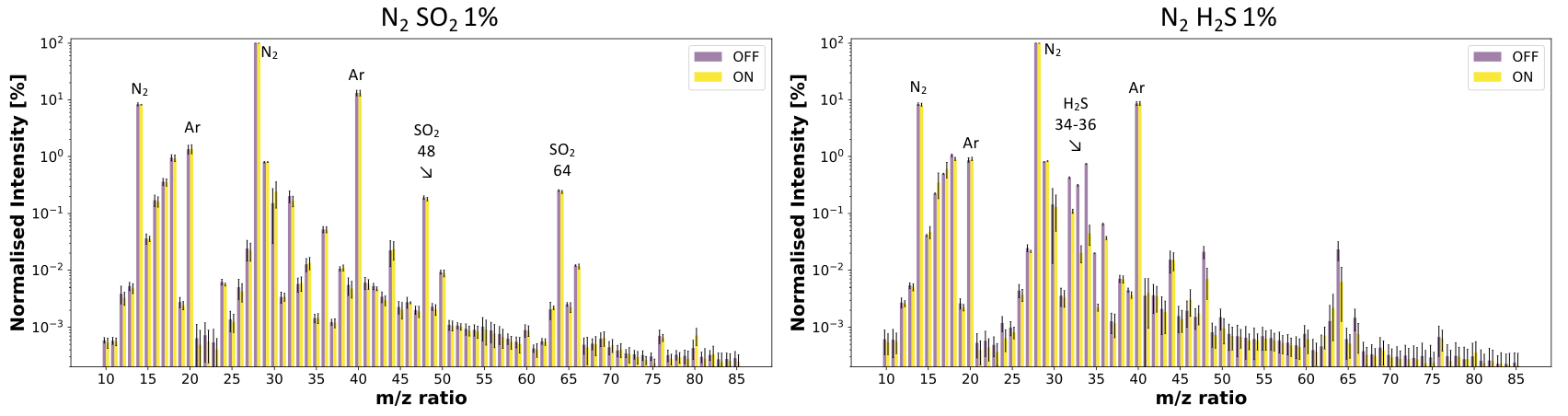}
\includegraphics[width=0.95\linewidth]{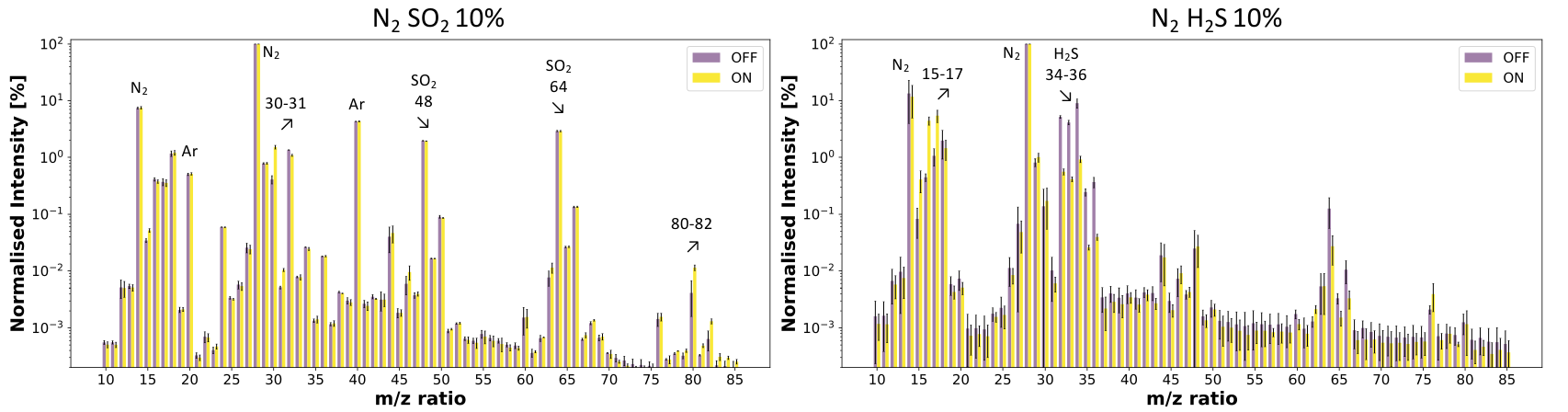}
\includegraphics[width=0.95\linewidth]{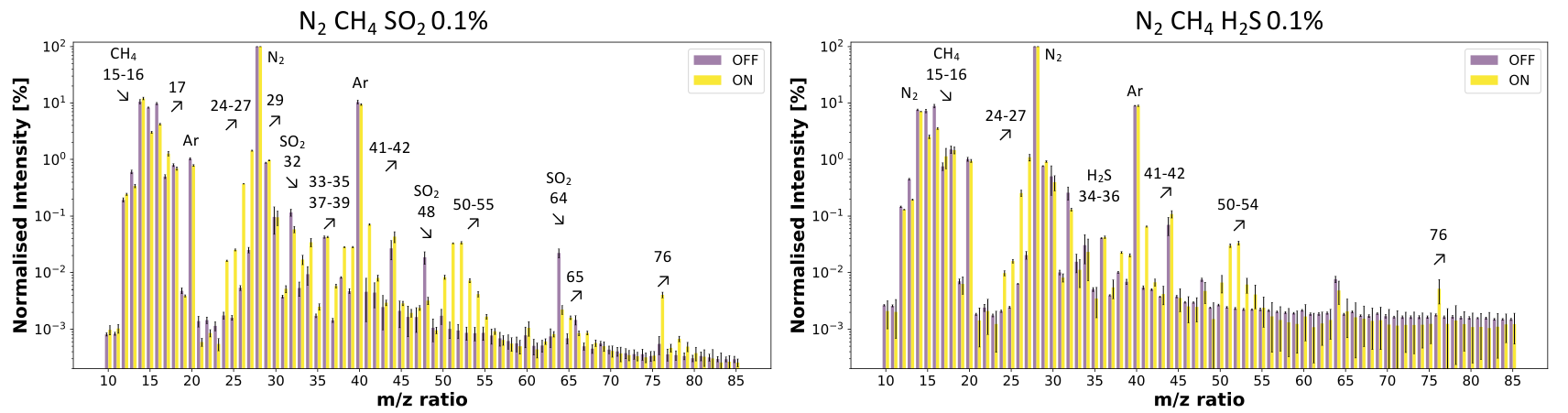}
\includegraphics[width=0.95\linewidth]{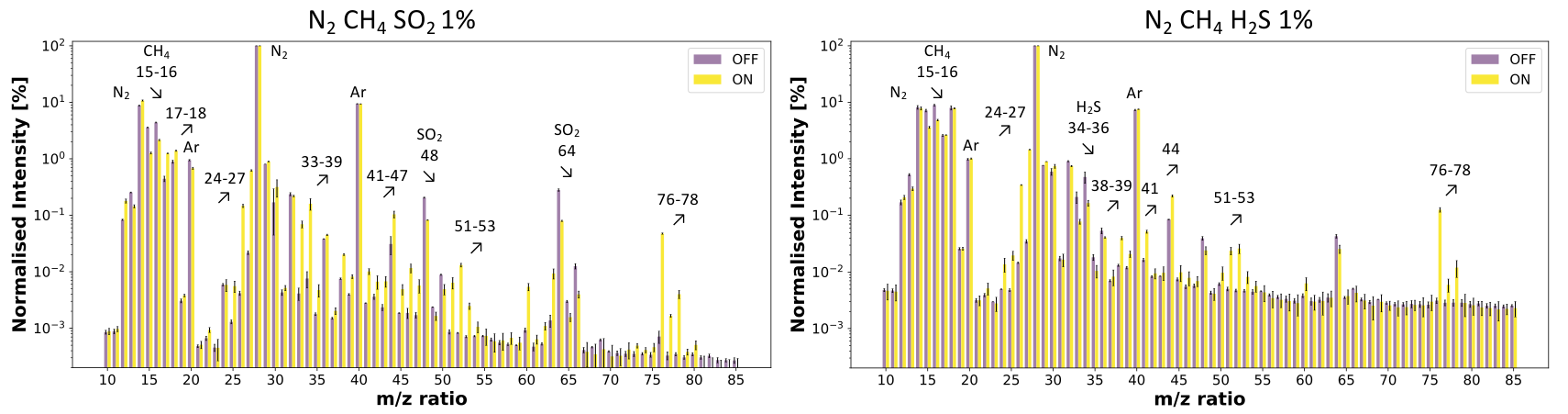}
\includegraphics[width=0.95\linewidth]{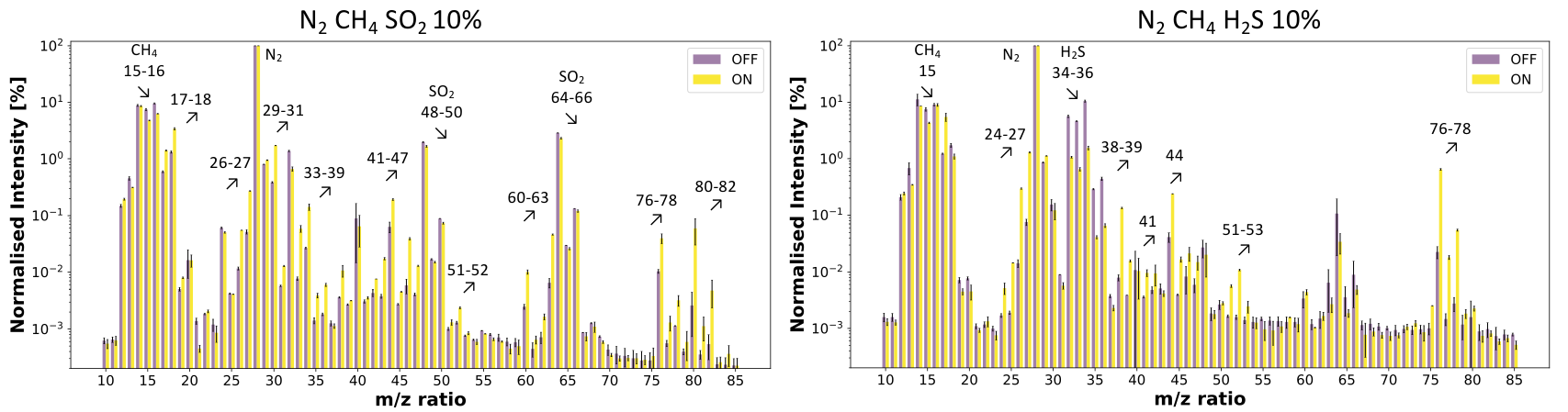}
\caption{Normalized mass spectra without additional transmission and noise corrections of the different main mixtures explored for this study (see Table 1).}
\label{figS-13}
\end{figure*}

\section{\centering Sulfur volatiles formation pathways in photochemical databases}

The table \ref{table:processes} details the neutral and ionic processes involved in the formation of CS$_2$, OCS, CH$_3$SH, and H$_2$S taken from three different reaction databases KIDA, UMIST, and NIST, as well as the photochemical model VULCAN. For OCS and H$_2$S, the pathways shown start from CH$_4$ and SO$_2$ as initial reactants. For CS$_2$ and CH$_3$SH, which are observed in both N$_2$-CH$_4$-SO$_2$ and N$_2$-CH$_4$-H$_2$S mixtures, the two cases are treated with CH$_4$ and SO$_2$ or CH$_4$ and H$_2$S as the initial reactants. 

 \begin{table*}[ht!]
\centering
\caption{CS$_2$, CH$_3$SH, OCS, and H$_2$S formation pathways}
\label{table:processes}
\resizebox{\textwidth}{!}{%
\scriptsize
\begin{tabular}[t]{l|l}
\toprule
\multicolumn{2}{c}{\textbf{CS$_2$}} \\
\midrule
\text{\makecell[l]{Neutral pathways \\from H$_2$S and CH$_4$}} & \text{\makecell[l]{Neutral pathways \\from SO$_2$ and CH$_4$}} \\
\midrule
$\text{CH}_4 \longrightarrow \text{CH}_3 + \text{H}$ &
$\text{CH}_4 \longrightarrow \text{CH}_3 + \text{H}$ \\[3pt]
$\text{CH}_4 \longrightarrow \text{CH}_2 + 2\text{H}$ &
$\text{CH}_4 \longrightarrow \text{CH}_2 + 2\text{H}$ \\[3pt]
$\text{CH}_2 + \text{H} \rightarrow \text{CH} + \text{H}_2$ &
$\text{CH}_2 + \text{H} \rightarrow \text{CH} + \text{H}_2$ \\[3pt]
$\text{H}_2\text{S} + \text{H} \rightarrow \text{SH} + \text{H}_2$ &
$\text{CH} + \text{H} \rightarrow \text{C} + \text{H}_2$ \\[3pt]
$\text{H}_2\text{S} \longrightarrow \text{SH} + \text{H}$ &
$\text{SO}_2 \longrightarrow \text{SO} + \text{O}$ \\[3pt]
$\text{SH} + \text{H} \rightarrow \text{S} + \text{H}_2$ &
$\text{SO} + \text{O} \rightarrow \text{S} + \text{O}_2$ \\[3pt]
$\text{CH}_3 + \text{S} \rightarrow \text{CH}_2\text{S} + \text{H}$ &
$\text{SO} + \text{SO} \rightarrow \text{SO}_2 + \text{S}$ \\[3pt]
$\text{CH}_2 + \text{S} \rightarrow \text{HCS} + \text{H}$ &
$\text{SO} \longrightarrow \text{S} + \text{O}$ \\[3pt]
$\text{CH}_2\text{S} + \text{H} \rightarrow \text{HCS} + \text{H}_2$ &
$\text{CH}_3 + \text{S} \rightarrow \text{CH}_2\text{S} + \text{H}$ \\[3pt]
$\text{HCS} + \text{SH} \rightarrow \text{CS} + \text{H}_2\text{S}$ &
$\text{CH}_2 + \text{S} \rightarrow \text{HCS} + \text{H}$ \\[3pt]
$\text{CH} + \text{S} \rightarrow \text{CS} + \text{H}$ &
$\text{CH}_2\text{S} + \text{H} \rightarrow \text{HCS} + \text{H}_2$ \\[3pt]
$\text{HCS} + \text{H} \rightarrow \text{CS} + \text{H}_2$ &
$\text{HCS} + \text{H} \rightarrow \text{CS} + \text{H}_2$ \\[3pt]
$\text{HCS} + \text{S} \rightarrow \text{CS}_2 + \text{H}$ &
$\text{CH} + \text{O} \rightarrow \text{CO} + \text{H}$ \\[3pt]
$\text{CS} + \text{SH} \rightarrow \text{CS}_2 + \text{H}$ &
$\text{C} + \text{SO} \rightarrow \text{CS} + \text{O}$ \\[3pt]
$\text{CS} + \text{S} + \text{M} \rightarrow \text{CS}_2 + \text{M} \:(\text{1})$ &
$\text{CH}_2 + \text{O} \rightarrow \text{CO} + 2\text{H}$ \\[3pt]
& $\text{CH} + \text{O} \rightarrow \text{CO} + \text{H}$ \\[3pt]
& $\text{CS} + \text{SO} \rightarrow \text{OCS} + \text{S}$ \\[3pt]
& $\text{S} + \text{CO} + \text{M} \rightarrow \text{OCS} + \text{M}$ \\[3pt]
& $\text{HCS} + \text{S} \rightarrow \text{CS}_2 + \text{H}$ \\[3pt]
& $\text{CS} + \text{S} + \text{M} \rightarrow \text{CS}_2 + \text{M} \:(\text{1})$ \\[3pt]
& $\text{OCS} + \text{CS} \rightarrow \text{CS}_2 + \text{CO} \:(\text{2})$ \\
\midrule
\multicolumn{2}{c}{\textbf{CH$_3$SH}} \\
\midrule
\text{\makecell[l]{Neutral pathways \\from H$_2$S and CH$_4$}} & \text{\makecell[l]{Neutral pathways \\from SO$_2$ and CH$_4$}} \\
\midrule
$\text{CH}_4 \longrightarrow \text{CH}_3 + \text{H}$ &
$\text{CH}_4 \longrightarrow \text{CH}_3 + \text{H}$ \\[3pt]
$\text{CH}_4 \longrightarrow \text{CH}_2 + 2\text{H}$ &
$\text{CH}_4 \longrightarrow \text{CH}_2 + 2\text{H}$ \\[3pt]
$\text{CH}_2 + \text{CH}_4 \rightarrow 2\:\text{CH}_3$ &
$\text{CH}_2 + \text{CH}_4 \rightarrow 2\:\text{CH}_3$ \\[3pt]
$\text{H}_2\text{S} + \text{H} \rightarrow \text{SH} + \text{H}_2$ &
$\text{H} + \text{H} \longrightarrow \text{H}_2$ \\[3pt]
$\text{H}_2\text{S} \longrightarrow \text{SH} + \text{H}$ &
$\text{SO}_2 \longrightarrow \text{SO} + \text{O}$ \\[3pt]
$\text{CH}_3 + \text{SH} + \text{M} \rightarrow \text{CH}_3\text{SH} + \text{M} \:(\text{3})$ &
$\text{SO} + \text{O} \rightarrow \text{S} + \text{O}_2$ \\[3pt]
& $\text{SO} + \text{SO} \rightarrow \text{SO}_2 + \text{S}$ \\[3pt]
& $\text{SO} \longrightarrow \text{S} + \text{O}$ \\[3pt]
& $\text{S} + \text{H}_2 \rightarrow \text{SH} + \text{H}$ \\[3pt]
& $\text{S} + \text{H} + \text{M} \rightarrow \text{SH} + \text{M}$ \\[3pt]
& $\text{CH}_3 + \text{SH} + \text{M} \rightarrow \text{CH}_3\text{SH} + \text{M} \:(\text{3})$ \\
\bottomrule
\end{tabular}
\hspace{1.5em}
\begin{tabular}[t]{l|l}
\toprule
\multicolumn{2}{c}{\textbf{OCS}} \\
\midrule
\text{\makecell[l]{Neutral pathways \\from SO$_2$ and CH$_4$}} & \text{\makecell[l]{Ionic pathways \\from SO$_2$ and CH$_4$}} \\
\midrule
$\text{CH}_4 \longrightarrow \text{CH}_2 + 2\:\text{H}$ &
$\text{N}_2 \longrightarrow \text{N}_2^+ + e^-$ \\[3pt]
$\text{CH}_4 \longrightarrow \text{CH}_3 + \text{H}$ &
$\text{N}_2^+ + \text{CH}_4 \rightarrow \text{N}_2 + \text{CH}_4^+$ \\[3pt]
$\text{H} + \text{H} \rightarrow \text{H}_2$ &
$\text{CH}_4 \longrightarrow \text{CH}_4^+ + e^-$ \\[3pt]
$\text{CH}_2 + \text{H} \rightarrow \text{CH} + \text{H}_2$ &
$\text{CH}_4 \longrightarrow \text{CH}_3 + \text{H}$ \\[3pt]
$\text{CH} + \text{H} \rightarrow \text{C} + \text{H}_2$ &
$\text{H} \longrightarrow \text{H}^+ + e^-$ \\[3pt]
$\text{SO}_2 \longrightarrow \text{SO} + \text{O}$ &
$\text{SO}_2 \longrightarrow \text{SO} + \text{O}$ \\[3pt]
$\text{SO} + \text{O} \rightarrow \text{S} + \text{O}_2$ &
$\text{CH}_4^+ + \text{H} \rightarrow \text{CH}_3^+ + \text{H}_2$ \\[3pt]
$\text{SO} + \text{SO} \rightarrow \text{SO}_2 + \text{S}$ &
$\text{CH}_4^+ + \text{O} \rightarrow \text{CH}_3^+ + \text{OH}$ \\[3pt]
$\text{SO} \longrightarrow \text{S} + \text{O}$ &
$\text{CH}_4 + \text{N}_2^+ \rightarrow \text{CH}_3^+ + \text{H} + \text{N}_2$ \\[3pt]
$\text{S} + \text{H}_2 \rightarrow \text{SH} + \text{H}$ &
$\text{CH}_4 + \text{H}^+ \rightarrow \text{CH}_3^+ + \text{H}_2$ \\[3pt]
$\text{S} + \text{H} + \text{M} \rightarrow \text{SH} + \text{M}$ &
$\text{CH}_3^+ + \text{SO} \rightarrow \text{OCSH}^+ + \text{H}_2$ \\[3pt]
$\text{CH}_2 + \text{O} \rightarrow \text{CO} + 2\:\text{H}$ &
$\text{OCSH}^+ + e^- \rightarrow \text{OCS} + \text{H}$ \\[3pt]
$\text{CH} + \text{O} \rightarrow \text{CO} + \text{H}$ & \\[3pt]
$\text{CH} + \text{S} \rightarrow \text{CS} + \text{H}$ & \\[3pt]
$\text{S} + \text{CO} + \text{M} \rightarrow \text{OCS} + \text{M} \:(\text{4})$ & \\[3pt]
$\text{CO} + \text{SH} \rightarrow \text{OCS} + \text{H} \:(\text{5})$ & \\[3pt]
$\text{CS} + \text{SO} \rightarrow \text{OCS} + \text{S} \:(\text{6})$ & \\
\midrule
\multicolumn{2}{c}{\textbf{H$_2$S}} \\
\midrule
\text{\makecell[l]{Neutral pathways \\from SO$_2$ and CH$_4$}} & \text{\makecell[l]{Ionic pathways \\from SO$_2$ and CH$_4$}}  \\
\midrule
$\text{CH}_4 \longrightarrow \text{CH}_3 + \text{H}$ &
$\text{SO}_2 \longrightarrow \text{SO}_2^+ + e^-$ \\[3pt]
$\text{H} + \text{H} \rightarrow \text{H}_2$ &
$\text{SO}_2 \longrightarrow \text{SO} + \text{O}$ \\[3pt]
$\text{SO}_2 \longrightarrow \text{SO} + \text{O}$ &
$\text{CH}_4 \longrightarrow \text{CH}_3 + \text{H}$ \\[3pt]
$\text{SO} + \text{O} \rightarrow \text{S} + \text{O}_2$ &
$\text{H} \longrightarrow \text{H}^+ + e^-$ \\[3pt]
$\text{SO} + \text{SO} \rightarrow \text{SO}_2 + \text{S}$ &
$\text{SO}_2^+ + \text{H} \rightarrow \text{SO}^+ + \text{OH}$ \\[3pt]
$\text{SO} \longrightarrow \text{S} + \text{O}$ &
$\text{SO}_2 + \text{H}^+ \rightarrow \text{SO}^+ + \text{OH}$ \\[3pt]
$\text{S} + \text{H}_2 \rightarrow \text{SH} + \text{H}$ &
$\text{SO} + \text{H}^+ \rightarrow \text{SO}^+ + \text{H}$ \\[3pt]
$\text{S} + \text{H} + \text{M} \rightarrow \text{SH} + \text{M}$ &
$\text{SO}^+ + \text{H} \rightarrow \text{S}^+ + \text{OH}$ \\[3pt]
$\text{S} + \text{S} + \text{M} \rightarrow \text{S}_2 + \text{M}$ &
$\text{N}_2^+ + \text{S} \rightarrow \text{N}_2 + \text{S}^+$ \\[3pt]
$\text{H} + \text{S}_2 \rightarrow \text{SH} + \text{S}$ &
$\text{S} + \text{H}^+ \rightarrow \text{S}^+ + \text{H}$ \\[3pt]
$\text{H} + \text{S}_2 \rightarrow \text{HS}_2$ &
$\text{H}_2 + \text{S}^+ \rightarrow \text{HS}^+ + \text{H}$ \\[3pt]
$\text{SH} + \text{H}_2 \rightarrow \text{H}_2\text{S} + \text{H}$ &
$\text{HS}^+ + \text{OH} \rightarrow \text{H}_2\text{S}^+ + \text{O}$ \\[3pt]
$\text{SH} + \text{H} + \text{M} \rightarrow \text{H}_2\text{S} + \text{M}$ &
$\text{HS}^+ + \text{H}_2 \rightarrow \text{H}_2\text{S}^+ + \text{H}$ \\[3pt]
$\text{SH} + \text{CH}_4 \rightarrow \text{H}_2\text{S} + \text{CH}_3$ &
$\text{S}^+ + \text{H}_2 \rightarrow \text{H}_2\text{S}^+$ \\[3pt]
$\text{SH} + \text{SH} \rightarrow \text{H}_2\text{S} + \text{S}$ &
$\text{H}_2\text{S}^+ + e^- \rightarrow \text{H}_2\text{S}$ \\[3pt]
$\text{SH} + \text{HS}_2 \rightarrow \text{H}_2\text{S} + \text{S}_2$ & \\[3pt]
$\text{H} + \text{HS}_2 \rightarrow \text{H}_2\text{S} + \text{S}$ & \\
\bottomrule
\end{tabular}}
\end{table*}
 
\end{appendix}
\end{document}